\documentclass{JFM-FLM_Au}

\newcommand{\algcomment}[1]{\hfill{\small\textit{$\triangleright$ #1}}}
\usepackage{bm}
\usepackage{amsmath}
\usepackage{arydshln}
\usepackage{tabu}
\usepackage{xcolor}

\usepackage{etoolbox}

\makeatletter
\patchcmd{\NAT@citex}{%
  \hyper@natlinkstart{#1}\NAT@open#2\NAT@close\hyper@natlinkend}%
  {\hyper@natlinkstart{#1}\textcolor{.}{\NAT@open}#2\textcolor{.}{\NAT@close}\hyper@natlinkend}%
  {}{}
\makeatother

\lefttitle{M.A. Bartolo, M.A. Haider, N.A. Hill, M.S. Olufsen}
\righttitle{Journal of Fluid Mechanics}

\title{Multiscale hemodynamics model for the pulmonary arteries, arterioles, capillaries, venules and veins}

\author{Michelle A Bartolo\aff{1,2}, Mansoor A Haider\aff{1}, Nicholas A Hill\aff{3} \and Mette S Olufsen\aff{1}}

\affiliation{\aff{1}Department of Mathematics, North Carolina State University, Raleigh, NC, USA
\aff{2}Department of Ophthalmology, Harvard Medical School, Massachusetts Eye and Ear Infirmary, Boston, MA, USA
\aff{3}School of Mathematics and Statistics, University of Glasgow, Glasgow, G12 8QQ, UK}

\corresau{Mette S Olufsen, \email{msolufse@ncsu.edu}}

\begin{document}
\maketitle

\begin{abstract}This study presents the first mathematical model of pulsatile hemodynamics that encompasses the complete pulmonary circulation, explicitly linking the large arteries, arterioles, capillaries, venules, and large veins. To overcome the limitations of previous models that exclude explicit capillary dynamics, we incorporate a one-dimensional structured-tree model of the pulmonary arteries and veins with a dynamic capillary sheet model. This approach establishes a recursive method for coupling the capillary sheets to the structured trees, connecting arterioles and venules in a ladder-like architecture. To evaluate the impact of incorporating this capillary structure, we compare simulated hemodynamics in a healthy control subject and a pulmonary hypertension (PH) patient. Results illustrate that including capillaries in the model significantly alters hemodynamic predictions by introducing downstream damping. In the healthy control subject, including the capillary network attenuates pulsatility, yielding the expected steady venous pressure and flow profiles, whereas omitting the capillaries produces unphysiologically high pulsatility that transmits into the venous system. The structural impact of the capillaries is even more pronounced in the PH patient, where explicitly modeling the capillary bed corrects an over-prediction in peak systolic pressure in the main pulmonary artery. Furthermore, unlike the healthy control subject, the remodeled PH microvasculature fails to isolate the venous system from arterial pulsations completely. Finally, we employ parametric analysis to investigate how specific biomechanical factors impact vascular remodeling, demonstrating the framework’s capability to quantify disease progression and severity.
\end{abstract}


\section{Introduction}\label{sec:Introduction}

Computational fluid dynamics (CFD) has emerged as a powerful tool for investigating cardiovascular hemodynamics, providing insights into disease progression that are inaccessible \textit{in vivo}. While three-dimensional (3D) flow models provide detailed information on local hemodynamics \citep{yang2019,botnar2000}, they are computationally intensive and therefore limited in application. One-dimensional (1D) models offer a much faster alternative that can readily predict hemodynamics across hundreds of vessels by averaging variables over the vascular cross-section \citep{olufsen2000,bartolo2022,richardson2024}. Importantly, 1D models demonstrate strong agreement with their 3D counterparts while offering greater computational efficiency \citep{blanco2018}. This study focuses on 1D pulmonary circuit models, which have been widely used to predict pressure and flow in the pulmonary vasculature under both normotensive and pathological conditions, e.g., \citep{bartolo2022,clark2010,clipp2012,colebank2021,ebrahimi2022,kachabi2024,kozitza2024,mynard2015}.

\begin{figure}
\centerline{\includegraphics[width=\textwidth]{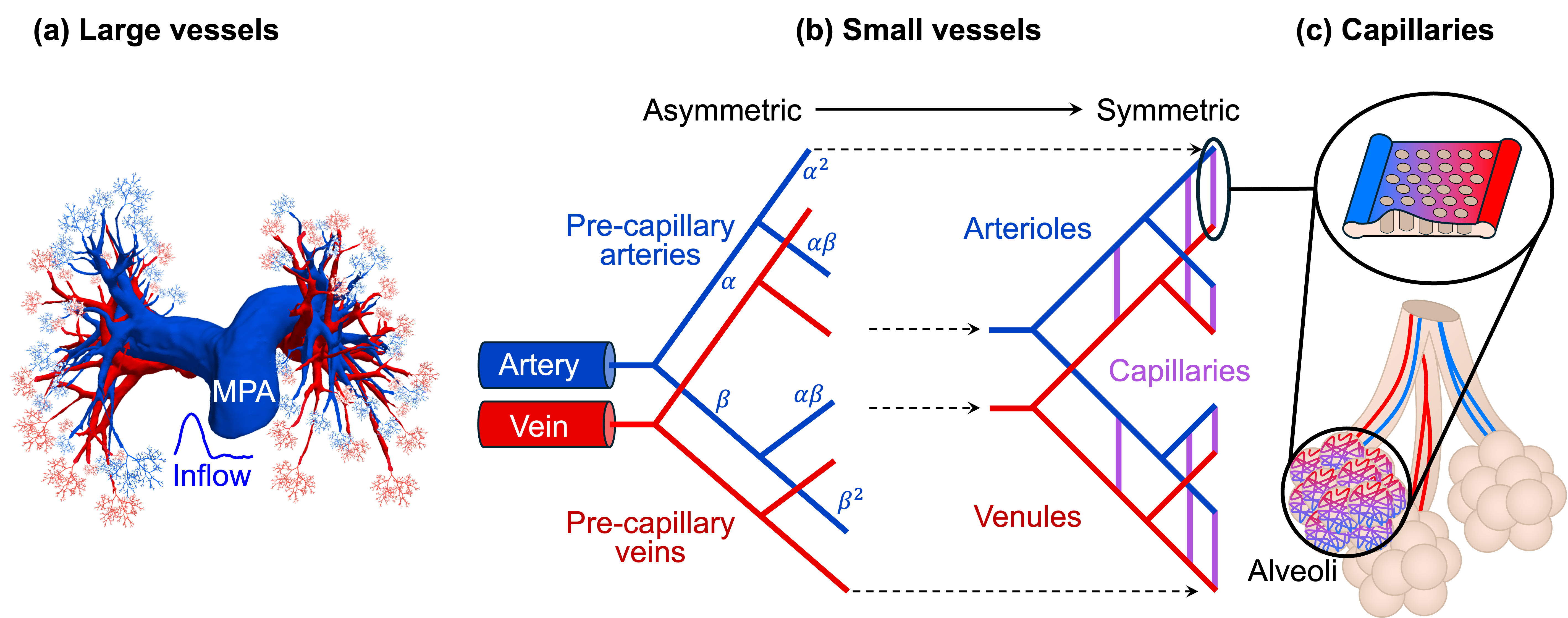}}
\caption{Overview of the multiscale pulmonary vascular model. (a) Large arteries and veins with geometry extracted from a computed tomography (CT) image. (b) Small vessels represented by structured trees, featuring asymmetric branching at the proximal level that shifts to symmetric branching where the capillaries bridge the network in a ladder-like manner. (c) Capillaries modeled as sheets that surround groups of alveoli. }
\label{fig:fullnetwork}
\end{figure}

The pulmonary circulation (Figure \ref{fig:fullnetwork}) is a vital low-pressure, low-resistance system that works alongside the respiratory tract to match ventilation and perfusion for efficient gas exchange \citep{townsley2011}. To achieve this balance, the network features an extensive alveolar surface area of nearly $130$ m$^2$ across roughly $10^8$ individual blood vessels \citep{weibel1984}. Anatomically, non-capillary vessels in the acinus region follow the branching pattern of the respiratory airways, eventually transitioning into a dense, interconnected capillary bed that maximizes the gas exchange area \citep{clark2010}. Hemodynamically, the network operates at a mean pulmonary arterial pressure of about 12 mmHg, which is far below the operating pressure of $100$ mmHg of the systemic arteries. Downstream, the pulmonary veins regulate capillary fluid filtration pressure before returning oxygenated blood to the left atrium to supply the rest of the body \citep{gao2005,weibel1984}. Quantifying flow resistance within this vast architecture is of significant clinical interest. While evidence suggests that the pre-capillary pressure drop occurs primarily in vessels with radii between $5$ and $150$\,$\mu$m \citep{levy2001,pries1995,zhuang1983}, the precise relative contributions of the arteries, veins and capillaries remain incompletely characterized \citep{gao2005,burton1965}, underscoring the need for comprehensive computational modeling. 

When dysfunction occurs within the pulmonary microvasculature, physiological balance is disrupted, and overall lung function is impaired. Changes in capillary permeability and contractility can lead to pulmonary hypertension (PH), a heterogeneous disease characterized by elevated blood lung blood pressure  \citep{ebrahimi2021,bartolo2022}. The most common form of PH is preceded by left heart failure (LHF), clinically known as World Health Organization Group 2 PH or PH-LHF \citep{dayeh2016}. Chronic LHF affects millions of adults worldwide, and among those, a majority will develop PH-LHF \citep{guazzi2012,ghio2001,lam2009,guglin2010,ramu2016,mozaffarian2016,world2017,national2016}.  Once PH-LHF develops, patients experience a severe decrease in quality of life and a significantly higher mortality risk, as there is no known cure and limited treatment options.

PH-LHF begins as a passive process called isolated post-capillary PH (Ipc-PH), characterized by a sustained increase in left atrial pressure that back-propagates through the venous and capillary networks to the pulmonary arteries \citep{ravi2013}. This pressure increase leads to vascular remodeling, starting in the microvasculature and progressively reaching the main pulmonary artery (MPA) \citep{ravi2013}. In some cases, Ipc-PH transitions to combined pre- and post-capillary PH (Cpc-PH), which increases pulmonary vascular resistance, which eventually leads to right ventricular failure. The transition to Cpc-PH significantly increases disease severity and worsens patient prognosis, yet the mechanobiological mechanisms driving this transition remain unknown  \citep{dayeh2016,gerges2015,miller2013}. Therefore, studying the pulmonary circulation during PH-LHF and identifying what drives the Ipc-PH to Cpc-PH shift is critical to improve patient outcomes. Biomechanical alterations within the smallest pulmonary blood vessels, such as arterioles, capillaries, and venules, are hypothesized to initiate this progression \citep{bartolo2022,allen2023}. However, directly measuring hemodynamics within these small vessels \textit{in vivo} is practically impossible. Physicians rely on right heart catheterization, which only captures pressure and flow in the largest vessels, making an alternative approach essential to characterize the microvascular forces driving PH-LHF.

Most numerical studies of the pulmonary vasculature \citep{clipp2012,colebank2021,kachabi2024,kozitza2024,qureshi2014} focus strictly on the pulmonary arteries using Windkessel \citep{kozitza2024,mynard2015} or structured-tree \citep{clipp2012,colebank2021,kachabi2024} outflow boundary conditions to simulate healthy hemodynamics \citep{clipp2012,mynard2015}, as well as disease states such as chronic thromboembolic PH \citep{colebank2021,kachabi2024}. More comprehensive approaches \citep{feng2021,bartolo2022,ebrahimi2022} use a two-sided artery-- vein structured tree framework originally based on \cite{olufsen2000} and adapted by \citet{qureshi2014}, which solves the linearized 1D Navier--Stokes equations to predict multiscale wave propagation. However, previous applications of this framework, including our own work investigating  pre- and post-capillary PH \citep{bartolo2022}, typically stop at the alveolar and venular levels. Because biomechanical changes in the pulmonary capillaries are hypothesized to drive the development of PH \citep{rafikova2019,guazzi2012}, excluding explicit capillary dynamics prevents a complete analysis of disease progression.

Integrating the capillary bed into these system-level models remains challenging. Anatomically, the dense capillary network surrounding an alveolus (Figure \ref{fig:fullnetwork}(c)) can be idealized either as a hexagonal network of short cylindrical tubes \citep{guntheroth1982,dhadwal1997,huang2001} or as a continuous sheet of fluid flowing between two membranes \citep{fung1969}. While sheet-flow models are more computationally efficient and anatomically representative \citep{sobin1970}, most studies isolate these capillary beds from the broader vasculature \citep{fung1969,dhadwal1997}. A few approaches have successfully integrated capillaries into a complete pulmonary circuit \citep{tawhai2011,clark2010,ebrahimi2022}. Still, they largely rely on steady-state blood flow models that fail to capture the temporal, pulsatile dynamics of pulmonary function. More recently, \citet{clark2018} incorporated a sheet model into the full pulmonary circulation using linear wave theory; however, this approach does not provide explicit pressure and flow predictions throughout the vasculature.

To overcome the limitations of the previous models and fully analyze pulmonary microcirculation, we have developed a comprehensive framework that incorporates a 1D structured-tree model of the pulmonary arteries and veins with a dynamic capillary sheet model. In this approach, the arterial and venous microcirculation is represented by a two-sided fractal network governed by the linearized 1D Navier--Stokes equations. To bridge these networks, we establish a recursive method for coupling the capillary sheet to structured trees, creating a novel boundary condition that connects arterioles and venules in a ladder-like architecture. To the best of our knowledge, this is the first study to combine a capillary sheet model with structured-tree boundary conditions, enabling explicit, time-varying hemodynamic predictions across the complete pulmonary circuit. We use this multiscale framework to demonstrate how biomechanical modulations within the pulmonary capillaries propagate to the upstream vasculature and contribute to PH-LHF progression.

\section{Methods}
This section outlines the development and application of our multiscale computational framework for the complete pulmonary circulation. First, we determine the physiological characteristics of the pulmonary circulation. Next, we define the geometric domains derived from medical images, structured trees, and capillary sheets. We then derive the 1D fluid dynamics model, which includes three subdomains for (i) large arteries and veins governed by the 1D Navier--Stokes equations, (ii) arterioles and venules represented by a linearized two-sided structured tree model, and (iii) capillaries modeled using sheet theory. To couple these domains, we detail a novel recursive method using admittance matrices. Finally, we evaluate our framework through a detailed parametric analysis comparing normotensive controls and PH-LHF patients.

\subsection{Physiological characteristics of the pulmonary circulation}
To establish a baseline for our computational framework, we first characterize the key physiological features that distinguish healthy controls from PH-LHF patients. Table \ref{tab:patientcharacteristics} summarizes average hemodynamics and biometric values compiled from literature \citep{boron2016,colunga2024,radchenko2020,sznajder2020,bredfelt2018}. Using these clinical averages, we define representative profiles for both the healthy and diseased states. Our model is initially solved using these baseline profiles, after which we systematically vary model parameters to investigate how targeted biomechanical changes impact PH-LHF progression.

\subsubsection{Pulmonary artery pressure}
The pathophysiological hallmark of PH is an elevated mean pulmonary arterial pressure (\textit{mPAP}) measured by right heart catheterization. PH is diagnosed when $m\!P\!A\!P > 20$ mmHg \citep{frost2019}. Typically, patients exhibit varying $m\!P\!A\!P$ values depending on disease severity, and in some cases, pulmonary pressures can reach systemic levels \citep{ghio2001,lam2009}.  As reported in several studies, pulmonary pressures can increase severely in patients with PH \citep{radchenko2020,sznajder2020,bredfelt2018,colunga2024}. Based on these literature values, we define a representative patient with PH with a diastolic pressure of $p_\text{dia} = 40$ mmHg, a systolic pressure of $p_\text{sys} = 80$ mmHg, and $m\!P\!A\!P = 50$ mmHg. For a representative normotensive control, we let $p_\text{dia} = 8$ mmHg, $p_\text{sys} = 20$ mmHg and $m\!P\!A\!P = 12$ mmHg \citep{boron2016,bartolo2022,colunga2024}.

\subsubsection{Left atrial pressure}
An increase in left atrial pressure $p_\text{la}$ (mmHg) is common in PH, particularly in PH-LHF \citep{ramu2016}. An elevated $p_\text{la}$ increases the pulmonary venous pressure, which injures the pulmonary capillary-alveolar membrane and progressively remodels the vasculature \citep{allen2023}. Although direct measurement of $p_\text{la}$ is invasive and clinically challenging, it is routinely estimated using the pulmonary capillary wedge pressure (PCWP). During right heart catheterization, a balloon-tipped catheter is wedged into a small pulmonary artery branch. Inflating the balloon temporarily blocks forward blood flow, which equalizes the pressure between the catheter tip and the left atrium. This equilibrium allows the PCWP to serve as a surrogate measurement for the pulmonary venous and left atrial pressure \citep{charalampopoulos2018}. 
Post-capillary PH is diagnosed when the mean pulmonary arterial pressure exceeds 20 mmHg and PCWP $>$ 15 mmHg. Patients with post-capillary PH are further classified by pulmonary vascular resistance (PVR), where Ipc-PH is defined by PVR $\leq$ 2 Wood units, and Cpc-PH by PVR $>$ 2 Wood units, indicating an additional pre-capillary component \citep{humbert2022}. Based on these clinical thresholds and our previous study \citep{bartolo2022}, we prescribed an average $p_\text{la}$ of $20$ mmHg for the PH patient profile and $4$ mmHg for the normotensive control subject. 

\subsubsection{Cardiac output}
Cardiac output (CO, mL/s) refer to the volume of blood pumped from the heart per unit of time, is a critical metric for evaluating the severity of PH \citep{colunga2024}. From a fluid dynamics perspective, sustained pulmonary vascular remodeling increases the downstream vascular resistance. As the right ventricle acts against this elevated pressure, the volume of blood ejected per contraction (stroke volume) progressively declines. Over time, the right ventricle cannot overcome this high resistance, reducing the system flow rate. To reflect this flow impairment, we assign the representative PH patient a reduced mean CO of 4.2 L/min. In comparison, the normotensive control maintains a healthy CO of 5.3 L/min \citep{colunga2024,boron2016,yang2019}.

\subsubsection{Cardiac cycle}
The length of the cardiac cycle, $T$, is closely tied to the drop in cardiac output during PH. When the right ventricle pumps less blood per beat, the body naturally tries to compensate by beating faster \citep{bouchery2017}. Mechanically, this faster heart rate shortens the overall cardiac cycle, leaving less resting time between pulses and changing how pressure and flow waves travel through the vascular network.  To capture this in our model, we assign the representative PH patient a shortened cycle period of $T=0.8$ s and the normotensive control $T=1$ s, corresponding to $75$ and $60$ beats per minute, respectively \citep{bouchery2017,kind2011}. However, for easier comparison in our parameter variation studies, including those on progressive disease severity, we keep the cardiac cycle length fixed at $T=1$ s.

\begin{table}
\caption{Average characteristics for normotensive control individuals and pulmonary hypertension (PH) patients.}
\centering
\begin{tabular}{c l l l }
\hline
\textbf{Metric} & \textbf{Description} & \textbf{Control} & \textbf{PH} \\
\hline
$m\!P\!A\!P$  & mean MPA BP (mmHg) & \ 12 \ \ \ \ [1,7] & 50 \ \ [2,3,4,7] \\
$p_\text{sys}$ & syst MPA BP (mmHg) & \ 20 \ \ \ \ [1,3,7]  & 80 \ \ [2,3,7]\\
$p_\text{dia}$ & diast MPA BP (mmHg) & \ 8 \ \ \ \ \ \ [1,3,7]  & 40  \ \ [3,7]\\
$p_\text{la}$ & left atrial BP (mmHg) & \ 4 \ \ \ \ \ \  [5,6] & 20 \ \ [4] \\
CO & cardiac output (L/min) & \ 5.25 \ \ [1] & 4.20  \ \ [2,6] \\
$T$ & cardiac cycle length (s) & \  1 \ \ \ \ \ \ [3] & 0.8 \ [3,8] \\ \hline
\multicolumn{4}{l}{\scriptsize 
[1]~\citet{boron2016}, 
[2]~\citet{chemla2015}, 
[3]~\citet{kind2011},} \\
\multicolumn{4}{l}{\scriptsize 
[4]~\citet{thenappan2011}, 
[5]~\citet{qureshi2013}, 
[6]~\citet{bartolo2022},}\\
\multicolumn{4}{l}{\scriptsize 
[7]~\citet{yang2019}, 
[8]~\citet{bouchery2017}}
\label{tab:patientcharacteristics}
\end{tabular}
\end{table}

\subsection{Geometric domain}
Our complete pulmonary vascular network consists of three components (Figure \ref{fig:fullnetwork}): (a) large arteries and veins with geometries extracted from medical images, (b) small vessels (arterioles and venules) modeled as structured trees, and (c) capillaries represented by continuous sheets. The structured trees branch from the terminal outlets of the large vessels, while the capillary sheets connect the arterioles and venules in a ladder-like manner. Using this multiscale framework, we evaluate acinar hemodynamics both with and without explicitly modeling the capillaries (Figure \ref{fig:capillaryschematic}).
 
\subsubsection{Large vessels}
A computed tomography image of the chest of a healthy $67$-year-old female volunteer from the Vascular Modeling Repository \citep{wilson2013} was analyzed to extract the large vessel geometry. The pulmonary arteries and veins were segmented through a combination of semi-automatic and manual methods in 3D Slicer, using the grow from seeds, threshold, paint, and scissor tools \citep{Federov12}. Image intensities were taken between $ 50$ and $ 3027$ Hounsfield units, with a voxel size of $0.68\times0.68\times 5.00$ mm$^3$. We segmented approximately 20 generations of pulmonary arteries and veins, but for this study we use only the first three arterial and first venous generations (Figure \ref{fig:ActualNetwork} and Table \ref{tab:AV_dim}). Consistent with our previous studies \citep{bartolo2022,qureshi2014,olufsen2012}, this streamlined geometry allows us to focus on model development rather than geometric extraction, which we detailed previously in \cite{bartolo2024}. The surface renderings were saved as standard tessellation language (STL) files, opened in ParaView, and converted into Visualization Toolkit (VTK) polygonal data files \citep{bartolo2024}. Finally, the renderings were processed through the Vascular Modeling Toolkit (VMTK) (\url{http://www.vmtk.org}) to generate centerlines and extract the length, radius, and connectivity of vessels in the network \citep{antiga2008,izzo2018}.

\begin{table}
\caption{Vessel network for large pulmonary arteries and veins based on image segmentation and centerline extraction.}
\centering
\begin{tabular}{l c c c c }\hline
\textbf{Vessel Name} & \textbf{Length (cm)} & \textbf{Radius (cm)} & \textbf{Parent} & \textbf{Daughters} \\ \hline
\textit{\textbf{Arteries}}  & & &\\
Main Pulmonary Artery (MPA)     & $3.58$ & $1.27$ & None & RPA, LPA\\
Right Pulmonary Artery (RPA)    & $5.58$ & $1.23$ & MPA  & RIA, RTA \\
Left Pulmonary Artery (LPA)     & $6.24$ & $1.19$ & MPA  & LIA, LTA\\
Right Interlobular Artery (RIA) & $2.25$ & $0.6$  & RPA  & RIV\\
Right Trunk Artery (RTA)        & $1.90$ & $0.8$  & RPA  & RSV \\
Left Interlobular Artery (LIA)  & $2.45$ & $0.5$  & LPA  & LIV\\
Left Trunk Artery (LTA)         & $2.01$ & $0.7$  & LPA  & LSV\\ \hline
\textit{\textbf{Veins}} & & & \\
Right Inferior Vein (RIV) & $1.22$ & $0.6$ & None &  RIA \\
Right Superior Vein (RSV) & $4.74$ & $0.8$ & None & RTA\\
Left Inferior Vein (LIV) & $1.22$ & $0.5$ & None & LIA \\
Left Superior Vein (LSV) & $2.42$ & $0.7$ & None & LTA \\ \hline
\end{tabular}
\label{tab:AV_dim}
\end{table}

\subsubsection{Small vessels}
To represent the vasculature that extends beyond the image-based large vessel domain, we define a small vessel network consisting of the downstream arterioles and venules. The geometry of these small vessels is generated using binary fractal trees connected to the outlets of the large pulmonary arteries and veins (Figure \ref{fig:fullnetwork}(b)). The structured venous tree mirrors the arterial tree, comprising the same number of branches and bifurcations; however, the veins are assigned distinct material properties, such as stiffness and length-to-radius ratio (Table \ref{tab:nominalparameters}).

Within each structured tree, a parent vessel ($p$) branches into two daughter vessels, $d_1$ and $d_2$. The vessel radii are governed by Murray's Law
\begin{align*}
r_p^\xi = r_{d_1}^\xi+r_{d_2}^\xi,
\end{align*}
along with the area ($\eta$) and asymmetry ($\gamma$) ratios relating the parent and daughter vessels
\begin{align*}
 \eta = \frac{r_{d_1}^2+r_{d_2}^2}{r_p^2}, \qquad
 \gamma = \left(\frac{r_{d_1}}{r_{d_2}}\right)^2.
\end{align*}
For vascular networks, the exponent $\xi$ typically ranges between $2.33$ (turbulent flow) and $3$ (laminar flow). As discovered by \cite{olufsen2000}, using these relationships, the daughter vessel radii can be expressed via scaling factors 
\begin{align*}
    r_{d_1} =\alpha r_p, \ r_{d_2}=\beta r_p, \ \text{or} \ r = \alpha^m \beta^n r_\text{root},
\end{align*}
where $\alpha, \beta < 1$. Here, $m$ and $n$ specify the vessel's position within the branching hierarchy, and $r_\text{root}$ is the radius of the tree's root vessel. The scaling factors $\alpha$ and $\beta$ can be derived from $\xi$ and $\gamma$ as
\begin{align}
\alpha = \left(1+\gamma^{(\xi/2)}\right)^{-1/\xi}, \qquad \beta = \alpha \sqrt{\gamma}.
\label{eq:alpha_beta}
\end{align}
As depicted in Figure \ref{fig:fullnetwork}, the structured trees bifurcate asymmetrically at the proximal pre-capillary level. As the vessels narrow toward the capillary bed, previous work by \citet{chambers2020} demonstrated that the bifurcations become nearly symmetric ($\alpha \approx \beta$). We incorporate this physiological feature by transitioning from asymmetric to symmetric branching at the level where the capillaries bridge the arterioles and venules.

Finally, as suggested by \cite{olufsen2000}, the vessel lengths are specified using a constant length-to-radius ratio
\begin{align*}
    l_{SA} = \ell_{rr}^{SA} r, \qquad l_{SV} = \ell_{rr}^{SV} r,
\end{align*}
where $l_{SA} \text{ and } l_{SV}$ are the lengths and $\ell_{rr}^{SA} \text{ and } \ell_{rr}^{SV}$ are the length-to-radius ratios of the arteries and veins, respectively. All parameter values are listed in Table \ref{tab:nominalparameters}.

\subsubsection{Capillaries}
The pulmonary acinus, the most distal portion of the airways, includes respiratory bronchioles, alveolar ducts, alveolar sacs, and alveoli, comprising an average of $9$ generations (ranging from $6$ to $12$) of symmetrically branching airways \citep{weinberger2017,haefeli1988}. Arterioles and venules closely follow the branching of the respiratory bronchioles, with capillary beds arising throughout this network. To represent this computationally, we model the capillary networks as a ladder-like structure connecting the midpoint of each arteriole to its corresponding venule \citep{clark2010}. Within this architecture, the arterioles act as a side rail of the ladder and the venules as the opposing rail. Blood flows between them across the capillary sheets, which act as the ladder rungs, enabling sequential perfusion of the acinus (Figures \ref{fig:fullnetwork}(c) and \ref{fig:capillaryschematic}). 

Unlike arteries and veins, pulmonary capillaries are densely packed and lack a distinct branching pattern. Following \citet{fung1969}, we model them as a sheet formed by two elastic membranes held apart by posts representing the septal tissue (Figure \ref{fig:fullnetwork}). Blood supplied by an arteriole flows through the sheet and drains into the corresponding venule. The sheet height is set to a single capillary diameter, and its length to the distance between the feeding arteriole and draining venule. Each sheet represents approximately $10$--$15$ alveoli supplied at one generation of the acinus \big(Figure \ref{fig:fullnetwork}(c)\big) \citep{clark2010}.  

\bigskip

\begin{table}
\caption{\small Parameter values for large vessels, small vessels, and capillaries in the pulmonary circulation for a healthy control subject and an average PH patient. These parameters were tuned to average patient characteristics from Table \ref{tab:patientcharacteristics}, using a sheet model in a ladder configuration to represent capillaries and structured trees to represent small arteries and veins.}
\centering
{\scriptsize
\begin{tabular}{l l l l l l} \hline
\textbf{Parameter} & \textbf{Description} & \textbf{Range} &\textbf{Control} & \textbf{PH} & \textbf{Reference}  \\ \hline 
\multicolumn{6}{l}{\textit{\textbf{Large vessels}}}  \\
$T$ & Cardiac cycle length (s) & -- & 1.0 & 0.8 & [1]\\
CO  & Cardiac output (L/min) &
  1.1 -- 5.3 &  5.3 &4.2 &[1]\\
LAP & Left atrial pressure & 
4 -- 20 & 4 & 10 &[1]\\
$r_{A}$ & Radius scale (N.D.) & 0.7 -- 1.2  & 1.0 & 0.8 & \\
$r_{V}$ & Radius scale (N.D.) & 0.7 -- 1.2  & 1.0 & 0.7 & \\
$k_3^{A}$ & Artery stiffness (g/cm/s$^2$) & 3.6 -- 8$\times 10^5$ & 4$\times 10^5$ & 8$\times 10^6$& [1,2,3,4]\\
$k_3^{V}$ & Vein stiffness (g/cm/s$^2$) & 3.6 -- 8$\times 10^5$ & 3.6$\times 10^5$ & 8$\times 10^6$& [1,2,3,4]\\
$\mu_L$ & Viscosity (g/cm/s) & -- & 0.032  & 0.032 & [5] \\ \hline
\multicolumn{6}{l}{\textit{\textbf{Small vessels}}}   \\
$r_{SV}$ & Radius scale (N.D.) & 0.85 -- 1.00 & 1.00 & 0.85 & \\
$r_{SA}$ & Radius scale (N.D.) & 0.85 -- 1.20 & 1.00 & 0.85 & \\
$k_1^{SA}$ & Artery stiffness (g/cm/s$^2$) &2.7 -- 6$\times10^5$& 3$\times 10^5$ & 6$\times 10^5$ & [1,2,4,6]\\
$k_2^{SA}$ & Artery stiffness (1/cm) &--& -15 & -15 & [1,2,4,6]\\
$k_3^{SA}$ & Artery stiffness (g/cm/s$^2$) & 0.9 -- 2$\times 10^5$& 1$\times 10^5$ &  2$\times 10^5$ & [1,2,4,6]\\
$k_1^{SV}$ & Vein stiffness (g/cm/s$^2$) &2.7 -- 6$\times 10^5$& 3.75$\times10^5$  & 6$\times10^5$  & [1,2,4,6]\\
$k_2^{SV}$ & Vein stiffness (1/cm) & --& -15  & -15 & [1,2,4,6]\\
$k_3^{SV}$ & Vein stiffness (g/cm/s$^2$) &0.9 -- 2$\times 10^5$& 1.25$\times10^5$  & 2$\times10^5$  & [1,2,4,6]\\
$\mu_S$ & Viscosity (g/cm/s) &--& Eq \eqref{eq:muS} & Eq \eqref{eq:muS} & [1,5] \\
$\ell_{rr}^{SA}$ & Length-to-radius ratio (N.D.) &--& 20 & 20 & [1,2,4,7,8]\\
$\ell_{rr}^{SV}$ & Length-to-radius ratio (N.D.) &--& 20 & 20 & [1,2,4,7,8]\\
$\xi$ &  Radius ratio (N.D.) & 2.3 -- 3.2& 2.76 & 2.4 & [1,2,7]\\
$\gamma$ &  Asymmetry ratio (N.D.) &--& 0.6252 & 0.6252 &  [1,2,7]\\
$\alpha$ & Radius ratio (N.D.) & Eq \eqref{eq:alpha_beta} & Eq \eqref{eq:alpha_beta}& Eq \eqref{eq:alpha_beta} & [1,2,7]\\
$\beta$  & Radius ratio (N.D.) & Eq \eqref{eq:alpha_beta}& Eq \eqref{eq:alpha_beta} & Eq \eqref{eq:alpha_beta} & [1,2,7]\\
$r_\text{min}$ & Minimum radius (cm) & -- &0.001 & 0.001 & [1,4,9]\\ \hline
\multicolumn{6}{l}{\textit{\textbf{Capillaries}}}  \\
$\hbar_0$ & Sheet height (cm) &0.001 -- 0.00035 &0.00035 & 0.00035 & [10,11,12,13]\\
$\alpha_c$ & Sheet compliance ((cm s)$^2$/g) & $1\times 10^{-9}$ -- 1$\times 10^{-7}$ &1.3$\times 10^{-8}$ &  1$\times 10^{-9}$ & 
[10,12,13,14]\\
$\mu_c$ & Capillary viscosity (g/cm/s) & -- &0.0192 & 0.0192 & [10,13,14,15,16]\\
$l_c$ &  Sheet length (cm) & --&0.1186 & 0.1186 & [12,17]\\
$\kappa$ & Sheet friction (N.D.) & 10 -- 100 &20& 20 & 
[10,13,14]\\
$r_{\textrm{ladder}}$ & Capillary start radius (cm) & 1.5 -- 50$\cdot r_{\textrm{min}}$&12$\cdot r_{\textrm{min}}$ & 12$\cdot r_{\textrm{min}}$ & [12,13,18]\\ \hline
\end{tabular}
\newline\raggedright{\scriptsize N.D.: non-dimensional}\\
\raggedright
[1]~\citet{bartolo2022}, 
[2]~\citet{colebank2024}, 
[3]~\citet{mynard2015}, 
[4]~\citet{qureshi2014}, 
[5]~\citet{pries1992}, 
[6]~\citet{paun2020}, 
[7]~\citet{chambers2020},
[8]~\citet{huang1996},
[9]~\citet{townsley2011}, 
[10]~\citet{fung2013C}, 
[11]~\citet{fung2013LT}, 
[12]~\citet{clark2018}, 
[13]~\citet{clark2010},
[14]~\citet{fung1972elast},
[15]~\citet{tawhai2011},
[16]~\citet{yen1988},
[17]~\citet{zhou2006},
[18]~\citet{haefeli1988}
}
\label{tab:nominalparameters}
\end{table}
\subsection{Fluid dynamics}
Throughout the pulmonary vasculature, we predict hemodynamics using a 1D fluid dynamics model that enforces mass conservation and momentum balance, coupled with a constitutive relationship for vessel wall elasticity. The computational framework spans three distinct domains, solving the unsteady 1D Navier--Stokes equations in the large arteries and veins, a linearized wave equation model in the small arteries and veins, and an analogy to the Stokes equations to characterize average flow through the capillary sheets. All model parameters for each domain are listed in Table \ref{tab:nominalparameters}.

\subsubsection{Large vessels}
Large arteries and veins are modeled as tubes with circular cross-sections and impermeable walls. Blood is assumed to be Newtonian and incompressible with laminar and axisymmetric flow without swirl. We compute volumetric flow, $q(x,t)$ (mL/s), blood pressure, $p(x,t)$ (mmHg) and cross-sectional area deformation, $A(x,t)$ (cm$^2$), using a 1D Navier--Stokes equations model, enforcing mass conservation and momentum balance 
\begin{align*}
    &\frac{\partial A}{\partial t} + \frac{\partial q}{\partial x}=0,\\
   &\frac{\partial q}{\partial t} +\frac{\partial}{\partial x}\left(\frac{q^2}{A}\right)+\frac{A}{\rho}\frac{\partial p}{\partial x}=-\frac{2\pi \nu R}{\delta} \frac{q}{A},
\end{align*}
where $\rho = 1.055$ g/mL is the blood density and $\nu= \mu_L/\rho$ (cm$^2$/s) is the kinematic viscosity. We assume that the fluid velocity profile, $u_x(r,x,t)$ (cm/s), has a boundary layer of thickness $\delta =\sqrt{\nu T/2\pi}$ (cm) \citep{olufsen2000} given by
\begin{align*}
     u_x = \begin{cases}
    \bar{u}_x, & \displaystyle 0 
    \leq r \leq R-\delta,\\[0.125cm]
    \bar{u}_x \dfrac{R-r}{\delta}, &  R-\delta \leq r \leq R,
    \end{cases}
\end{align*}
where $\bar{u}_x=q/A$ is the average velocity. To close this system of equations, we impose a linear elastic relationship between pressure and cross-sectional area, assuming that the vessels are thin-walled and homogeneous \citep{valdez2010}. Thus, 
\begin{align}
    p(x,t) = p_0 +\frac43 \frac{Eh}{r_0} \left( \sqrt{\frac{A}{A_0}}-1\right), \ \ \ \frac{Eh}{r_0} = k_1 e^{-k_2 r_0} + k_3\label{eq:pressurearea},
\end{align}
where $p_0$ (mmHg) is the reference pressure at which $A=A_0=\pi r_0^2$, $E$ (mmHg) is Young’s modulus in the circumferential direction, and $h$ (cm) is the wall thickness. We assume that vascular stiffness, $Eh/r_0$, is constant for large arteries and veins ($k_1=0$) \citep{wiener1966,paun2020,maloney1970} \big(Table \ref{tab:nominalparameters}\big).

Boundary conditions are imposed at the vessel inlet at $(x=0)$ and outlet at $(x=L)$. At the inlet of the MPA, we prescribe a volumetric flow waveform digitized from phase-contrast magnetic resonance imaging (PC-MRI) flow data obtained from the SimVascular repository (\citep{updegrove2017}. We use this waveform to define the physiological shape of the flow pulse, and scale its overall amplitude so that the time-averaged flow over one cardiac cycle matches our target CO for the representative normotensive control subject and PH patient. At the venous outlets (draining into the left atrium), a constant mean left atrial pressure, $p_{la}$, is specified. \\
\begin{figure}
\centerline{\includegraphics[width=0.6\textwidth]{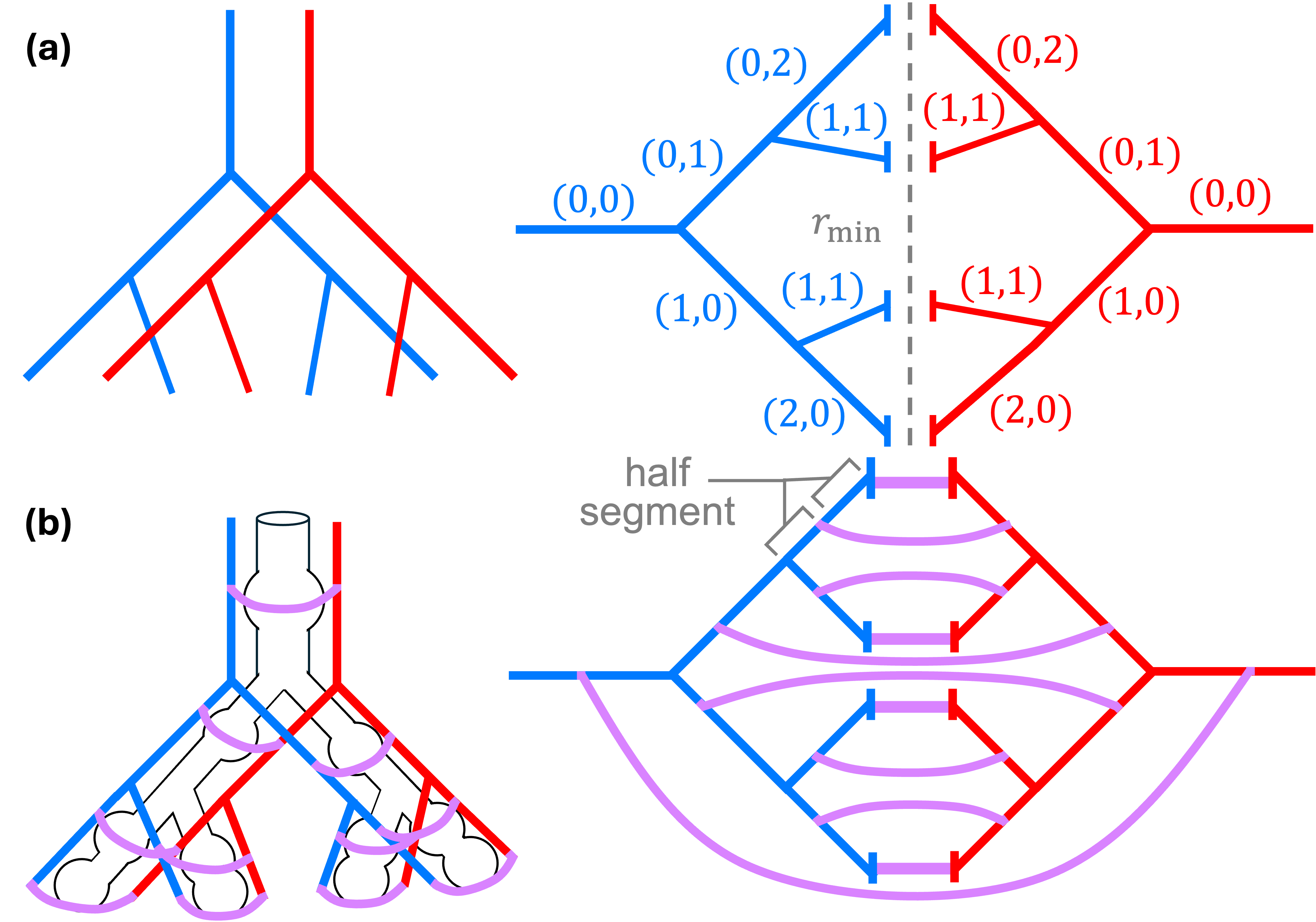}}
  \caption{Schematic of the structured tree models for arterioles (blue) and venules (red), illustrating configurations without (a) and with (b) capillaries. The left-hand diagrams depict the structural overlap of the vessels, while the right-hand diagrams display the corresponding flattened networks to map their connectivity. Panel (a) shows the isolated arterial and venous trees generated down to a prescribed minimum radius, $r_\text{min}$ (dashed line). Panel (b) incorporates capillary sheets (purple) bridging the arterial and venous networks in a ladder-like manner. Vessels are divided into ``half-segments'' so that the capillary sheets connect at the midpoint of each corresponding arteriole and venule.}
  \label{fig:capillaryschematic}
\end{figure}

\subsubsection{Small vessels}
As in the large vessel domain, we model the arterioles and venules as axisymmetric tubes with circular cross-sections with Newtonian, incompressible fluid. However, because flow velocities are significantly lower than the Moens--Korteweg wave speed, the Navier--Stokes equations in the small blood vessels can be linearized. Furthermore, since the propagating wavelengths are much greater than the small vessel radii, axial derivatives in the viscous terms are negligible compared to radial derivatives \citep{pedley1980}. Thus, the momentum balance and mass conservation equations reduce to
\begin{align*}
   &\frac{\partial u_x}{\partial t}+\frac{1}{\rho} \frac{\partial p}{\partial x} = \frac{\nu}{r}\frac{\partial}{\partial r}\left(r\frac{\partial u_x}{\partial r}\right),\\[0.125 cm]
    &C\frac{\partial p}{\partial t}+\frac{\partial q}{\partial x} = 0, \qquad C=\frac{\partial A}{\partial p}=\frac{3A_0r_0}{2Eh}\left(1-\frac{3r_0(p- p_0)}{4Eh}\right)\approx\frac{3A_0r_0}{2Eh},
\end{align*}
where $C$ is the vessel compliance derived from linearizing the pressure-area relationship \eqref{eq:pressurearea}. To solve these equations, we represent pressure and flow as periodic, discrete, frequency-domain functions, denoted $P(x,\omega_k)$ and $Q(x,\omega_k)$. For each frequency $\omega_k =2\pi k/T$, where $k \in \mathbb{Z}$, the linearized equations can be rewritten as
\begin{align}
    i\omega_k CP + \frac{\partial Q}{\partial x}&=0, \label{eq:cont}\\[0.125 cm]
 i\omega_k Q+ \frac{A_0 (1-F_J)}{\rho} \frac{\partial P}{\partial x} &=0, \qquad F_J = \frac{2J_1(w_0)}{w_0J_0(w_0)}, \label{eq:momeq}
\end{align}
where $F_J$ represents the quotient of Bessel functions of the first and zeroth orders, and $w_0^2 = i^3 r_0^2 \omega_k/ \nu$ is the non-dimensional Womersley number. 

Differentiating the linearized continuity equation (\ref{eq:cont}) with respect to $x$ and substituting the result into the linearized momentum equation (\ref{eq:momeq}) yields the wave equations
\begin{align*}
    \frac{\omega_k^2}{c^2}Q+\frac{\partial^2Q}{\partial x^2} =0\ \ \text{ or } \ \  \frac{\omega_k^2}{c^2}P+\frac{\partial^2P}{\partial x^2} =0,
\end{align*}
In this equation, $c=\sqrt{A_0(1-F_J)/\rho C}$ is the wave propagation speed. For a given axial location along the vessel, $x$, and a given frequency, $\omega_k$, the solutions are
\begin{align*}
    Q(x,\omega_k) &= a\cos(\omega_k x/c)+b\sin(\omega_k x/c),\\
    P(x,\omega_k) &= \frac{i}{g_\omega}\left(-a\sin(\omega_kx/c)+b\cos(\omega_kx/c)\right),
\end{align*}
where $a$ and $b$ are arbitrary integration constants and $g_\omega=\sqrt{CA_0(1-F_J)/\rho}$. We use $P$ and $Q$ to form an admittance matrix that relates pressure and flow at the proximal and distal ends of the vessels. The proximal flow and pressure $Q_1 = Q(0,\omega_k)$ and $P_1 = P(0,\omega_k)$, and the distal pressure is $P_2 = P(L,\omega_k)$. We define the distal flow as the flow entering the vessel at $x=L$, $Q_2 = -Q(L,\omega_k)$, so that both $Q_1$ and $Q_2$ are flows into the vessel. As reported in our previous study \cite{qureshi2013,qureshi2014}, an admittance matrix, $\bm{Y}(\omega_k)$, relates these quantities, specifically
\begin{align*}
       \begin{pmatrix}
        Q_1\\
        Q_2
    \end{pmatrix} = \bm{Y}(\omega_k)   \begin{pmatrix}
        P_1\\
        P_2
    \end{pmatrix}.
\end{align*}
When $\omega_k \neq 0$, the admittance matrix is
\begin{align}
    \mathbf{Y}(\omega_k) =  \frac{i g_\omega}{\sin\left(\omega_k L/c\right)}
    \begin{pmatrix}
         -\cos\left(\omega_k L/c\right) & 1\\
        1 &  -\cos\left(\omega_k L/c\right)
    \end{pmatrix},
     \label{eq:admittance_n0}
\end{align}
and when $\omega_k = 0,$
\begin{align}
    \mathbf{Y}(0) = \frac{\pi r^4}{8\mu_S L}
    \begin{pmatrix}
        1 & -1 \\
        -1 & 1
    \end{pmatrix}.  \label{eq:admittance_0}
\end{align}
For arterioles and venules connected at the midpoint by capillaries (the ladder architecture outlined in Section \ref{sec:multiscale} and Figure \ref{fig:capillaryschematic}), we must compute the admittance for half vessels. By calculating the admittance between $x=0$ to $x=L/2$ and $x=L/2$ to $x=L$, we equivalently obtain
\begin{align}
\mathbf{Y}_\text{half}(\omega_k) &=  \frac{i g_\omega}{\sin\left(\omega_k L/2c\right)}
    \begin{pmatrix}
         -\cos\left(\omega_k L/2c\right) & 1\\
        1 &  -\cos\left(\omega_k L/2c\right)
    \end{pmatrix}, \label{eq:admittanceHALF_n0}\\ 
    \mathbf{Y}_\text{half}(0) &= \frac{\pi r_0^4}{4\mu_S L}
    \begin{pmatrix}
        1 & -1 \\
        -1 & 1
    \end{pmatrix}. \label{eq:admittanceHALF_0}
\end{align}
The elastic modulus $E$ and wall thickness $h$ change with vessel radius and, as in (\ref{eq:pressurearea}), we model the wall-stiffness $Eh/r_0$ as an empirical exponential function
\begin{align*}
    \frac{Eh}{r_0} = k^i_1 \exp(k^i_2 r_0) + k^i_3, \ \ i=S\!A,S\!V,
\end{align*}
where $k^i_1, k^i_2,$ and $k^i_3$ are parameters defined as stated in our previous studies \citep{colebank2021,bartolo2022}. The parameter $i \in \{ S\!A, S\!V \}$ represent the small arteries and veins, respectively.

Finally, the apparent viscosity of blood becomes highly dependent on the vessel radius in the microcirculation due to the  F{\aa}hr{\ae}us--Lindqvist effect \citep{chebbi2015dynamics}. Thus, we employ a non-linear viscosity function as suggested by \citet{pries1992}
\begin{align}
    \mu_S(r_0) &= \frac{1}{3.2} \left(1+(\mu_{0.45}-1)\left(\frac{2r_0}{2r_0-1.1}\right)^2\right)\left(\frac{2r_0}{2r_0-1.1}\right), \label{eq:muS}\\
    \mu_{0.45}(r_0) &=6e^{-0.17r_0}+3.2-2.44e^{-0.12r_0^{0.645}}, \nonumber
\end{align}
where $\mu_{0.45}$ is the relative viscosity at a hematocrit of 45\%, and the constant 3.2 is the apparent viscosity limit.

\subsubsection{Capillaries}
Following the theory developed by \citet{fung1969}, we represent the capillary beds as sheets consisting of two elastic membranes separated by posts with blood flowing between them (Figure \ref{fig:fullnetwork}(c)). 
We orient the sheet such that blood enters from a feeding arteriole and drains into a venule, which run parallel to one another \citep{maloney1969,wangensteen1970}, with the $x$-axis running between them and the $y$-axis parallel to the vessels. We therefore assume the flow is uniform along the $y$-axis, so all quantities depend only on $x$ and $t$ \citep{fung1969}. We assume that blood is incompressible, i.e. $\nabla \cdot \mathbf{u} =0$, where  $\mathbf{u} = (u,v,w)$ is the flow velocity of blood. Since the Reynolds number in the capillaries is small (approximately $10^{-2}$), fluid dynamics are dominated by viscous forces  \citep{fung2013C,fung1969}. Consequently, we simplify the Navier--Stokes equations to the Stokes equations, $\nabla p = \mu_c \nabla^2 \bm{u}$. 

Focusing on a region of the sheet that is large relative to individual posts, we characterize the average behavior of blood flow. Based on the experimental and dimensional analyses established  by \citet{fung2013C}, the momentum balance is given by a Darcy-type law
\begin{align}
     \frac{\partial{p}}{\partial{x}} =  -\frac{\mu_c {U}}{\hbar^2} \kappa, \label{eq:phrelation}
\end{align}
where $\mu_c$ is the apparent viscosity of blood in the capillaries, $U$ is the average flow velocity, $\hbar(x,t)$ is the sheet height, and $\kappa$ is a friction coefficient capturing geometric properties of the sheet, including red blood cell deformation, post dimensions, and post arrangement. Detailed derivations of $\kappa$ can be found in \cite{fung2013C}, \cite{fung2013LT}, \cite{fung1969}, and \cite{yen1973}.

To account for sheet membrane compliance, we impose a linear elastic relationship between height and pressure
\begin{align}
    \hbar = \hbar_c + \alpha_c (p - p_\text{alv}),
    \label{eq:thickness}
\end{align}
where $\hbar_c$ is the resting height when $p=p_\text{alv}$, and $\alpha_c$ is the sheet membrane compliance. When the lung inflates, the stretching of alveoli causes the sheet membrane to deform. The sheet height increases when the blood pressure, $p$, exceeds the air pressure, $p_\text{alv}$, takes its resting height $\hbar_c$ when $p=p_\text{alv}$, and decreases when $p<p_\text{alv}$.

Defining the flow through the sheet as  $q = \hbar l_c U$ (mL/s) and combining Equations \eqref{eq:phrelation} and \eqref{eq:thickness} yields a direct relationship between flow and height
\begin{align}
   q =-\frac{\hbar^3 l_c}{\mu_c} \kappa \alpha_c \frac{\partial \hbar}{\partial x}. \label{eq:flowthick}
\end{align}
Mass within a control volume changes through transient changes in height, $\partial \hbar/\partial t$, and through fluid transfer across the blood tissue barrier, governed by Starling's hypothesis, $K_p(p-p^*)$. Here, $K_p$ is the filtration coefficient and $p^*$ is the constant interstitial hydrostatic pressure plus the difference in osmotic pressure across the alveolar wall. Using Equations \eqref{eq:thickness} and \eqref{eq:flowthick}, and enforcing mass conservation, yields 
\begin{align*}
 \frac{\partial^2 \hbar^4}{\partial x^2} &=  4\alpha_c\mu_c \kappa\left(
     \frac{\partial \hbar}{\partial t}+\frac{2K_p}{\alpha_c \rho}\left(\hbar-\hbar^*\right)\right),
\end{align*}
where $\rho$ is blood density and $\hbar^*=\hbar_c+\alpha_c(p^*-p_{\text{alv}})$ is the sheet height corresponding to the reabsorption pressure $p^*$. Because fluid transfer to the extravascular space is negligible compared to intravascular flow, we neglect filtration and take $K_p=0$ \citep{fung1972theor,clark2018}. This gives \begin{align}
    \frac{\partial^2 \hbar^4}{\partial x^2} &=  4\alpha_c\mu_c \kappa
     \frac{\partial \hbar}{\partial t}. \label{eq:goveq}
\end{align}

We linearize \eqref{eq:phrelation}, \eqref{eq:thickness}, and \eqref{eq:goveq} by assuming only small perturbations in blood pressure and sheet height occur at the capillary level \citep{fung1972theor,clark2018}. We express the variables as an expansion of powers of a small parameter $\varepsilon$, where $0<\varepsilon \ll 1$ 
\begin{align}
    \hbar(x,t) &= \hbar_0(x)+\varepsilon \hbar_1(x,t)+\mathcal{O}(\varepsilon^2), \nonumber \\
    p(x,t) &= p_0(x)+\varepsilon p_1(x,t)+\mathcal{O}(\varepsilon^2),  \nonumber\\
   q(x,t) &= q_{0}(x)+\varepsilon q_{1}(x,t)+\mathcal{O}(\varepsilon^2),  \nonumber
\end{align}
where $\hbar_0,p_0,$ and $q_0$ represent the steady components and $\hbar_1,p_1$, and $q_1$ the unsteady components. We linearize about a uniform reference sheet height $\hbar_0$ \citep{tawhai2011}, with corresponding reference pressure $p_0 = (\hbar_0-\hbar_c)/\alpha_c + p_\text{alv}$. This approximation holds when variations in height across the sheet are small relative to $\hbar_0$, i.e., $\alpha_c \Delta p \ll \hbar_0$. The mean pressure drop and mean flow through the sheet are then captured by the $\omega_k=0$ component of the first-order problem.
The first-order unsteady expressions satisfy 
\begin{align}
   & \hbar_0^3\frac{\partial^2\hbar_1}{\partial x^2} =  \alpha_c\mu_c \kappa \frac{\partial \hbar_1}{\partial t}, \label{eq:linear_h1}
   \qquad p_1 = \frac{1}{\alpha_c}\hbar_1,
   \qquad q_1 = -\frac{\hbar_0^3 l_c}{\mu_c\kappa\alpha_c}\frac{\partial \hbar_1}{\partial x}.
\end{align}
Assuming all unsteady perturbations are periodic, we map pressure, flow, and height into the frequency domain using complex periodic Fourier series. For each frequency $\omega_k$, Equations \eqref{eq:linear_h1} are transformed to
\begin{align*}
     \hbar_0^3 \frac{\partial^2 H}{\partial x^2} =  i\omega_k \alpha_c\mu_c \kappa H,
     \qquad P =\frac{1}{\alpha_c}H,
    \qquad Q = -\frac{\hbar_0^3 l_c}{\mu_c\kappa\alpha_c}\frac{\partial H}{\partial x}. \nonumber
\end{align*}
Solving for the frequency-domain sheet height $H(x,\omega_k)$ at any position $x$ across the capillary bed yields
\begin{align*}
  H(x,\omega_k) &= c_1 e^{sx} + c_2 e^{-sx}, \ \  s = \sqrt{i \omega_k \alpha_c\mu_c\kappa / \hbar_0^3}, 
\end{align*}
where $c_1, c_2$ are constants. The associated solutions for $P(x,\omega_k)$ and $Q(x,\omega_k)$ are
\begin{align}
    P(x,\omega_k) &= \frac{1}{\alpha_c}\left(c_1 e^{sx} + c_2 e^{-sx}\right), \label{eq:Pref}
    \qquad Q(x,\omega_k) &= -\frac{\hbar_0^3 l_c s}{\mu_c} \kappa \alpha_c \left(c_1e^{sx}-c_2 e^{-sx}\right).
\end{align}
Finally, the admittance matrix is constructed by relating the flow and pressure at the proximal and distal edges of the capillary sheet (i.e., the junction points where the feeding arteriole and draining venule connect to the network).  Evaluating Equations \eqref{eq:Pref} at $x=0$ and $x=l_c$ (where $l_c$ is the sheet length) yields the capillary admittance matrix, $\mathbf{Y}^C$, such that 
\begin{align*}
    \begin{pmatrix} Q(0,\omega_k) \\ -Q(l_c,\omega_k) \end{pmatrix}
    = \mathbf{Y}^C(\omega_k)
    \begin{pmatrix} P(0,\omega_k) \\ P(l_c,\omega_k) \end{pmatrix},
\end{align*}
where both entries on the left are flows entering the sheet, consistent with the vessel admittances. For $\omega_k \neq 0$,
\begin{align}
    \mathbf{Y}^C(\omega_k) =  \sqrt{\frac{i\omega_k \hbar_0^3 l_c^2 \alpha_c}{\mu_c \kappa}}\frac{1}{e^{-sl_c}-e^{sl_c}} 
    \begin{pmatrix}
       -(e^{sl_c}+e^{-sl_c}) & 2 \\
        2 &   -(e^{sl_c}+e^{-sl_c}) 
    \end{pmatrix},
    \label{eq:YCW1}
\end{align}
and for $\omega_k=0$, we have
\begin{align}
    \mathbf{Y}^C(0) =\frac{\hbar_0^3 l_c}{\mu_c\kappa l_c}
    \begin{pmatrix}
        1 & -1 \\
       -1 & 1
    \end{pmatrix}. \label{eq:YCW0}
\end{align}

\subsection{Multiscale coupling} \label{sec:multiscale}
To construct the complete multiscale framework, the equations governing the three domains must be dynamically coupled and solved. The large vessel equations are solved numerically in the time domain, while the small vessel and capillary equations are solved analytically in the frequency domain using admittance matrices. Ultimately, the combined small vessel and capillary network serves as a boundary condition for the large vessels, explicitly relating pressure and flow between the arterial and venous systems.

\subsubsection{Numerical method}
The hyperbolic partial differential equations governing the large vessels are solved numerically using the two-step Lax--Wendroff method \citep{olufsen2000}. The simulation runs over successive cardiac cycles until convergence. In the initial cycle, the system is solved at all spatial and temporal points from a baseline state. In subsequent cycles, the simulation uses the pressure, flow, and area from the preceding cycle as initial conditions. Following the criterion established by \citet{mackenzie2021}, the simulation is considered to have reached a steady periodic state when the changes in pressure and flow are less than 1$\times 10^{-6}$ over five consecutive cycles. 

The small vessels and capillaries are coupled to the terminal large arteries and veins via admittance matrices, which are computed recursively in the frequency domain. Once we find the total admittance of the structured tree, a convolution integral is applied to link the frequency and time domains. This establishes the dynamic boundary condition, relating the pressure and flow at the outlet of each large terminal artery to the inlet of the corresponding large terminal vein.

To derive this grand admittance, we first calculate a unique admittance matrix for each vessel. See Supplemental Figure 1 for an illustration of each step in the calculation. Using electrical circuit theory, we connect the corresponding vessels in parallel and in series. The recursive algorithms for connecting admittances for structured trees without capillaries, with terminal capillaries, and with ladder-like capillaries are detailed in the following sections and in Algorithms \ref{alg:admit} and \ref{alg:admit_ladder_capillary}.

\subsubsection{Admittance matrix for two vessels in parallel}
To derive this grand admittance, we first calculate a unique admittance matrix for each vessel. See Supplemental Figure 1 for an illustration of each step in the calculation. Using electrical circuit theory, we connect the corresponding vessels in series and parallel. The recursive algorithms for connecting admittances for structured trees without capillaries, with terminal capillaries, and with ladder-like capillaries are detailed in the following sections and in Algorithms \ref{alg:admit} and \ref{alg:admit_ladder_capillary}.
\begin{align}
    \begin{pmatrix}
        Q_1\\
        Q_2
    \end{pmatrix} =
    \mathbf{Y}^{\|}  \begin{pmatrix}
        P_1\\
        P_2
    \end{pmatrix}, \qquad  \mathbf{Y}^{\|} = \mathbf{Y}^A +  \mathbf{Y}^B,
\label{eq:parallel} 
\end{align}
where $\mathbf{Y}^A$ and $\mathbf{Y}^B$ are the admittance matrices for the vessels $A$ and $B$, and $\mathbf{Y}^{\|}$ is the admittance matrix for the vessels connected in parallel.

\subsubsection{Admittance matrix for two vessels in series}
Two vessels are joined in series when they do not share a parent, e.g., at the ends of the arterial and venous trees (Algorithm \ref{alg:series1}). By conservation of flow, the outflow of one vessel is the inflow of the other. For vessels $A$ and $B$, which do not share a parent, this means $Q_2^A = -Q_1^B,$ where the subscript `$2$' indicates the outflow and `$1$' the inflow \citep{qureshi2014}. In addition, the pressure continuity condition produces $P_2^A = P_1^B$. From these principles, we derive a matrix relationship to connect vessels $A$ and $B$ in series, viz.~
\begin{align} 
\begin{pmatrix}
        Q_1^A\\
        Q_2^B
    \end{pmatrix} &= \mathbf{Y}^{\Leftrightarrow}  
    \begin{pmatrix}
        P_1^A\\
        P_2^B
    \end{pmatrix}, \ \ 
\mathbf{Y}^{\Leftrightarrow}  =   
    \frac{1}{Y_{22}^A+Y_{11}^B} 
\begin{pmatrix}
 \det(\mathbf{Y}^A)+Y_{11}^AY_{11}^B & - Y_{12}^AY_{12}^B\\
    -Y_{21}^A Y_{21}^B & \det(\mathbf{Y}^B) + Y_{22}^A Y_{22}^B
\end{pmatrix}, \label{eq:series}
\end{align}
where subscripts `$11$', `$12$', `$21$', and `$22$' denote the matrix elements, `det' represents the determinant, and $\mathbf{Y}^{\Leftrightarrow} $ is the admittance matrix for vessels connected in series.

\subsubsection{Linking arterial and venous trees without capillaries}
To join the arterial (denoted by $A$) and venous (denoted by $V$) networks, the structured trees are generated down to a minimum radius, $r_{\min}$. Each matched artery--vein pair is indexed by $(m,n)$, denoting a vessel of radius $r(m,n)=\alpha^m \beta^n r_\text{root}$, where $r_\text{root}$ is the radius of the structured tree's root vessel. The admittance matrix of each vessel is computed from  Equations \eqref{eq:admittance_n0} and \eqref{eq:admittance_0}. The admittance of the connected network is assembled recursively, beginning at the smallest arteries and veins and moving toward the root, by combining individual admittance matrices in parallel ($+$, Equation \eqref{eq:parallel}) and series ($\Leftrightarrow $, Equation \eqref{eq:series}), as described in Algorithm \ref{alg:admit}.  For a single arterial and venous bifurcation in which each terminal artery connects directly to its corresponding vein (Figure \ref{fig:capillaryschematic}(a)), the  total admittance is given by
\begin{align}
    \mathbf{Y}^A(0,0) \Leftrightarrow \Bigg( \Big( \mathbf{Y}^A(1,0) \Leftrightarrow \mathbf{Y}^V(1,0) \Big) + \Big(\mathbf{Y}^A(0,1) \Leftrightarrow \mathbf{Y}^V(0,1)\Big)\Bigg) \Leftrightarrow \mathbf{Y}^V(0,0). \label{eq:Y_nocap}
\end{align}

\begin{algorithm}[ht!]
\caption{Recursive function computing the grand admittance matrix of the subtree rooted at branch $(m,n)$, with terminal arteries connected directly to terminal veins (no capillaries). Here, $Y$ denotes a subtree admittance, and $Y^A$ and $Y^V$ denote single-vessel admittances.}
\label{alg:admit}
\textbf{Recursive Function \texttt{admit(m,n)}}\\[0.25cm]
\textbf{if} $r(m+1,n) < r_\text{min}$ \algcomment{$\alpha$-daughter is terminal}\\
\hspace*{0.25cm} \textbf{for} $i = A,V$\\
\hspace*{0.5cm} $Y^k(m+1,n)$ = Eq.~\eqref{eq:admittance_n0} if $\omega_k \neq 0$; \ $Y^k(m+1,n)$ = Eq.~\eqref{eq:admittance_0} if $\omega_k = 0$\\
\hspace*{0.5cm} \textbf{end}\\
\hspace*{0.5cm}$Y(m+1,n) = \texttt{series}\left(Y^A(m+1,n),Y^V(m+1,n)\right)$ \\ \hspace*{\fill}{\small\textit{$\triangleright$ Join terminal artery to vein, Eq.~\eqref{eq:series}}}\\
\textbf{else} \algcomment{$\alpha$-daughter branches further}\\
\hspace*{0.5cm} $Y(m+1,n) = \texttt{admit}(m+1,n)$ \algcomment{Recurse into $\alpha$-subtree}\\
\textbf{end}\\ 
\vspace*{0.5cm}
\textbf{if} $r(m,n+1) < r_\text{min}$ \algcomment{$\beta$-daughter is terminal}\\
\hspace*{0.25cm} \textbf{for} $i = A,V$\\
\hspace*{0.5cm} $Y^k(m,n+1)$ = Eq.~\eqref{eq:admittance_n0} if $\omega_k \neq 0$; \ $Y^k(m,n+1)$ = Eq.~\eqref{eq:admittance_0} if $\omega_k = 0$\\
\hspace*{0.25cm} \textbf{end}\\
\hspace*{0.5cm}$Y(m,n+1) = \texttt{series}\left(Y^A(m,n+1),Y^V(m,n+1)\right)$ \\ \hspace*{\fill}{\small\textit{$\triangleright$ Join terminal artery to vein, Eq.~\eqref{eq:series}}}\\
\textbf{else}  \algcomment{$\beta$-daughter branches further}\\
\hspace*{0.5cm} $Y(m,n+1) = \texttt{admit}(m,n+1)$  \algcomment{Recurse into $\beta$-subtree}\\
\textbf{end}\\
\vspace*{0.25cm}
$Y_\text{mid}(m,n) = Y(m+1,n)+Y(m,n+1)$ \algcomment{$\alpha$- and $\beta$-subtrees in parallel, Eq.~\eqref{eq:parallel}}\\
\vspace*{0.25cm}
\textbf{for} $k=A,V$\\
\hspace*{0.5cm} $Y^k(m,n)=$ Eq.~\eqref{eq:admittance_n0} if $\omega_k \neq 0$; \ $Y^k(m,n)=$ Eq.~\eqref{eq:admittance_0} if $\omega_k = 0$\\
\textbf{end}\\
\vspace*{0.25cm}
$Y(m,n) = \texttt{series}\left(\texttt{series}\left(Y^A(m,n),Y_\text{mid}(m,n)\right),Y^V(m,n)\right)$ \\ \hspace*{\fill}{\small\textit{$\triangleright$ Parent artery $\Leftrightarrow$ subtrees $\Leftrightarrow$ parent vein, Eq.~\eqref{eq:series}}}
\end{algorithm}

\subsubsection{Linking arterial and venous trees with capillaries}
Algorithm \ref{alg:admit} assumes that each terminal artery connects directly to its corresponding vein.  Physiologically, however, blood passes from arterioles to venules through capillaries, which within the acinus are not confined to the vessel ends but instead form sheets connecting arterioles and venules along their length. We therefore extend Algorithm \ref{alg:admit} to include capillary admittances, $\mathbf{Y}^C$, computed from Equations \eqref{eq:YCW1} and \eqref{eq:YCW0}. The direct artery--vein connection at each terminal pair ($r<r_\text{min}$) is replaced by a terminal capillary sheet. Below a threshold radius $r_\text{ladder}$ (Table \ref{tab:nominalparameters}), an additional capillary sheet bridges the midpoint of each arteriole to the midpoint of the corresponding venule, so that the arterial and venous trees are joined like the rungs of a ladder (Figure \ref{fig:capillaryschematic}(b)). Vessels with $r\geq r_\text{ladder}$ carry no capillaries and are treated as in Algorithm \ref{alg:admit}. Attaching a capillary at the vessel midpoint requires splitting each ladder vessel into proximal ($0\leq x \leq L/2$) and distal ($L/2 \leq x \leq L$) halves, with half-admittances $\mathbf{Y}^A_{\text{half}}$ and $\mathbf{Y}^V_{\text{half}}$ computed from Equations \eqref{eq:admittanceHALF_n0} and \eqref{eq:admittanceHALF_0}. Because the two halves are identical, a single half-admittance is computed per vessel.

The recursion (Algorithm \ref{alg:admit_ladder_capillary}) distinguishes three cases. For a terminal pair, the distal halves are joined through the terminal capillary, giving $\mathbf{Y}^{ACV}(m,n) = \mathbf{Y}^A_{\text{half}}(m,n) \Leftrightarrow \mathbf{Y}^C(m,n) \Leftrightarrow \mathbf{Y}^V_\text{half}(m,n)$. This pathway is placed in parallel with the midpoint capillary sheet and then connected in series with the proximal halves 
\begin{align}
  \mathbf{Y}(m,n) = \mathbf{Y}^A_\text{half}(m,n) \Leftrightarrow \Big( \mathbf{Y}^{ACV}(m,n) + \mathbf{Y}^C(m,n) \Big) \Leftrightarrow \mathbf{Y}^V_\text{half}(m,n). \label{eq:Y_terminal}
\end{align}
For a non-terminal ladder vessel ($r_\text{min}\leq r<r_\text{ladder}$), the terminal capillary is replaced by the parallel combination of the daughter subtrees, $\mathbf{Y}_\text{mid}(m,n)=\mathbf{Y}(m+1,n)+\mathbf{Y}(m,n+1)$
\begin{equation}
\begin{split}
   \mathbf{Y}(m,n) = \mathbf{Y}^A_\text{half}(m,n) \Leftrightarrow \Big( & \big(\mathbf{Y}^A_\text{half}(m,n) \Leftrightarrow \mathbf{Y}_\text{mid}(m,n) \Leftrightarrow \mathbf{Y}^V_\text{half}(m,n)\big) \\
    &+ \mathbf{Y}^C(m,n) \Big) \Leftrightarrow \mathbf{Y}^V_\text{half}(m,n).
    \end{split}\label{eq:Y_ladder}
\end{equation}
For $r \ge r_\text{ladder}$, the admittance $\mathbf{Y}(m,n) = \mathbf{Y}^A(m,n) \Leftrightarrow \mathbf{Y}_\text{mid}(m,n) \Leftrightarrow \mathbf{Y}^V(m,n)$, as stated in Algorithm \ref{alg:admit}. For example, for a single bifurcation with terminal daughters (Figure \ref{fig:capillaryschematic}(b)), the total admittance is $\mathbf{Y}^A(0,0) \Leftrightarrow \big( \mathbf{Y}(1,0) + \mathbf{Y}(0,1) \big) \Leftrightarrow \mathbf{Y}^V(0,0)$, with $\mathbf{Y}(1,0)$ and $\mathbf{Y}(0,1)$ given by Equation \eqref{eq:Y_terminal}.

\begin{algorithm}[ht!]
\caption{Recursive function computing the grand admittance matrix of the subtree rooted at branch $(m,n)$, including the midpoint (ladder) and terminal capillary sheets. Here, $Y$ denotes a subtree admittance, and $Y^A$, $Y^V$, and $Y^C$ denote single-vessel or capillary admittances.}
\label{alg:admit_ladder_capillary}
\textbf{Recursive Function \texttt{admit\_capillary}(m,n)}\\[0.25cm]
\textbf{if} $r(m,n) < r_\text{min}$ \algcomment{Terminal pair}\\
\hspace*{0.25cm} \textbf{for} $i = A,V$\\
\hspace*{0.5cm} $Y_\text{half}^k$ = Eq.~\eqref{eq:admittanceHALF_n0} if $\omega_k \neq 0$; Eq.~\eqref{eq:admittanceHALF_0} if $\omega_k = 0$\\
\hspace*{0.25cm} \textbf{end}\\
\hspace*{0.25cm} $Y^C$ = Eq.~\eqref{eq:YCW1} if $\omega_k \neq 0$; Eq.~\eqref{eq:YCW0} if $\omega_k = 0$\\
\hspace*{0.25cm} $Y^{ACV} = \texttt{series}\left(\texttt{series}\left(Y_\text{half}^A,Y^C\right),Y_\text{half}^V\right)$ \\ \hspace*{\fill}{\small\textit{$\triangleright$ Join distal halves via terminal capillary, Eq.~\eqref{eq:series}}}\\
\hspace*{0.25cm} $Y_U = Y^{ACV} + Y^C$ \algcomment{Add midpoint capillary in parallel, Eq.~\eqref{eq:parallel}}\\
\hspace*{0.25cm} $Y = \texttt{series}\left(\texttt{series}\left(Y_\text{half}^A,Y_U\right),Y_\text{half}^V\right)$ \algcomment{Proximal halves in series, Eq.~\eqref{eq:Y_terminal}}\\[0.25cm]
\textbf{else if} $r(m,n) < r_\text{ladder}$ \algcomment{Ladder vessel}\\
\hspace*{0.25cm} $Y_\text{mid} = \texttt{admit\_capillary}(m+1,n) + \texttt{admit\_capillary}(m,n+1)$ \\ \hspace*{\fill}{\small\textit{$\triangleright$ Recurse and connect subtrees in parallel}}\\
\hspace*{0.25cm} \textbf{for} $i = A,V$\\
\hspace*{0.5cm} $Y_\text{half}^k$ = Eq.~\eqref{eq:admittanceHALF_n0} if $\omega_k \neq 0$; Eq.~\eqref{eq:admittanceHALF_0} if $\omega_k = 0$\\
\hspace*{0.25cm} \textbf{end}\\
\hspace*{0.25cm} $Y^C$ = Eq.~\eqref{eq:YCW1} if $\omega_k \neq 0$; Eq.~\eqref{eq:YCW0} if $\omega_k = 0$\\
\hspace*{0.25cm} $Y_U = \texttt{series}\left(\texttt{series}\left(Y_\text{half}^A,Y_\text{mid}\right),Y_\text{half}^V\right) + Y^C$ \\ \hspace*{\fill}{\small\textit{$\triangleright$ Connect subtrees via distal halves and midpoint capillary in parallel}}\\
\hspace*{0.25cm} $Y = \texttt{series}\left(\texttt{series}\left(Y_\text{half}^A,Y_U\right),Y_\text{half}^V\right)$ \algcomment{Proximal halves in series, Eq.~\eqref{eq:Y_ladder}}\\[0.25cm]
\textbf{else} \algcomment{Above ladder region, no capillaries}\\
\hspace*{0.25cm} $Y_\text{mid} = \texttt{admit\_capillary}(m+1,n) + \texttt{admit\_capillary}(m,n+1)$ \\ \hspace*{\fill}{\small\textit{$\triangleright$ Recurse, connect subtrees in parallel}}\\
\hspace*{0.25cm} \textbf{for} $i = A,V$\\
\hspace*{0.5cm} $Y^k$ = Eq.~\eqref{eq:admittance_n0} if $\omega_k \neq 0$; Eq.~\eqref{eq:admittance_0} if $\omega_k = 0$\\
\hspace*{0.25cm} \textbf{end}\\
\hspace*{0.25cm} $Y = \texttt{series}\left(\texttt{series}\left(Y^A,Y_\text{mid}\right),Y^V\right)$ \\ \hspace*{\fill}{\small\textit{$\triangleright$ Parent artery $\Leftrightarrow$ subtrees $\Leftrightarrow$ parent vein, Eq.~\eqref{eq:series}}}\\
\textbf{end}
\end{algorithm}

\begin{algorithm}
\caption{Function used to combined two admittance matrices in series.}
\label{alg:series1}
\textbf{Function \texttt{series}($\bm{Y^S,Y^T}$)}\\[0.25cm]
$\mathrm{det\_S} = Y_{11}^S Y_{22}^S - Y_{12}^S Y_{21}^S$\\
$\mathrm{det\_T} = Y_{11}^T Y_{22}^T - Y_{12}^T Y_{21}^T$\\
\vspace*{0.25cm}
$Y_{11} = \big(\mathrm{det\_S}+Y_{11}^SY_{11}^T\big) / \big(Y_{22}^S+Y_{11}^T\big)$\\
$Y_{12} = \big(-Y_{12}^SY_{12}^T\big)/ \big(Y_{22}^S+Y_{11}^T\big)$\\
$Y_{21} = \big(-Y_{21}^SY_{21}^T\big)/ \big(Y_{22}^S+Y_{11}^T\big)$\\
$Y_{22} = \big(\mathrm{det\_T}+Y_{22}^SY_{22}^T\big) / \big(Y_{22}^S+Y_{11}^T\big)$\\ 
\end{algorithm}

\subsection{Quantities of interest (biomarkers)}
We calculate several quantities of interest (biomarkers) to gain insight into the PH-LHF progression, previously identified by \citet{bartolo2022}.

\subsubsection{Wall shear stress (WSS)}
  In the proximal vasculature, which consists of large arteries and veins, we predict the time-series and average pressure $p(x,t)$, flow $q(x,t)$, area $A(x,t)$, and wall shear stress
\begin{align}
    \tau = -\mu_L  \frac{\partial u_x}{\partial r}\big|_{r=R} = \mu_L \bar{u}_x/\delta.
    \label{eq:shearL}
\end{align}
In the microcirculation, consisting of arterioles, venules, and capillaries, the wall shear stress is defined as \citep{pedley1980,olufsen2000}
\begin{align*}
    \tau = \frac{4 \mu}{A_0 r_0}\left(q+\frac{r_0^2}{24\nu}\frac{dq}{dt}+O(w_0^4)\right).
\end{align*}

\subsubsection{Cyclic stretch (CS)}
In addition, we calculate the cyclic stretch, defined as the change in radius from systole to diastole divided by the diastolic radius \citep{bartolo2022}
\begin{align*}
    C\!S = \frac{\text{max}\big(R(x,t)\big) - \text{min}\big(R(x,t)\big)}{\text{min}\big(R(x,t)\big)}.
\end{align*}

\subsubsection{Wave intensity}
Another important quantity for evaluating computational predictions, particularly for assessing PH, is wave intensity, a measure of the power transmitted by the pulse wave. In PH, the stiffening of proximal arteries and the remodeling of the distal microvasculature create impedance mismatches that generate abnormally large backward-traveling waves, significantly increasing right ventricular afterload. As discussed in our previous study \citep{qureshi2015}, we can predict wave intensity by decomposing the pressure $p(x,t)$, flow $q(x,t)$, and area $A(x,t)$ waves into their forward and backward components. Using pressure and flow predictions, we estimate the pulse wave velocity \begin{align*}
c = \sqrt{\frac{A}{\rho} \frac{\partial p}{\partial A}},
\end{align*}
where $\rho = 1.055$ (g/mL) is the blood density. To separate the propagating waves into their forward and backward components, we compute the difference
\begin{align*}
d\mathcal{X}=\mathcal{X}(x,t+\delta t)-\mathcal{X}(x,t)
\end{align*}
for the pressure, flow, and area, where $\mathcal{X} = (p,q,A)$. Using these approximations, the forward and backward waves are computed as
\begin{align*}
dp_\pm = \frac12 \left(dp\pm\rho c\, du\right), \qquad  du_\pm = \frac12\left(du\pm \frac{dp}{\rho c}\right)
\end{align*}
and the associated forward and backward wave intensities are
\begin{align*}
dI_+ = dp_+ \, du_+ = \frac{1}{4\rho c}\left(dp + \rho c\, du\right)^2, \quad dI_- = dp_{-} \, du_+ = -\frac{1}{4\rho c}\left(dp - \rho c\, du\right)^2.
\end{align*}
Forward-propagating waves correspond to $dI_+ > 0$, whereas for backward-propagating waves $dI_- < 0$. Furthermore, waves are classified based on the sign of the pressure change such that  compression waves correspond to  $dp_\pm > 0$ and expansion (or decompression) waves correspond to $dp_\pm < 0$.

\subsection{Simulations}

\begin{figure}
\centerline{\includegraphics[width=0.9\textwidth]{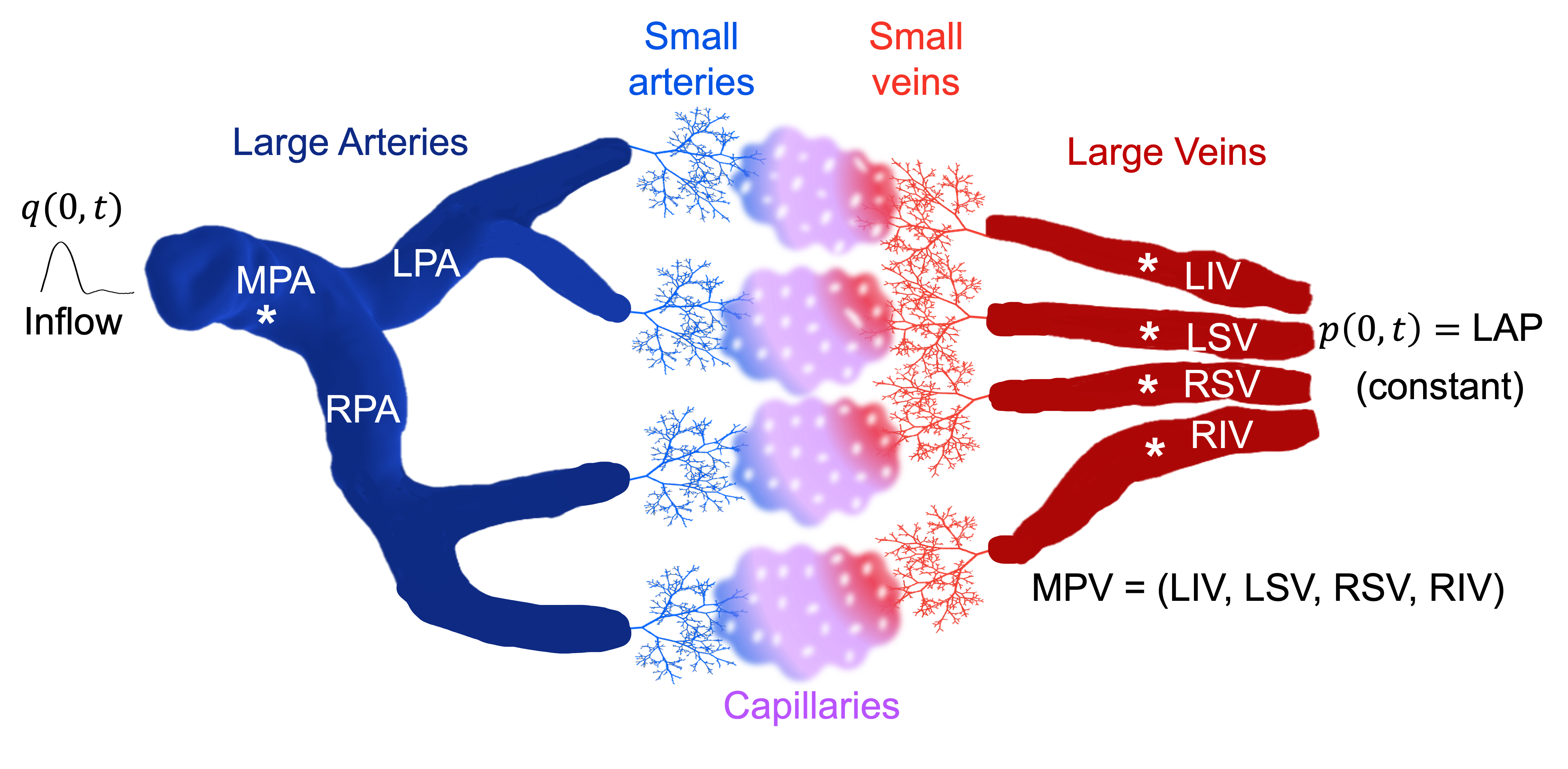}}
\caption{Network used for fluid dynamics simulations. Inflow to the main pulmonary artery (MPA) is prescribed from a PC-MRI flow waveform, and the pressure at the end of each of the four pulmonary veins (LIV, LSV, RSV, RIV) is set to a constant left atrial pressure. The asterisks (*) indicate the locations where macrovascular results are extracted. Large arterial results are reported at the midpoint of the MPA, and large venous results are averaged across the center of the four main pulmonary veins (MPV).}
\label{fig:ActualNetwork}
\end{figure}

The focus of this study is to evaluate the hemodynamic impact of adding a pulmonary capillary model to an existing multiscale framework of the pulmonary vasculature. As shown in Figure \ref{fig:ActualNetwork}, the large vessel domain contains 7 large arteries and 4 major veins. The dimensions of these vessels (listed in Table \ref{tab:AV_dim}) are extracted from a CT image obtained from a healthy human subject \citep{bartolo2022}. We systematically evaluate this multiscale framework by comparing predictions under different microvascular structural configurations. To facilitate easy comparison of results, we used the same inflow waveform for all simulations. For most simulations, we set the cardiac cycle time $T = 1$ s. The exception is the simulation comparing control vs. PH. For the latter, we account for the higher heart rate by setting the cardiac cycle length to 0.8 s.The parameter values and patient characteristics used for these simulations are reported in Tables \ref{tab:patientcharacteristics} and \ref{tab:nominalparameters} as well as Figure \ref{fig:severitychanges}. Parameter values for the structured tree model and capillary sheet are aggregated from the following studies: Structured tree model \citep{bartolo2022,bartolo2024,olufsen2012,qureshi2019,olufsen2000,colebank2024,Colebank2021sens} and the capillary sheet models \citep{clark2010,tawhai2011,clark2018,dhadwal1997}. Specifically, we compare baseline hemodynamics with and without explicitly including capillaries. For each configuration, we assess system dynamics by varying the model parameters using the values listed in Table \ref{tab:nominalparameters}.
Next, we simulate hemodynamics for a healthy control subject and a patient with PH to investigate the biomechanical differences associated with disease. Finally, to understand model behavior as the disease progresses, we conduct a parametric study by varying one parameter at a time and set up a disease-progression simulation. Parameter values for these simulations are reported in Figure~\ref{fig:severitychanges}.

\section{Results}
This study presents the first mathematical model of pulsatile hemodynamics that encompasses the complete pulmonary circulation, explicitly linking the large arteries, arterioles, capillaries, venules, and large veins. In the following sections, we evaluate the physiological importance of this expanded architecture by comparing simulated hemodynamics between a healthy control and a PH patient. We then utilize parametric analysis to quantify how specific biomechanical factors affect hemodynamics in response to vascular remodeling and disease progression.

\subsection{Hemodynamic effects of the capillary network in the pulmonary circulation model}

\begin{figure}
\centerline{\includegraphics[width=\textwidth]{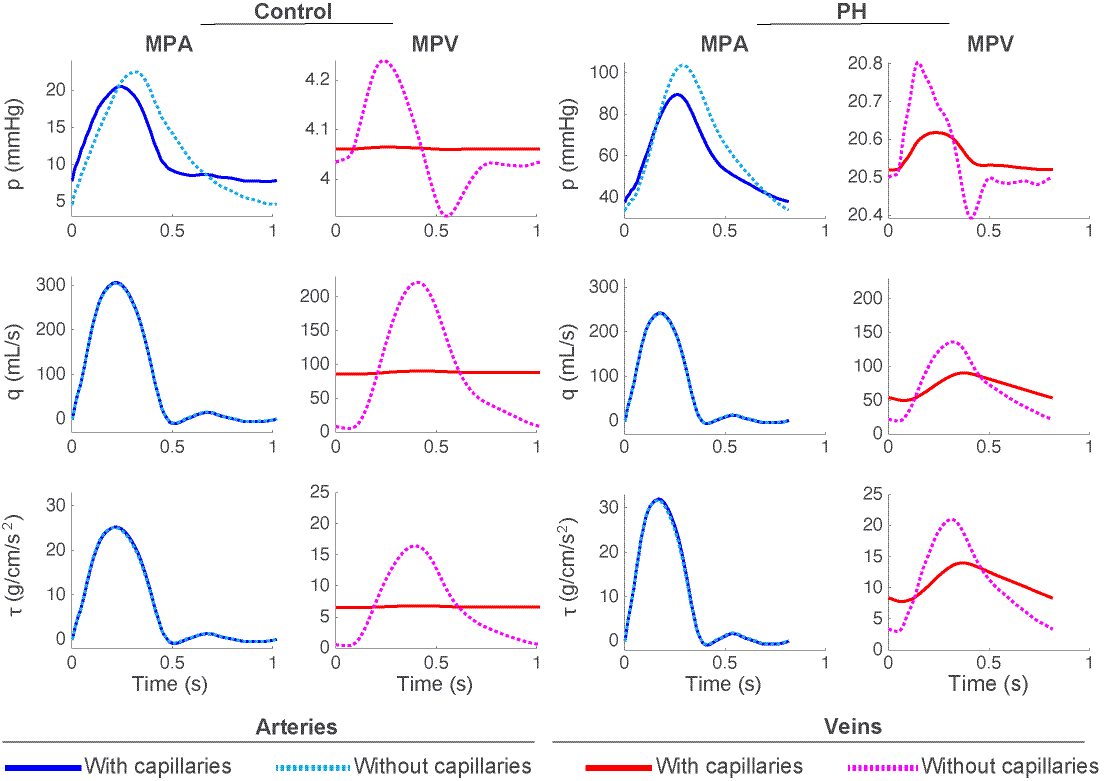}}
\caption{Simulations with (solid) and without (dashed) capillaries for a control (left) and hypertensive (right) subject. Arteries are denoted by blue/cyan lines and veins by red/magenta lines. Results in arteries are from the main pulmonary artery (MPA) and averaged over the four main pulmonary veins (MPV). Note the change in waveform shape and damping of oscillations with the capillary network.}
\label{fig:controlPHcap}
\end{figure}

To study the impact of incorporating a capillary structure into our multiscale pulmonary circulation model, we report simulation results for a healthy control subject and a PH patient. For both subjects, simulations are conducted using the network depicted in Figure \ref{fig:capillaryschematic} with the parameters listed in Table \ref{tab:nominalparameters}. Large arterial waveforms are extracted from the midpoint of the MPA. Venous pressure ($p$) and wall shear stress ($\tau$) are averaged across the centers of the four main pulmonary veins (MPV), while total venous volumetric flow ($q$) is computed as the sum across these four vessels. As illustrated in Figure \ref{fig:controlPHcap}, including the capillary bed significantly alters hemodynamic predictions by introducing downstream damping. Because identical inlet flow waveforms are prescribed across all simulations, the mean MPA and total venous flow remain constant due to mass conservation; however, the capillary network profoundly attenuates flow pulsatility. In the healthy control, omitting the capillaries allows unphysiological pulsatility to transmit into the venous system, whereas their inclusion yields the expected steady venous profiles (solid red lines).

\begin{figure}
\centerline{\includegraphics[width=\textwidth]{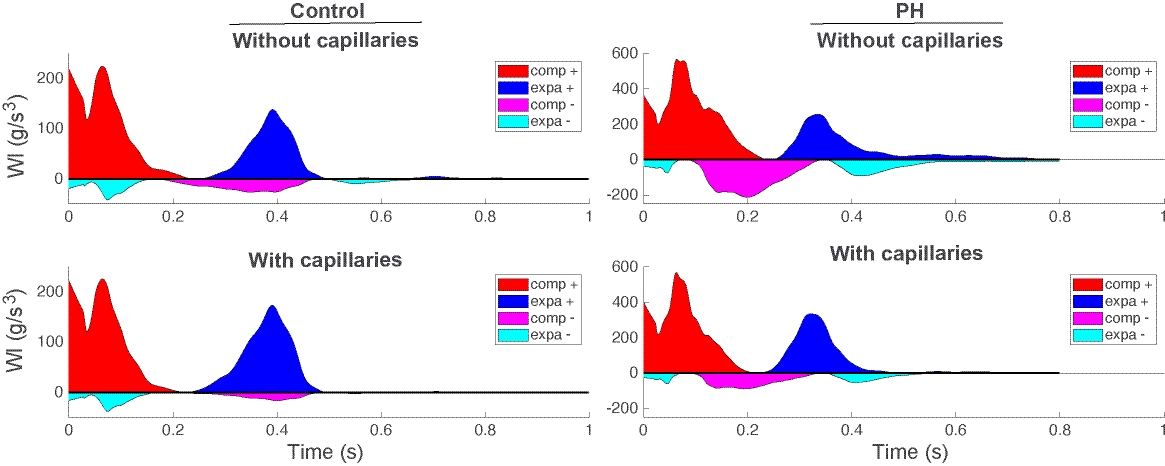}}
\caption{Wave intensity simulations at the proximal end of the MPA: the red and dark blue colors denote the forward $(+)$ compression and expansion waves, while the magenta and cyan colors refer to the backward  $(-)$ compression and expansion waves. The backward-propagating waves arrive later and are reflected from the first bifurcation into the left and right pulmonary arteries.}
\label{fig:waveintensity}
\end{figure}

In contrast, the remodeled PH microvasculature fails to completely isolate the venous system, transmitting residual pulsatility into the MPV even with capillaries included. Wall shear stress waveforms computed from Equation \eqref{eq:shearL} display similar damping effects. In the MPA, the compliant nature of the capillary network generates a superposition of reflected waves that refines the pressure-waveform morphology. Specifically, it shifts the systolic peak earlier in the cardiac cycle and creates a more rapidly decaying tail during diastole. For the healthy subject, the compliance of the system ensures that total resistance does not change significantly, resulting in nearly identical mean arterial pressures with and without capillaries (Table \ref{tab:w_woCap}). Conversely, in the stiffer PH network, both pulse and mean pressures are massively over-predicted when capillaries are excluded; explicitly modeling the capillary bed corrects the peak systolic pressure from roughly 130 mmHg down to approximately 90 mmHg. Wave intensity analysis in the MPA (Figure \ref{fig:waveintensity}) reveals that backward-traveling waves are reduced in networks containing capillaries compared to those without, most significantly in the case of PH pathophysiology. This highlights the biomechanical importance of the capillaries in disease.

\begin{table}
\caption{Mean, minimum, and maximum pressure (mmHg), flow (mL/s), and shear stress (g/cm/s$^2$) predictions in arterial (blue) and venous (red) networks. \\[-0.2cm]}
\centering
\label{tab:w_woCap}
\begin{tabular}{l |c c : c c | l | c c : c c} \hline
& \multicolumn{2}{c:}{\textbf{With Capillaries}} & \multicolumn{2}{c|}{\textbf{Without Capillaries}} &
& \multicolumn{2}{c:}{\textbf{With Capillaries}} &
\multicolumn{2}{c}{\textbf{Without Capillaries}} \\[0.25cm]
\textbf{Control} & \textbf{MPA} & \textbf{Veins} & \textbf{MPA} & \textbf{Veins} &
\textbf{\hspace{0.35cm}PH} & \textbf{MPA} & \textbf{Veins} & \textbf{MPA} & \textbf{Veins}\\ \hline 
Mean BP & 11.9 & 4.06 & 12.1 & 4.06& 
Mean BP & 59.3 & 20.6 &65.4 & 20.6 \\
Max BP  & 20.1 & 4.07 & 20.5 & 4.25 &
Max BP  & 89.6 & 20.6 & 104 & 20.8 \\
Min BP  & 7.73 & 4.06 & 5.13 & 3.92&
Min BP  & 37.9 & 20.5 & 33.9 &20.4\\
Mean BF & 87.5 & 87.5 & 87.5 & 87.5 &
Mean BF & 70.0 & 70.0 & 70.0 & 70.0 \\
Max BF  & 306  & 89.7 & 305 & 204 &
Max BF  & 243 & 90.5 & 241& 136\\
Min BF  & -11.5 & 85.9 & -10.9 & 11.9 & 
Min BF  & -6.21 & 49.7& -5.40& 19.1 \\ \hline
\end{tabular}
\end{table}

\begin{figure}
\centerline{\includegraphics[width=\textwidth]{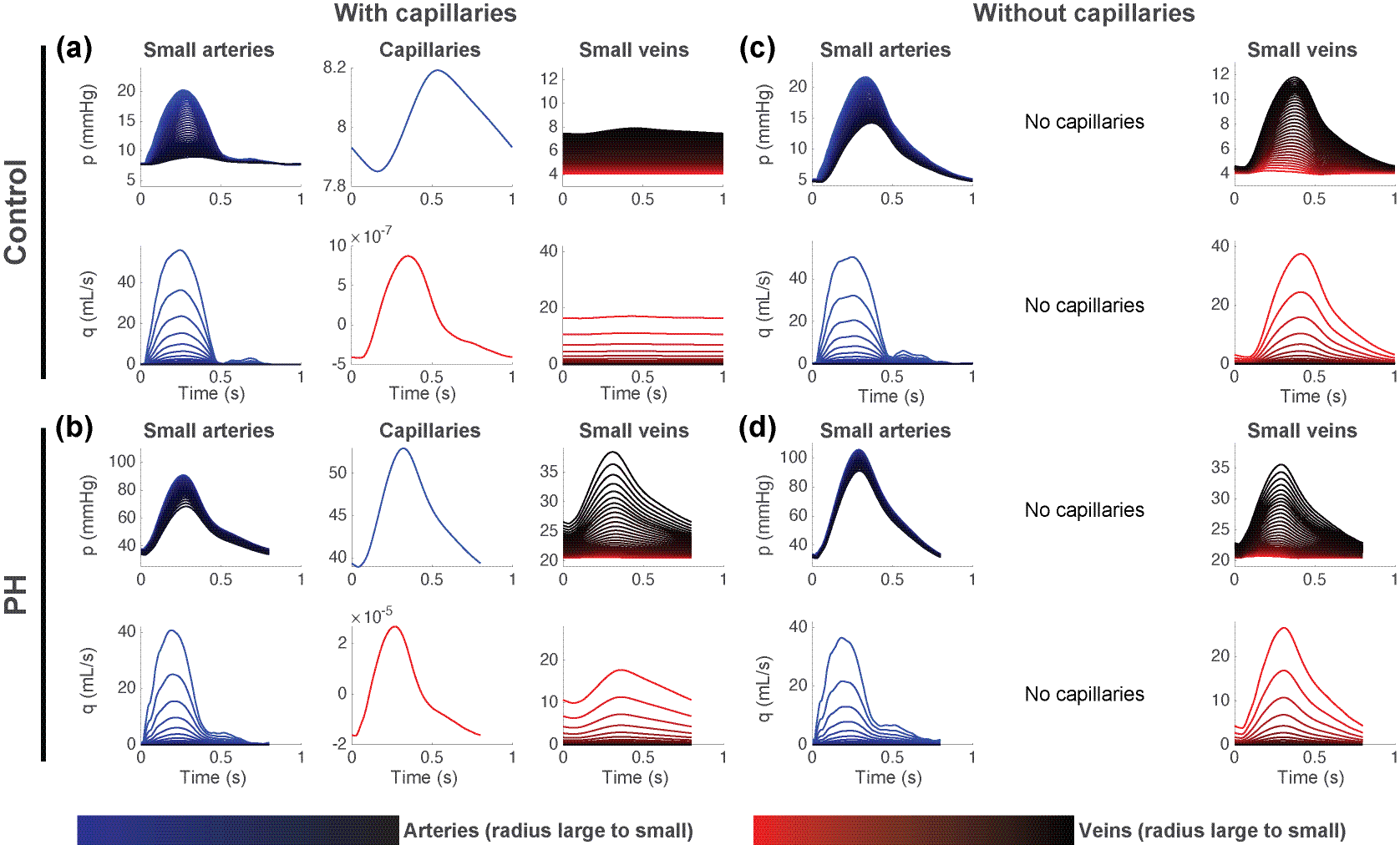}}
\caption{Simulations of pressure and flow waveforms with (a,b) and without (c,d) capillaries for a control (upper) and hypertensive (lower) subject. All results are computed along the $\alpha$-branch (the longest branch)of the structured trees.} Small arteries are denoted by blue/cyan lines and small veins by red/magenta lines. Note the change in waveform shape and damping of oscillations with the capillary network.
\label{fig:micro_resultsPulse}
\end{figure}

\begin{figure}
\centerline{\includegraphics[width=\textwidth]{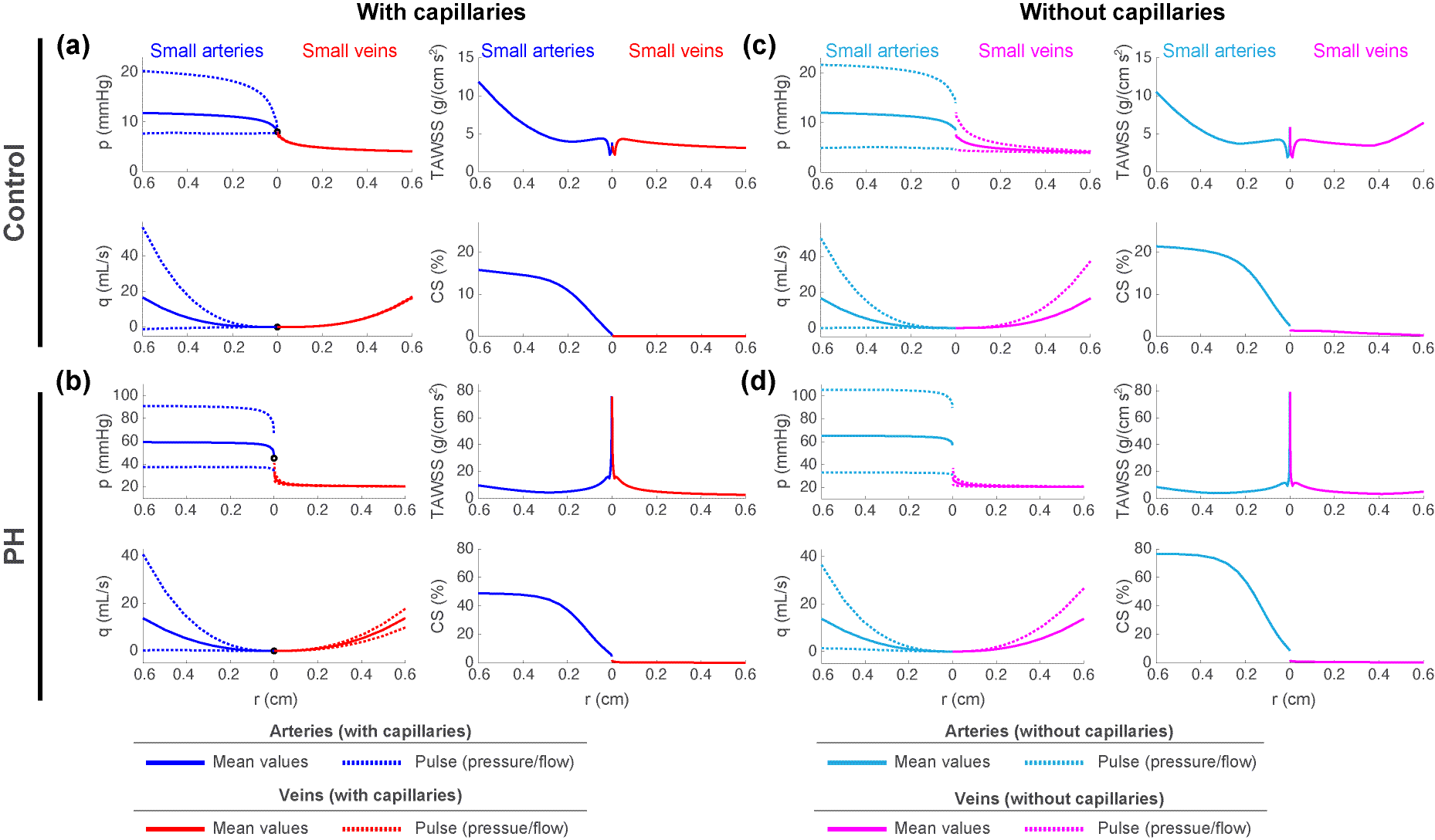}}
\caption{Variation in the mean, maximum and minimum pressure $p$, and flow $q$ as a function of vessel radius $r$, with (a,b) and without (c,d) capillaries, for a control (upper) and hypertensive (lower) subject. Arteries are denoted by blue/cyan lines and veins by red/magenta lines. Mean values are denoted by solid lines and systolic/diastolic values by dashed lines. Results are generated at the beginning of each vessel along the $\alpha$-pathway, the pathway with the most vessels. We also show key biomarkers: time-averaged wall shear stress (TAWSS) and cyclic stretch (CS).}
\label{fig:micro_resultsMean}
\end{figure}

To characterize how adding capillary sheets affects hemodynamics in the microcirculation, Figures \ref{fig:micro_resultsPulse} and \ref{fig:micro_resultsMean} show the pulsatile and mean pressure, flow, and wall shear stress along the longest path in the structured tree (the $\alpha$-branch). Except the capillary bed, all predictions are computed at the beginning of each vessel. Values are computed in the center of the capillary connecting the arterial and venous $\alpha$-branch. Results are plotted as a function of vessel radius ($r$) on the arterial (blue/cyan) and venous (red/magenta) vessels along the $\alpha$-branch, with color gradients shifting from light to dark to denote the progression from large to small vessels. The pulsatile waveforms (Figure \ref{fig:micro_resultsPulse}) demonstrate that the capillary bed attenuates pulsatility in the distal microvasculature. In the healthy control model (Figure \ref{fig:micro_resultsPulse}(a)), arterial pressure ($p$) and flow ($q$) oscillations decrease as the vessel radius decreases. 

The capillary bed dampens the remaining pulsatility, resulting in steady profiles in the small veins. When capillaries are omitted (Figure  \ref{fig:micro_resultsPulse}(c)), this dampening is absent, and pulsatility propagates into the venous system. In the PH model, distal stiffening elevates systolic pressures in the small arteries to nearly 100 mmHg (Figure \ref{fig:micro_resultsPulse}(b)). The capillary bed provides resistance but does not fully isolate the venous system, allowing pulsatility to transmit into the small veins. Omitting the capillaries in the PH model (Figure \ref{fig:micro_resultsPulse}(d)) overestimates this transmitted venous pulsatility. Figure \ref{fig:micro_resultsMean} illustrates these spatial trends as a function of vessel radius ($r$). In the healthy model with capillaries (Figure \ref{fig:micro_resultsMean}(a)), the pulse pressure,  narrows along the arterial tree and is fully attenuated at the capillary bed ($r=0$). The mean pressure drops gradually across the arteries and continues to fall across the capillaries. Without capillaries (Figure \ref{fig:micro_resultsMean}(c)), the pulse pressure remains wide and transmits directly into the venous tree. In the PH model (Figure \ref{fig:micro_resultsMean}(b)), mean arterial pressure remains elevated in the larger vessels and exhibits a sharp drop near the capillary bed. Finally, structural biomarkers such as time-averaged wall shear stress (TAWSS) peak sharply at the capillary junction ($r \to 0$) due to the reduced radius. At the same time, cyclic stretch (CS) decreases in the smaller distal vessels.

\subsection{Parametric analysis}
To gain insight into the interplay of arterial, capillary, and venous dynamics, we performed a parametric analysis on the complete multiscale model. This analysis evaluates how variations in key geometric and mechanical parameters (listed in Table \ref{tab:nominalparameters}) affect hemodynamics across the complete pulmonary circuit. 
For the large arteries and veins, we modulated left atrial pressure (LAP, mmHg), vessel radii ($r_i$, cm), vessel wall stiffness ($Eh/r_0$, mmHg), and cardiac output (CO, mL/min), keeping the heart rate constant at $T=1$ s. In the small vessel structured trees, we examined the effects of a global radius scaling factor ($rs_i$ cm), vessel stiffness ($Eh/r_0$ mmHg), and the radius exponent ($\xi$). In the capillary network, we studied the effects of changing sheet height, compliance ($\alpha_c$, (cm s)$^2$/g),  post density ($\kappa$, dimensionless), and a maximum radius ($r_{\textrm{ladder}}$, cm) specifying vessels for which ladders should be included. 

The results of these simulations (shown in the Supplement) demonstrate how each factor impacts flow and pressure in the arteries and veins. The parameters impacting dynamics the most are LAP and CO. An increase in LAP leads to increases in both mean and pulse pressure throughout the system, and inverse effects are observed in response to a decrease in CO. Changes in LAP do not affect flow, but changes in CO affect both flow and pressure. Changes in large vessel stiffness and radii have minor effects, likely since most of the vasculature is represented by structured trees. 

To simulate changes in microvascular vessel caliber, we include a scaling factor ($rs_i$), which uniformly modulates the vessel radii throughout the structured tree. Decreasing $rs_i$ across both the arterioles and venules leads to an increase in mean pressure, pulse pressure, and pulse flow within the MPA. The most influential structured tree parameter is the radius exponent ($\xi$). Smaller values of $\xi$ lead to less dense trees. Smaller (less dense) trees have a higher pulse pressure, mean pressure, and venous pulse flow, yet the mean flow stays constant. This finding agrees with results from our previous study \cite{qureshi2014}.

Within the capillary network, sheet height ($h$), membrane compliance ($\alpha_c$), and post density ($\kappa$) all independently impact hemodynamics. Increasing sheet height or decreasing post density increases venous pulse pressure and flow but does not alter mean values. However, changes in these parameters remain localized and do not propagate upstream to impact arterial dynamics. In contrast, altering the ladder threshold radius ($r_{\text{ladder}}$), which dictates the vessel size at which the capillary bed bridges the arterioles and venules, has an effect on the full system. Modulating $r_{\text{ladder}}$ has a nonlinear effect on the venous side, where the pulse pressure and flow initially decrease and then increase.  Additionally, changing $r_{\text{ladder}}$ decreases the  MPA pulse and mean pressure.

Finally, we explore effects of symmetry in the microcirculation. Asymmetric branching is known to reduce wave reflections, generating more realistic waveforms. Here we explore whether this phenomenon is preserved when we include a capillary network, which is already symmetric. Results of these simulations are shown in Figure 2 of the Supplement. Again, the morphometry of the network has a significant impact on predictions. This result agrees with our findings in \cite{qureshi2014} and the observation that the radius scaling factor $\xi$ is influential.

These parameters are all related to PH progression, so varying them helps isolate how specific structural changes impact hemodynamics. For example, while upstream arterioles undergo muscularization, remodeling at the capillary level is instead driven by endothelial cell proliferation and wall thickening. In our model, we represent this by decreasing both the effective sheet height and membrane compliance. These mechanical restrictions limit capillary recruitment and alter local flow patterns, which captures the broader microvascular decline seen in PH, such as capillary rarefaction, reduced post density, and increased vessel tortuosity (though tortuosity is not explicitly modeled here) \citep{dayeh2016,gerges2015,miller2013}.

\subsection{Disease progression}
\begin{figure}
\centerline{\includegraphics[width=\textwidth]{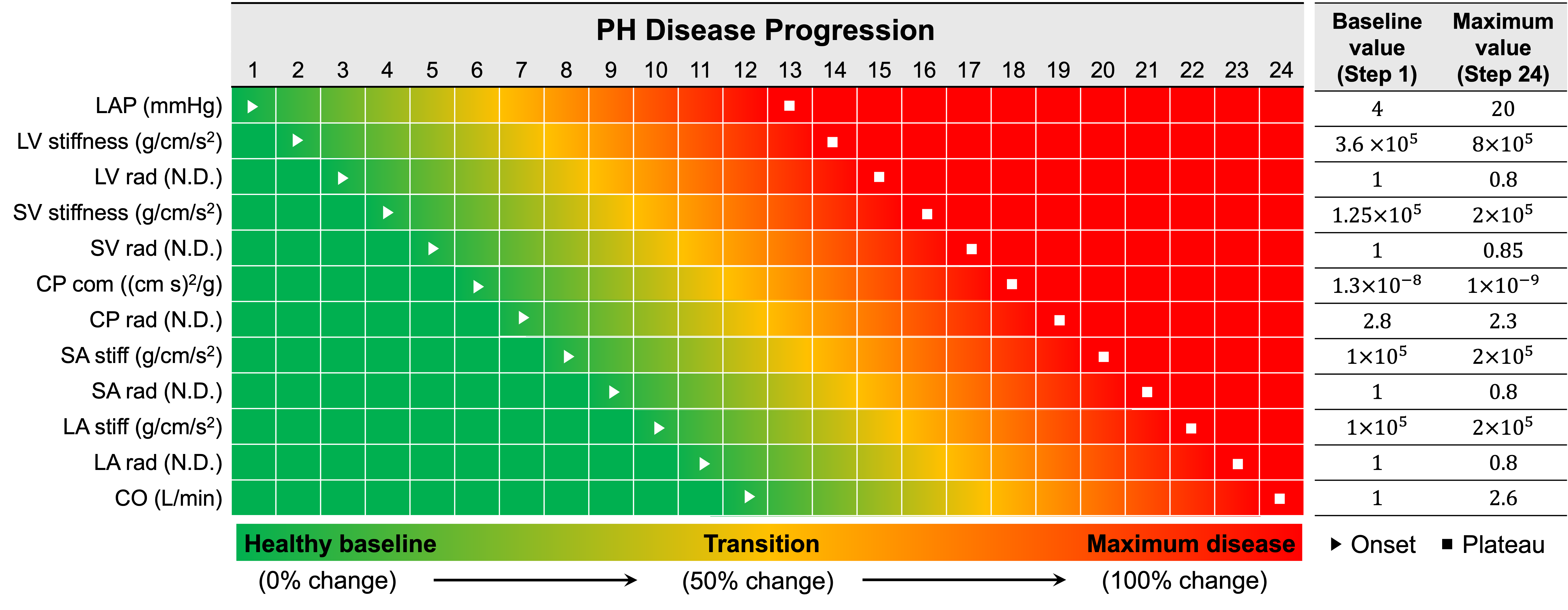}}
\caption{Disease progression is simulated by changing model parameters associated with PH from their lower to upper bounds, listed on the right side of the Figure}. These changes are split into equidistant steps, and at each step an additional factor is imposed. The light-colored bars indicate when the factor is initiated and when it reaches its maximum value. The parameter in question remains at its maximum value (indicated by a dark bar).
\label{fig:severitychanges}
\end{figure}

We assessed how the vasculature responds to disease progression. Following the approach of \citet{ebrahimi2021} and the hypothesis that the transition from Ipc-PH to Pcp-PH involves remodeling that starts in the venous system, transitions to the capillaries, finally affecting the right ventricle \citet{Huston2022}, we modeled disease severity by introducing gradual pathological changes illustrated in Figure \ref{fig:severitychanges}. This study is set up to demonstrate that the multiscale model proposed here can examine effects of incremental changes. The parameters changed likely all are affected by remodeling, but as addressed in the introduction, the exact order and mechanisms involved are not fully understood \citep{allen2023}. Parameters are incrementally scaled from their healthy baseline to a diseased upper bound in equidistant steps, with a new parameter alteration initiated at staggered intervals. Once a parameter reaches its upper bound, it remains fixed at that value for the remainder of the simulation. In Figure \ref{fig:severitychanges}, triangles ($\blacktriangleright$) mark the step at which each parameter begins to change, and squares ($\blacksquare$) mark the step at which it reaches its maximum-disease value, after which it remains constant. The range over which parameters are varied is listed in the right column of the figure.

\begin{figure}
\centerline{\includegraphics[width=0.8\textwidth]{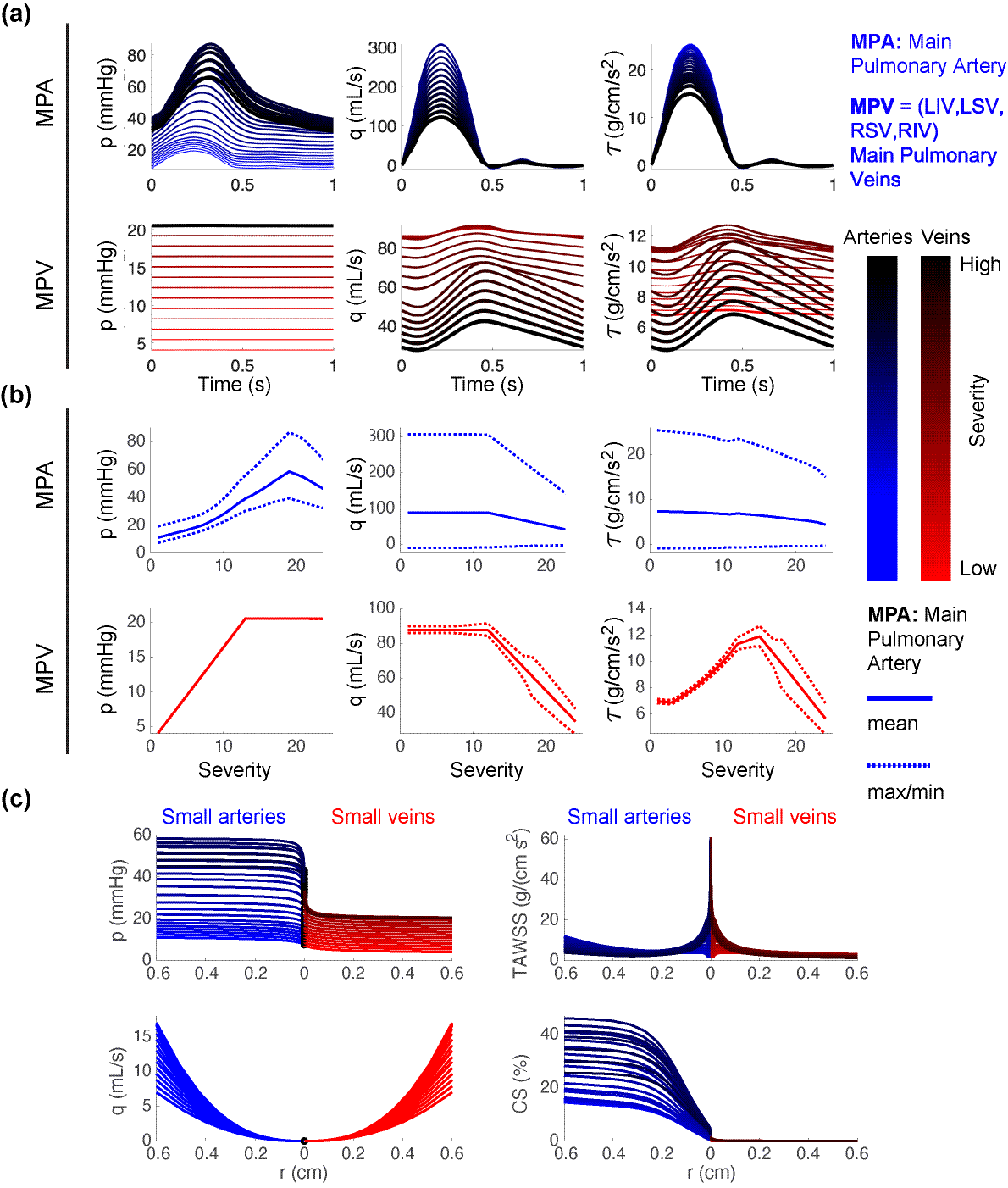}}
\caption{Simulations depicting the effects of disease severity in the large and small arteries and veins. (a) Large artery and vein pulse pressure, flow, and shear stress, with increasing disease severity. (b) Large vessel mean values (blue (arteries)/ red (veins)) and pulse pressure/flow  (dashed lines) with increased disease severity. Simulations include 24 steps, marked 1--24 on the x-axis. (c) Microvascular predictions. Each panel shows the time-averaged feature as a function of radius along the alpha branch. Values to the left (blue) of 0 denote predictions along the arterial structured tree, while values to the right (red) denote responses in the venous structured tree. }
\label{fig:severityPressureFlow}
\end{figure}

Results (Figure \ref{fig:severityPressureFlow}) show that the mean pulmonary arterial pressure steadily increases with disease progression. Once the left atrial pressure (LAP) reaches its upper bound (panel b), the rise in mean pressure slows, but the pulse pressure (cyan lines) continues to climb. Interestingly, when cardiac output begins to decline, mimicking the onset of heart failure, both the mean and pulse pulmonary arterial pressures start to drop. In the large arteries, shear stress falls slightly over the course of the disease. Venous shear stress, however, initially increases before dropping in tandem with the declining cardiac output.

In the microcirculation (Figure \ref{fig:severityPressureFlow}c), mean pressure increases in both the arteries and veins, while the mean flow decreases. Cyclic stretch also increases. Time-averaged shear stress drops in the arteries but rises in the veins. One critical observation is that in the smallest vessels, the time-averaged wall shear stress increases significantly.

\section{Discussion}
In this study, we present a pulse wave propagation model of the pulmonary circulation that integrates three submodules: (1) large arteries and veins with dimensions extracted from CT images, (2) arterioles and venules represented by structured trees, and (3) capillaries modeled as elastic sheets. This framework extends existing pulmonary circulation models that either neglect the pulmonary capillaries \citep{vaughan2010,mackenzie2021,bartolo2022,qureshi2014,colebank2021phd}, rely on an idealized representation of the vascular structure \citep{zhou2006,tawhai2011}, consider the arterial tree in isolation \citep{Colebank2021sens,olufsen2000}, or only investigate steady-state dynamics \citep{burrowes2005,clark2010}. By addressing these limitations and solving the full wave-propagation equations across compliant vessels, our model accurately captures the physiological wave damping observed throughout the system. Our findings demonstrate that explicitly including the capillary bed is essential for tracking wave propagation through the pulmonary system and studying how hemodynamics shifts with disease severity. This is crucial for accurately predicting the evolution of mechanical forces in the microcirculation, where the pathophysiology of PH-LHF begins.

Previous studies have repeatedly emphasized the need to develop a pulmonary capillaries model and couple it with a 1D arterial-venous model to accurately predict pulmonary hemodynamics \citep{vaughan2010,qureshi2013,mackenzie2021}. For instance, when calibrating a coupled 1D arterial-venous model to healthy hemodynamics, \citet{vaughan2010} encountered significant limitations when attempting to represent the smallest vessels using only a structured tree. Specifically, extending the structured tree down to a minimum radius of  $r_\text{min}=0.001$ cm, which includes vessels of near-capillary size, caused the predicted upstream arterial pressures to rise well above physiological bounds, demonstrating that the structured tree formulation alone is not suitable for describing capillary hemodynamics. To produce physiologically realistic pressures, that model had to be truncated at $r_\text{min}=0.005$ cm, with the caveat that explicitly modeling the capillary bed would be necessary to accurately capture the resistance and pressure drop across the smallest vessels. This limitation was echoed in \citet{qureshi2013} and \citet{mackenzie2021}, who noted that truncating the arterial and venous trees at $0.005$ cm prevents the model from providing detailed spatial profiles of the microvascular pressure drop patterns. As hypothesized in these prior works, incorporating a specialized capillary model, such as the sheet architecture developed here, successfully bridges this gap, significantly improving the accuracy and resolution of microvascular pressure predictions. 

In Figures \ref{fig:waveintensity}--\ref{fig:micro_resultsMean}, we compare hemodynamic predictions between models with and without explicit capillary networks. While mass conservation ensures the mean volumetric flow is identical across all configurations, the ladder-like capillary structure yields lower microvascular pressure. This pressure reduction fundamentally decreases mechanical stress on the system by lowering overall vascular resistance ($p/q$) \citep{clark2010}. Furthermore, our results demonstrate that the addition of the capillary bed alters the effective compliance of the system, which directly modulates wave intensity profiles, especially in the disease state. It should be noted, however, that inserting the capillary sheets inherently adds resistance and compliance, meaning part of the observed pressure shift stems from changes in net downstream impedance rather than capillary dynamics alone. While structured tree parameters could be retuned to match baseline resistance between setups, we kept them fixed to evaluate the structural addition directly. Beyond the improvements in hemodynamic predictions, the ladder-shaped capillary sheet structure is important because it provides a more accurate morphological representation of the pulmonary microvasculature, aligning with anatomical imaging studies \citep{gao2005,weibel1984,clark2010}. 

Previous studies have modeled the pulmonary acinus by restricting capillaries to connect only to the final terminal arteriole and venule \citep{dhadwal1997,huang1996}. Although these models offer valuable initial insights, this simplified structure neglects the intricacies of the true capillary network by underestimating the gas-exchange surface area and disregarding the hierarchical branching of the acinus. As a result, models relying solely on terminal capillaries tend to artificially overestimate the pressure and flow distribution across the network, limiting their predictive accuracy, particularly under pathological conditions. Furthermore, models that omit explicit capillary representation entirely suffer from a fundamental loss of spatial and functional detail, leading to misrepresentations of global hemodynamic profiles. Thus, we follow the structure first proposed by \citep{clark2010} that models capillaries in a ladder-like formation. By bridging the arterioles and venules at multiple generations within the acinus, this formulation facilitates a physiologically realistic distribution of blood flow throughout the system. 

Even when models with and without capillaries predict similar mean values in the large vessels, their pulsatile dynamics differ significantly. Models without capillaries can still adequately predict overall hemodynamics for both healthy and  PH phenotypes within accepted physiological ranges \citep{allen2014,bartolo2022}. However, they fail to capture the natural wave damping provided by the capillary bed. The primary advantage of the complete framework presented here is its ability to provide high-resolution insights into the microcirculation. Despite this added structural complexity, the model remains computationally efficient because it is solved via recursion. It is also highly practical because its parameters are based directly on measurable physiological quantities \citep{fung1969,fung1972elast,fung2013C,clark2018}. As \textit {in vivo} imaging continues to improve, new experimental data can easily be fed into the model to refine it \citep{grothausmann2017,willfuhr2015}. For example, \citet{grothausmann2017} recently used fluorescence microscopy to reconstruct highly detailed 3D models of the pulmonary acinus by embedding lung tissue slices in resin. This study validates the ``sheet-and-post'' description of the capillaries and provides empirical estimates of capillary density that directly inform our mathematical framework.

In addition to being the first to integrate a sheet model into a structured tree framework, this study is also the first to systematically evaluate how specific capillary structural properties influence global pulmonary hemodynamics. Previous models \citep{tawhai2011,clark2018,clark2010} have generally relied on fixed baseline values reported by \cite{fung2013C} without exploring how varying these parameters impacts the wider system. However, performing a parametric analysis is essential when mathematically representing a physiological system. Biological traits are inherently variable across individuals, and fixed mathematical values are ultimately approximations of a complex \textit{in vivo} reality. Our analysis demonstrates that exploring this parameter space is crucial, as each geometric and mechanical factor uniquely alters flow and pressure dynamics throughout the system. Furthermore, connecting these specific parameters to PH progression can illuminate disease mechanisms and potential therapeutic targets. Because microvascular remodeling is widely hypothesized to be the initiating event in PH \citep{guazzi2014}, explicitly modeling these early capillary changes is a vital step toward understanding the pathogenesis of the disease. 

We evaluated baseline sheet heights ($\hbar_0$) ranging from $2$ to $20 \ \mu$m, with a nominal sheet height of $3.5 \ \mu$m, to investigate the theoretical and physiological implications of this parameter. The lower limit, $2 \ \mu$m, is based on the size of a red blood cell, which has a diameter of $6-8 \ \mu$m and a thickness of $2-2.5 \ \mu$m. Because resting capillaries are naturally narrower than a red blood cell by about 25\%, the cells must physically deform to pass through, a mechanism that increases surface area contact and enhances oxygen transfer to tissues \citep{snyder1999}. This minimum bound is further supported by  \textit{ex vivo} lung photomicrographs, which confirm sheet heights as narrow as $2.5 \ \mu$m, depending on the type of species and the difference between air and blood pressure \citep{fung2013C,sobin1979}.  While \citet{sobin1979} reported a maximum sheet height of $7.7 \ \mu$m, we intentionally extended our upper bound to $20 \ \mu$m. This matches our minimum structured tree radius ($r_\text{min} = 10 \ \mu$m), ensuring a continuous geometric transition (a diameter of 20 $\mu$m) between the terminal vessels and capillary bed. 

Hemodynamically, varying the sheet height produces clear systematic effects.  Because a greater sheet height directly lowers microvascular resistance, increasing $h_0$ reduces overall pressure and area deformation while maintaining flow and WSS in the MPA. On the venous side, a greater sheet height increases pulsatility through the bed, leading to a slight increase in pressure and WSS in the LSV. In the microcirculation, an increased sheet height results in a lower WSS in the $\alpha$-path compared to the $\beta$-path, likely due to differences in flow distribution. In the PH phenotype, these mechanical relationships related to changing $h_0$ become especially pronounced. Because the progression of PH often includes vasoconstriction and microvascular occlusion, reducing the capillary sheet height in our model successfully replicates the mechanical consequences of this pathological narrowing \citep{bartolo2022,humbert2022}.

We also investigated the effects of altering the friction factor $\kappa$, which serves as a proxy for sheet post density, between $12$ and $80$.  The minimum bound of $\kappa=12$ represents the theoretical limit of an idealized sheet without posts \citep{fung1969}. The upper bound of $\kappa=80$ was established based on experiments carried out by \citep{yen1973}, who calculated this maximum friction factor by measuring the pressure drop of a homogeneous silicone fluid with a known viscosity across a simulated capillary bed. As shown in the Supplement Figures 10 and 25, a lower value of $\kappa$ correlates with slightly elevated pressures and WSS, particularly in the simulated PH patient. Within the microcirculation, reducing $\kappa$ increases pressure along the pre-capillary arterial pathways, whereas post-capillary venous pressures remain largely unaffected. Biomechanically, a reduced number of posts per unit area yields a less rigid capillary sheet structure, which effectively lowers local microvascular resistance. The overall hemodynamic impact of this structural change is more pronounced in the PH case, where the pulmonary vascular system is already compromised, making it more sensitive to structural changes.

The membrane compliance $\alpha_c$ influences the ability of capillary sheets to deform in response to transmural pressure changes. Based on established physiological measurements \citep{clark2010,clark2018,fung1969}, we varied $\alpha_c$ across a range of $1\times 10^{-9}$ and $1\times 10^{-7}$ cm\,s$^2$/g. Lower values of $\alpha_c$ reflect a stiffer capillary membrane that actively resists distension. Because these rigid walls cannot expand to absorb pulsatility, the network exhibits higher overall pressures and elevated peak flow rates. Conversely, a higher $\alpha_c$ yields a more compliant membrane that readily distends to accommodate incoming blood volume. This dynamic expansion effectively lowers microvascular resistance and buffers the cardiac pulse, leading to a decrease in the pulsatility of both pressure and flow on the post-capillary venous side.

We established a physiological range for the capillary sheet length $l_c$  of 300 -- 2000\,$\mu$m to evaluate its impact on spatial pressure gradients across the pulmonary network. This range encompasses experimental measurements by \citet{sobin1970}, who reported an average arteriole-to-venule path length of $556 \ \pm 285 \ \mu$m, and theoretical estimates of $1186 \ \mu$m by \citet{zhou2006}. Moreover, analyzing the anatomical circumference of alveolar clusters confirms that this 300 -- 2000\,$\mu$m range is physiologically accurate \citep{weibel1963,weibel1984}.  Increasing the sheet length produces a linear rise in pre-capillary arterial pressures. This elevation is determined by the cumulative viscous resistance that blood encounters over the extended path. This geometry-motivated increase in resistance becomes more significant in the PH case, where vascular remodeling and stiffening further impede blood flow.

The number of capillary generations controlled by $r_{\textrm{ladder}}$ dictates the branching complexity and effective density of the microvascular network. By varying this parameter from $ 1.5$ to $ 50 \times r_{\textrm{min}}$, generates between $1$ and $15$ generations, we simulated the hemodynamic consequences of both physiological capillary recruitment and pathological rarefaction. Adjusting $r_{\textrm{ladder}}$ also tests the necessity of our ladder-like formulation by comparing the dynamics of a complex, hierarchical architecture against the model with no capillaries. Reducing the number of generations elevates upstream arterial pressures. Because fewer generations provide fewer parallel pathways for blood transit, a lower factor fundamentally increases the global microvascular resistance. This resulting pressure burden on the arterial tree underscores the physiological requirement for a dense, highly branched capillary network to distribute blood volume and minimize upstream mechanical stress efficiently.

A limitation of our model is the lack of sufficient data for validation. Due to the small dimensions of the alveoli and capillaries, direct \textit{in vivo} measurements of their physical properties are rare. The data used to calibrate the model were derived from larger-scale experiments by \citet{fung1969,fung1972pulm,fung2013C}. Although these experiments provide insight into the general behavior of capillary networks, \textit{in vivo} studies are essential to capture the complex, dynamic behavior of this system fully. Another limitation is that we imposed the same flow waveform for all simulations. This quantity is clearly patient-specific, but can be extracted from imaging studies. In addition, there are no existing pressure or flow measurements in the capillaries or veins to validate our results directly. Instead, we rely on reported MPA data to calibrate the model parameters and predictions.

Future work should focus on integrating the sheet model described here with airway models to study the interplay between gas exchange, oxygen transport, and blood flow. This will enable a comprehensive analysis of how ventilation and perfusion are related and how they are affected by diseases such as PH, pulmonary atelectasis, acute respiratory distress syndrome, and ventilator-induced lung injury. An example of a study that connects blood flow and alveolar airflow is \citet{si2022}. In this study, they consider the interactions between pulmonary ventilation, membrane diffusion, and capillary perfusion in an elastic 3D alveolar–capillary geometry. However, they considered only one alveolus in isolation from the pulmonary circulation, used an idealized geometry, and assumed constant capillary pressures. In contrast, employing our 1D model coupled with a sheet model will decrease the computational cost and allow the impacts of capillary dynamics to be studied across the entire system.
Another promising avenue is leveraging high-resolution 3D imaging, such as hierarchical phase-contrast computed tomography \cite{Yagis2024}, to measure properties of the capillary network. These datasets could directly inform sheet parameters, such as dimensions and vascular density, which will allow us to generate higher fidelity patient-specific models with data-derived parameters.
\subsection{Conclusion}
In this study, we have developed a novel pulse wave propagation model of large arteries and veins with geometry derived from images, small arterioles and venules modeled with a structured tree, and capillaries represented by a sheet. This comprehensive approach allows us to predict pressure, flow, wall shear stress, and area deformation throughout the pulmonary circulation, which can be used to predict the impact of vascular remodeling on PH, advancing our understanding of its pathophysiology. Including the capillary sheet model is crucial because it accurately represents the microcirculation, where PH-LHF initiates, and improves on previous models that either neglected capillaries or used simplified descriptions. In addition, parametric analysis further enhances our understanding of how structural changes in the capillaries impact overall pulmonary hemodynamics, particularly in disease. The findings that the PH vasculature is more sensitive to changes in the capillaries may inform future therapeutic strategies. In addition, our model and algorithm are flexible. They can be coupled to different descriptions of the proximal vasculature and airflow models, allowing researchers to study ventilation-perfusion dynamics in the lung.  

\begin{bmhead}[Acknowledgements.]
The authors wish to acknowledge ongoing collaborations and many fruitful discussions on the pulmonary vasculature in health and disease with Dr. Naomi Chesler and her group at the University of California, Irvine.  We also acknowledge Ms. Liza Molchanova for the initial development of the algorithm to calculate admittance in ladder-like structured trees describing the arterioles and capillaries, as part of her undergraduate MSci project supervised by NAH.
\end{bmhead}
\begin{bmhead}[Funding.]
MAB and MSO were supported in part by the National Institutes of Health (R01HL147590). NAH was supported by a Research Fellowship (RF-2020-496\textbackslash 9) from the Leverhulme Trust.
\end{bmhead}
\begin{bmhead}[Declaration of Interests.]
The authors report no conflict of interest.\end{bmhead}

\end{document}


\maketitle

\vspace{1cm}
\noindent This document includes supplementary figures detailing computational results not included in the main text.

\section{Supplementary Methods}
\begin{figure}[ht]
\centerline{\includegraphics[width=\textwidth]{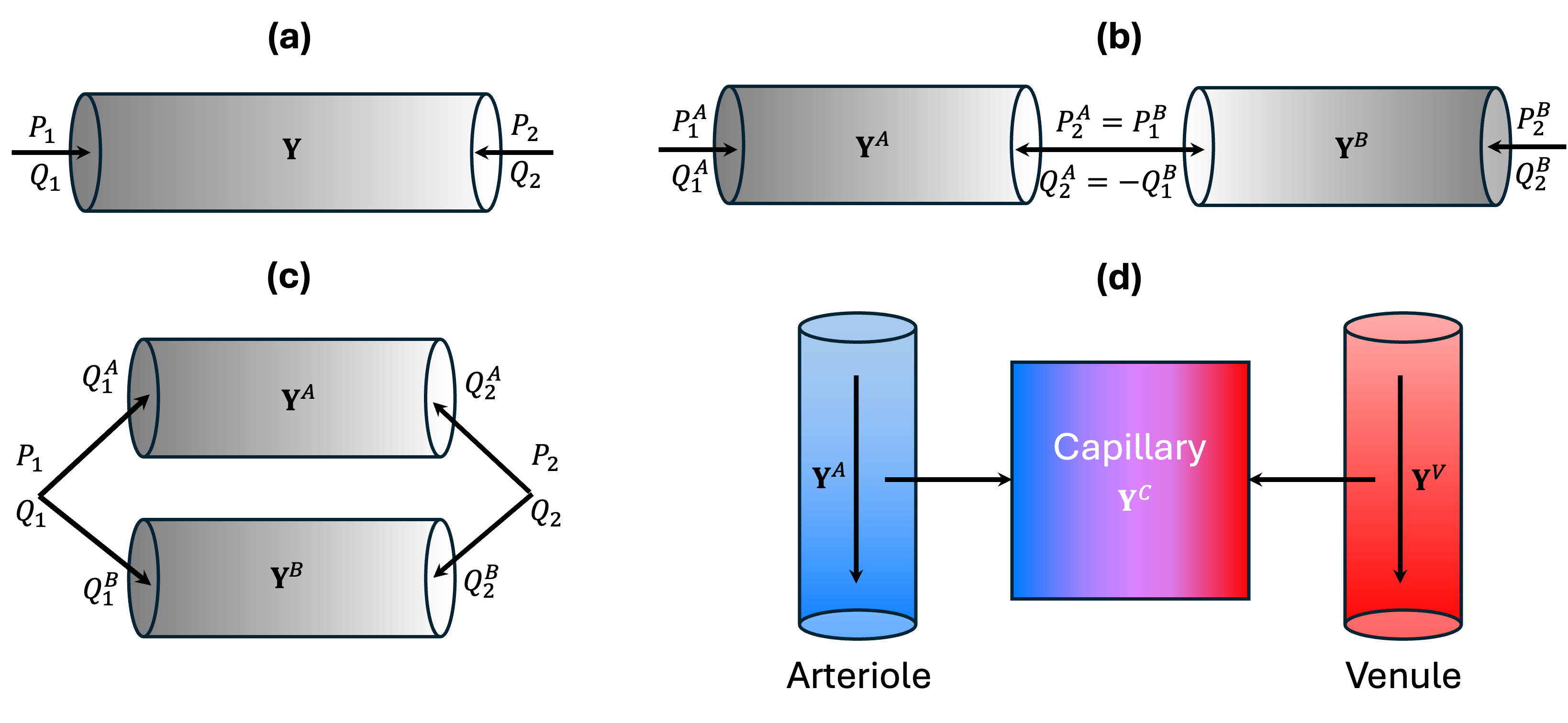}}
\caption{Relationship between pressure and flow using admittance matrices, where (a) represents the admittance of a single vessel, (b) represents vessels $A$ and $B$ connected in parallel, and (c) represents vessels $A$ and $B$ connected in series. Panel (d) represents the connection between an arteriole and venule joined at the midpoint by a capillary. The arrows represent the direction of blood flow.}
\label{fig:admittance_setup}
\end{figure}

Figure~\ref{fig:admittance_setup} illustrates how flow and pressure at the end of each small vessel is connected to form the admittance matrix for: (a) a single vessel, (b) two vessels in series, and (c) two vessels in parallel. (c) shows how this matrix is set up connecting an arteriole, capillary, and venule. 

\section{Supplementary Results}
\begin{figure}[ht]
\centerline{\includegraphics[scale=0.35]{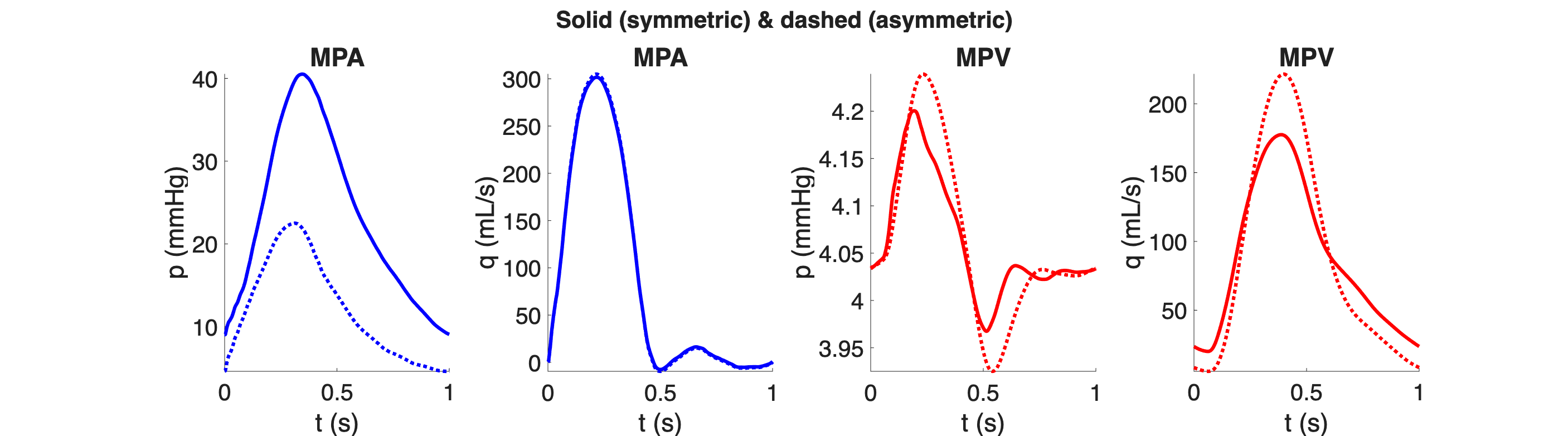}}
\caption{Simulations depicted with solid lines are from a symmetric network, and those with dashed }
\label{fig:symmetry}
\end{figure}

Figure \ref{fig:symmetry} show impact of imposing asymmetry in the vascular network.

\section{Parameter variation - effects on flow and pressure waveforms}
\begin{figure}[ht]
\centering
\includegraphics[scale=0.35]{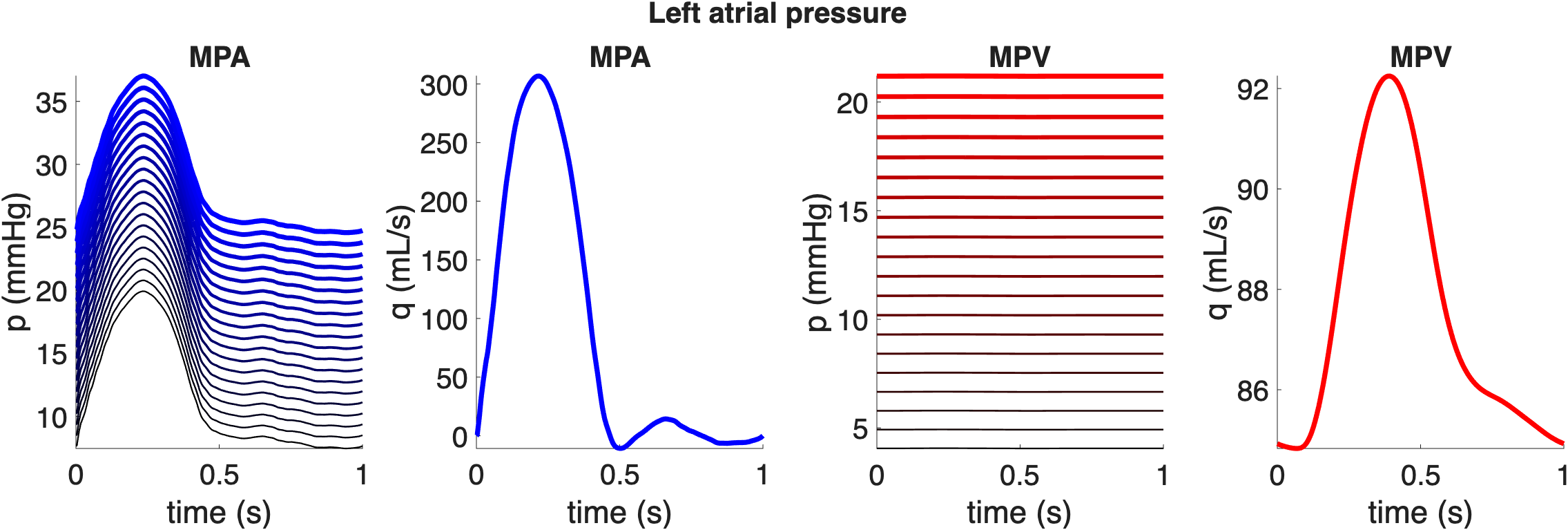}
\caption{Parameter variations: effects on pressure and flow wave forms to variations in left atrial pressure (LAP $= 4-20$ mmHg). }
\label{fig:LAPpulse}
\end{figure}

\begin{figure}[ht]
\centering
\includegraphics[scale=0.35]{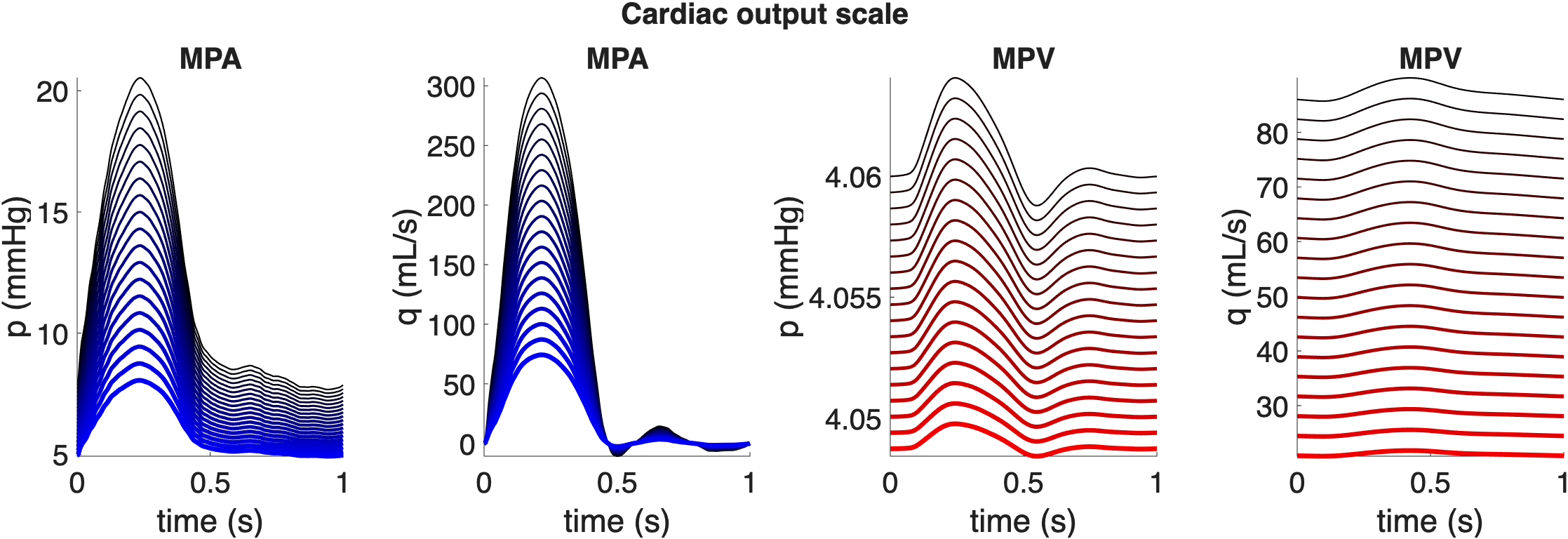}
\caption{Parameter variations: effects on pressure and flow wave forms to variations in  cardiac output scaling (CO$ = 5.3 - 1.1$  L/min).}
\label{fig:COpulse}
\end{figure}

Figures \ref{fig:LAPpulse}-\ref{fig:COpulse} show impact on flow and pressure waveforms from varying parameters one-at-the-time. For each parameter varied, the range over which they are varied is listed in Table 3 (in the main manuscript). Note the cardiac cycle length and fluid dynamics parameters including boundary layer thickness $\delta$, viscosity $\mu$, and density $\rho$ are not varied.

Results show that {\bf the main pulmonary arterial pressure} is impacted the most by left atrial pressure (LAP), radii of the small arteries (rsa) and veins (rsv),  the number of vessels with ladders, the radius scaling factor $\xi$,  and cardiac output.

The {\bf the main pulmonary arterial flow} is only affected by cardiac output. This can be explained by the flow being specified by the boundary condition. 

The {\bf venous pressure} is fixed at the left atrium, therefore similar to the pulmonary arteries, this pressure is primarily affected by LAP. But it should be noted that the venous pressure waveform, is affected by: venous radius, capillary compliance and height, the radius scaling factor $\xi$, and cardiac output. 

The {\bf venous flow} varies significantly with capillary compliance and height, the number of ladders (corresponding to the number of capillaries), the radius scaling factor $\xi$, and cardiac output.

\begin{figure}[ht]
\centering
\includegraphics[scale=0.35]{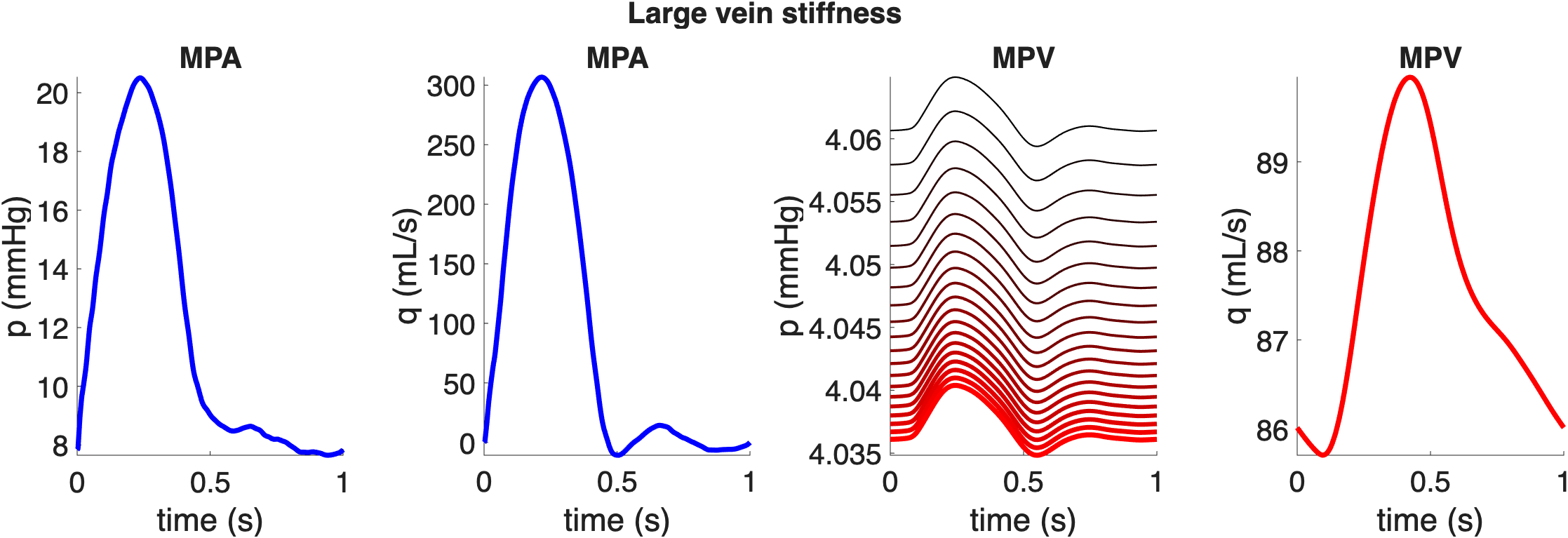}
\caption{Parameter variations: effects on pressure and flow wave forms to variations in large venous stiffness (LVstiff $= 3.6-8\times 10^5$ g/cm/s$^2$). }
\end{figure}

\begin{figure}[ht]
\centering
\includegraphics[scale=0.35]{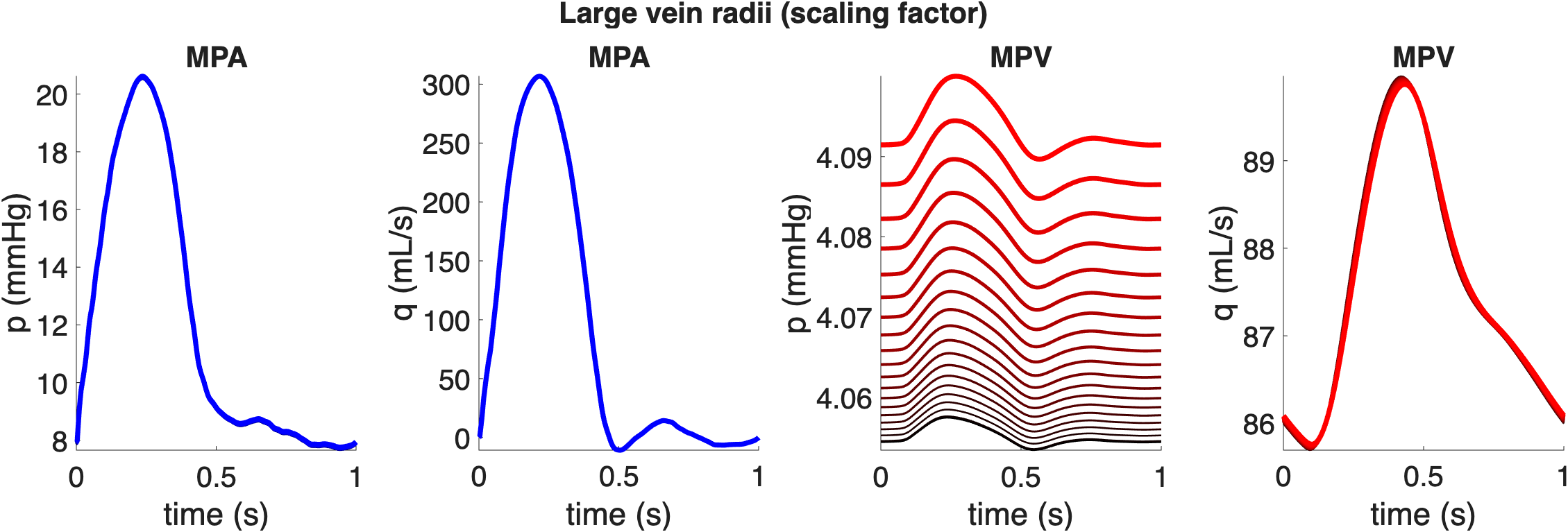}
\caption{Parameter variations: effects on pressure and flow wave forms to variations in large vein radii scaling ($r_v = 1.2 - 0.7$ non.dim.). }
\end{figure}

\begin{figure}[ht]
\centering
\includegraphics[scale=0.35]
{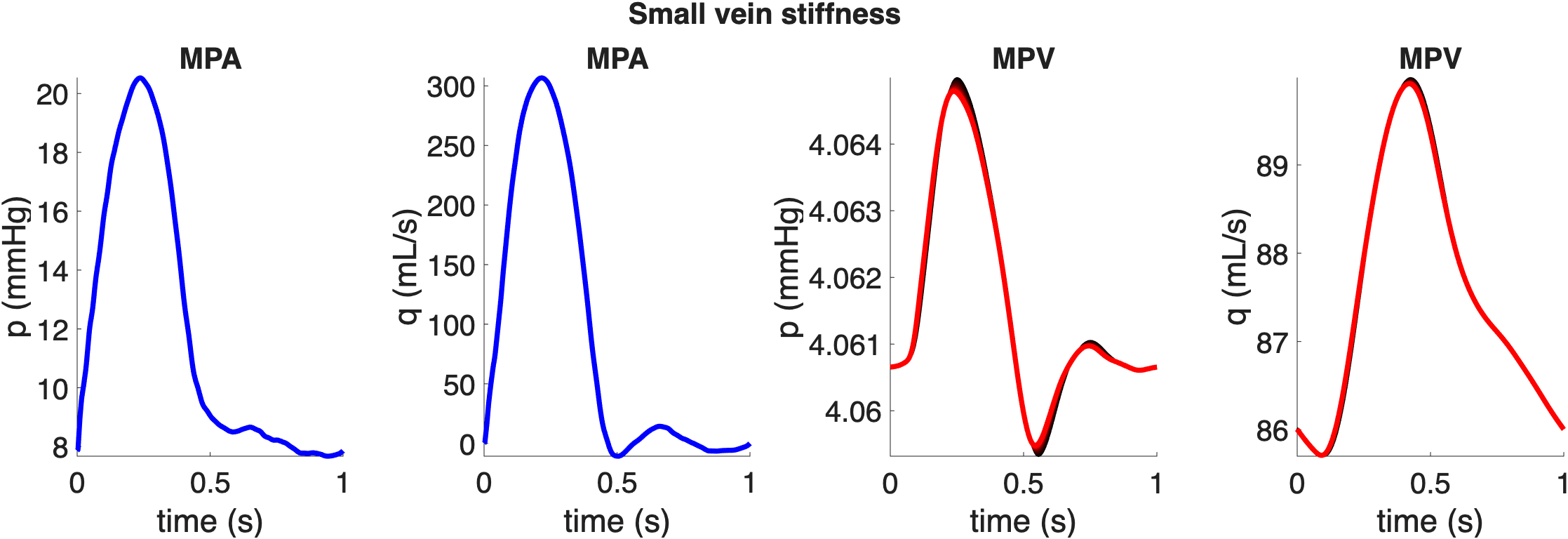}
\caption{Parameter variations: effects on pressure and flow wave forms to variations in the scale multiplying $k_{s1}$ and $k_{s3}$. The small venous stiffness scaling factor (SVstiff$_s$ $= 0.9-2$ non.dim.) giving $k_{s1} = 2.7 - 6 \times 10^5$  g/cm/s$^2$ and $k_{s3} = 0.9 - 2 \times 10^5$ g/cm/s$^2$. }
\end{figure}

\begin{figure}[ht]
\centering
\includegraphics[scale=0.35]{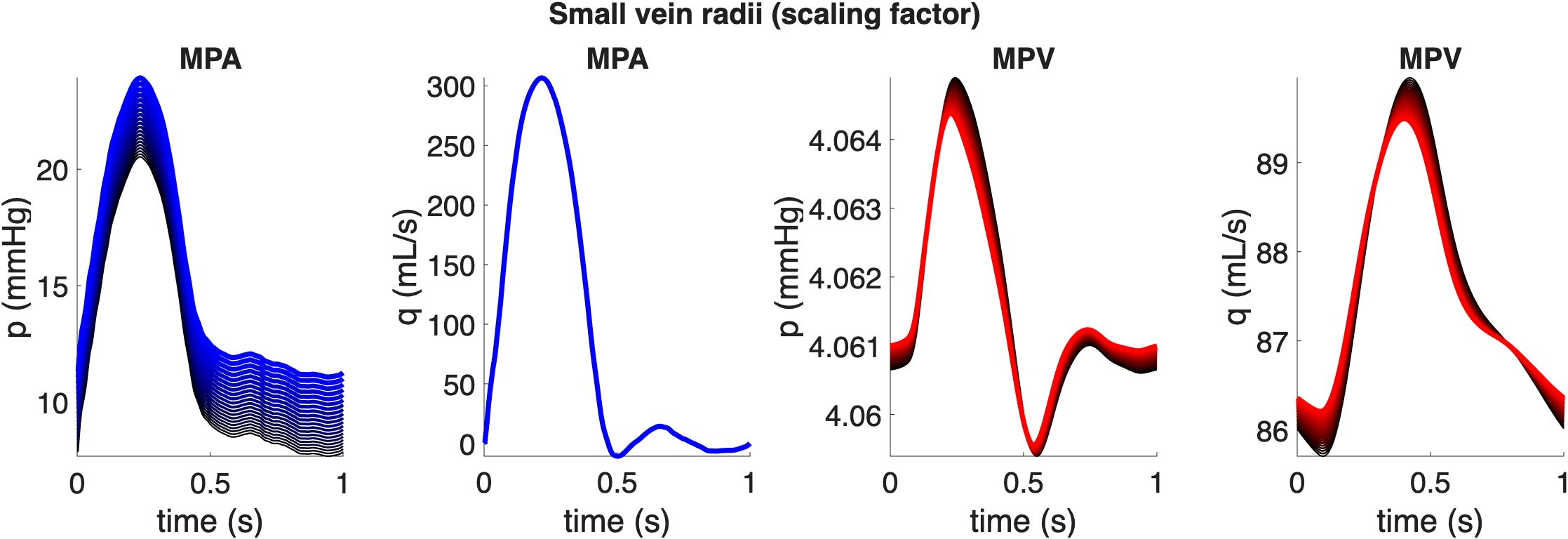}
\caption{Parameter variations: effects on pressure and flow wave forms to variations in small vein radii scaling ($r_{sv} = 1 - 0.85$ non.dim.). }
\end{figure}

\begin{figure}[ht]
\centering
\includegraphics[scale=0.35]{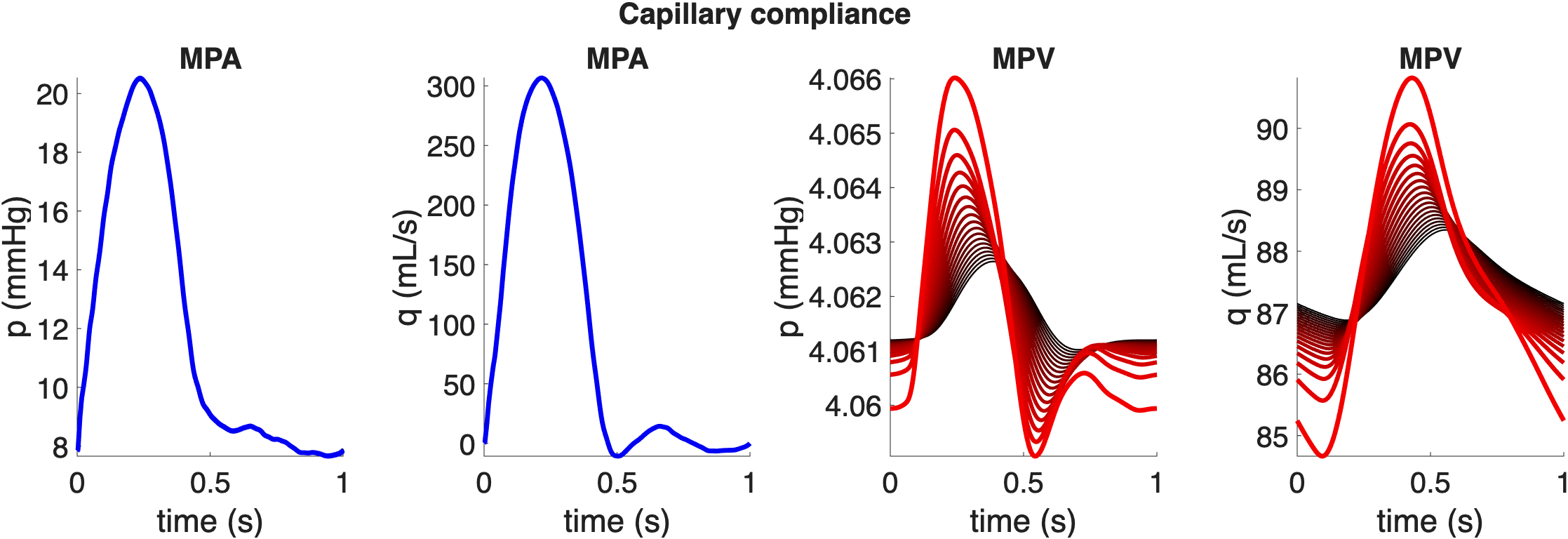}
\caption{Parameter variations: effects on pressure and flow wave forms to variations in capillary compliance (sheet compliance $\alpha_c = 1\times 10^{-7}-1\times 10^{-9}$ (cm s)$^2$/g). }
\end{figure}

\begin{figure}[ht]
\centering
\includegraphics[scale=0.35]{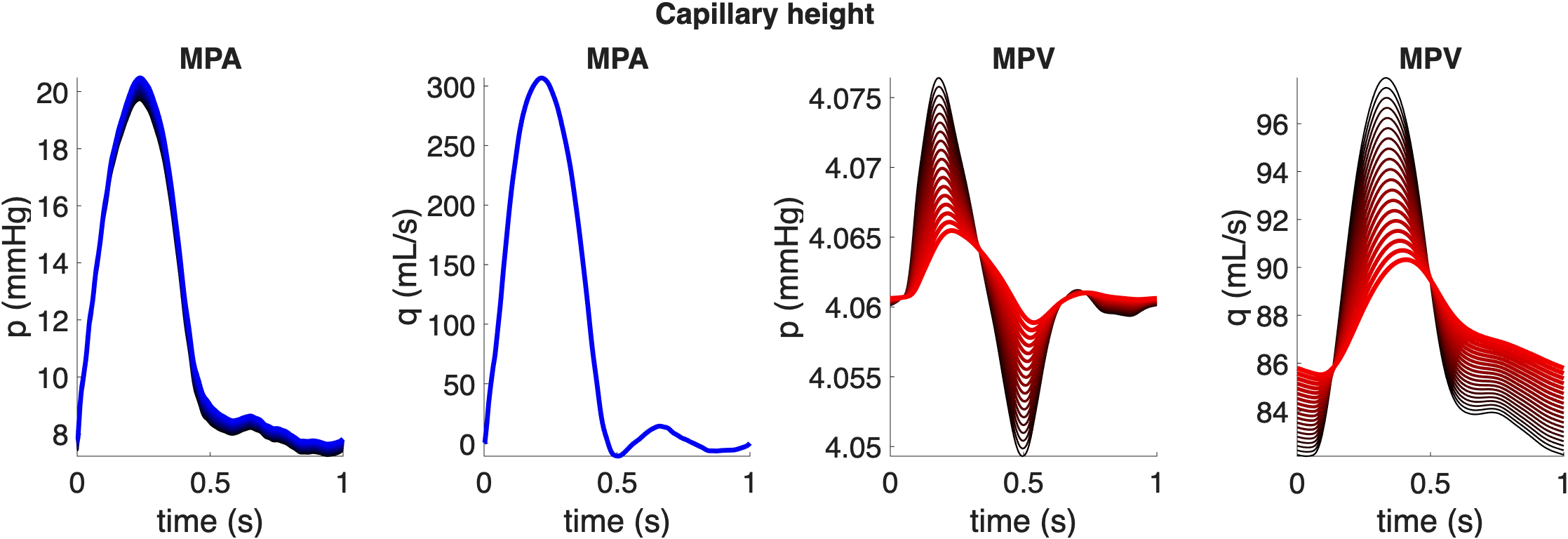}
\caption{Parameter variations: effects on pressure and flow wave forms to variations in capillary height ($h = 0.001 - 0.00035$ cm). }
\end{figure}

\begin{figure}[ht]
\centering
\includegraphics[scale=0.35]{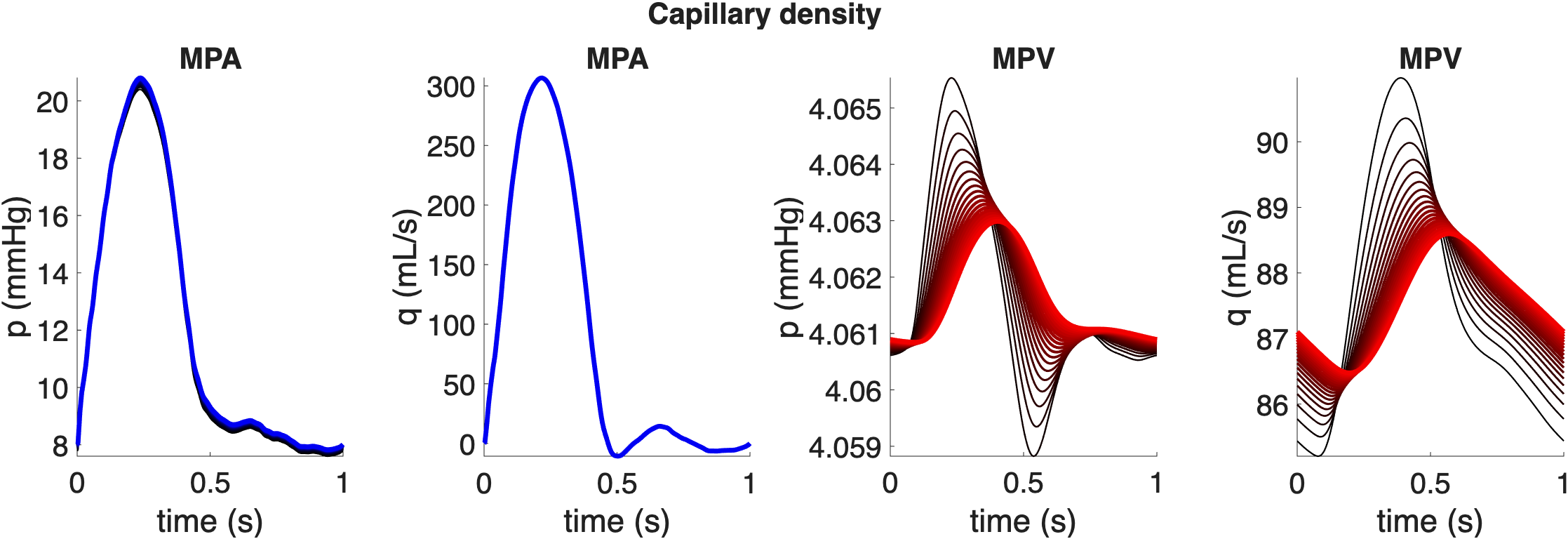}
\caption{Parameter variations: effects on pressure and flow wave forms to variations capillary density 
($\kappa = 100-10$ non.dim.). }
\end{figure}

\begin{figure}[ht]
\centering
\includegraphics[scale=0.35]{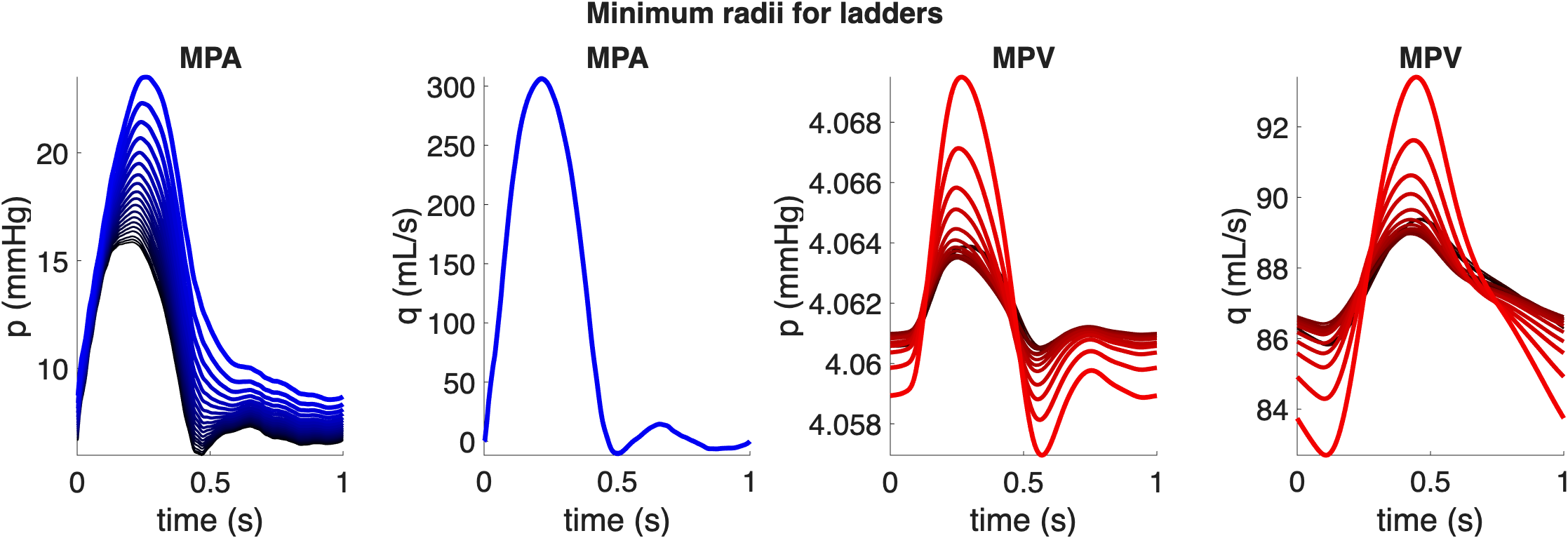}
\caption{Parameter variations: effects on pressure and flow wave forms to variations minimum radius at which ladders are generated. Indirectly modulating the number of generations with capillary ladders ($r_{\text{ladder}} = 50 - 1.5 \ r_{\text{min}}$ cm). }
\end{figure}

\begin{figure}[ht]
\centering
\includegraphics[scale=0.35]{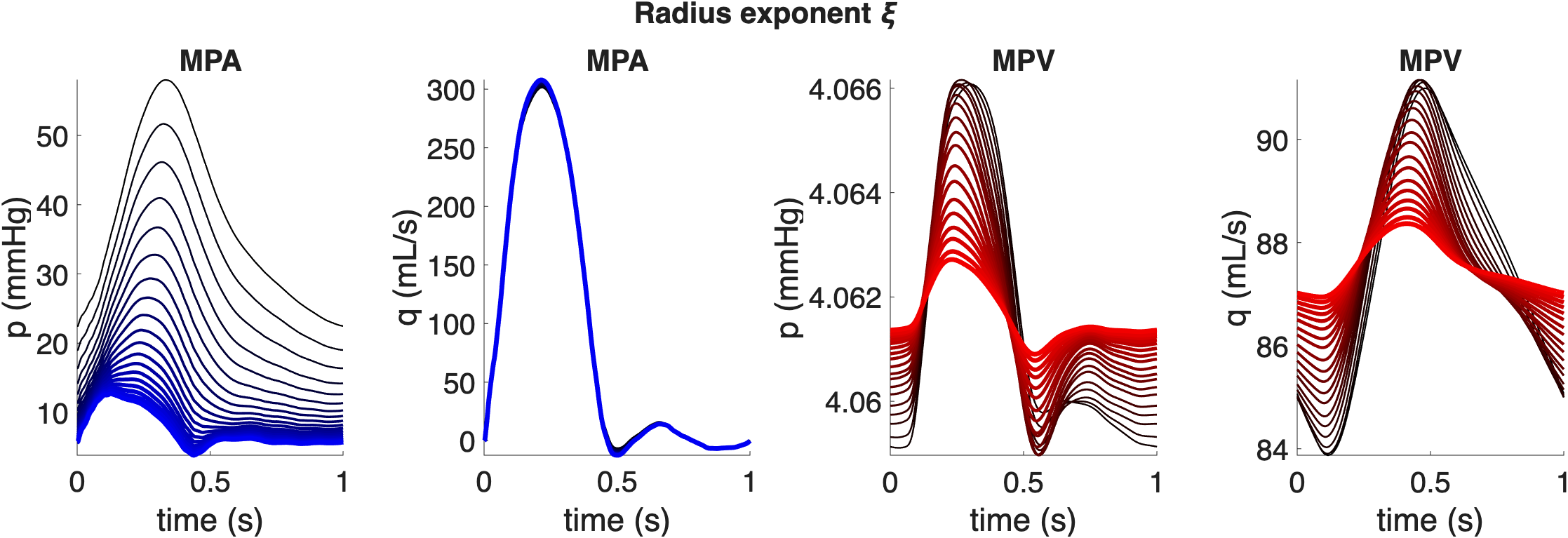}
\caption{Parameter variations: effects on pressure and flow wave forms to variations in the radius exponent $\xi$, indirectly modulating $\alpha$ and $\beta$.  ($\xi = 2.3-3.2$ non.dim.). }
\end{figure}

\begin{figure}[ht]
\centering
\includegraphics[scale=0.35]{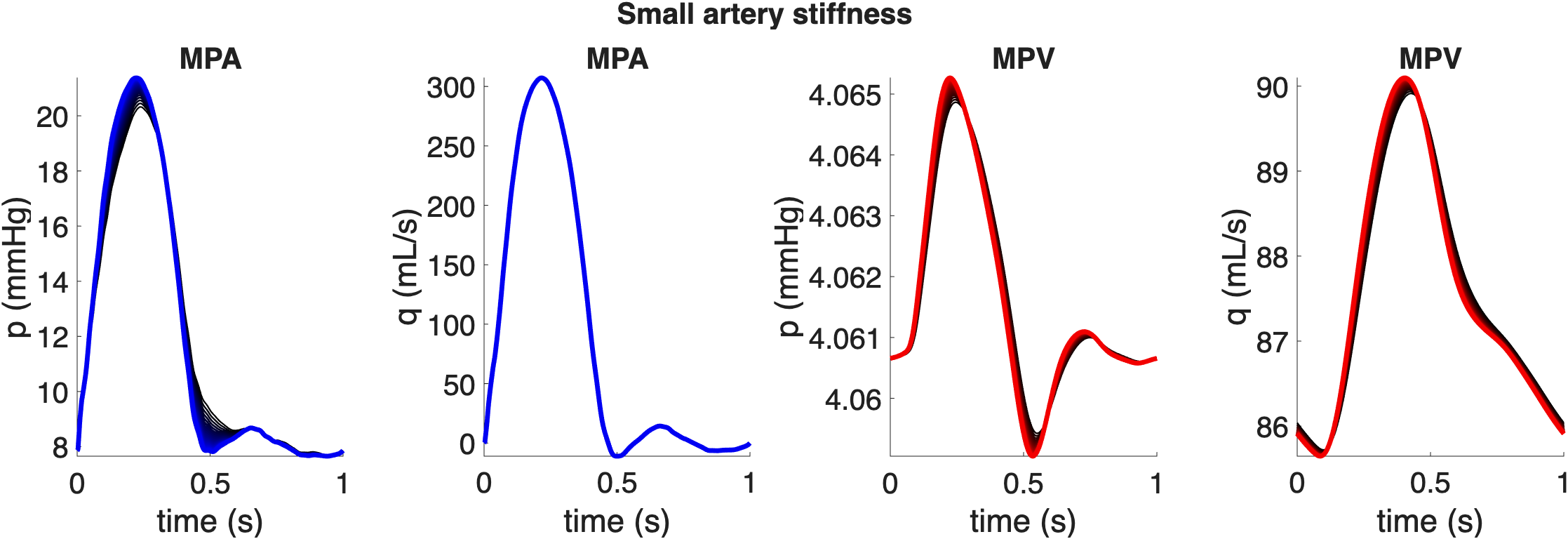}
\caption{Parameter variations: effects on pressure and flow wave forms to variations in small artery stiffness. The scaling factor is applied to both $k_{s1}$ and $k_{s3}$.  (SAstiff$_s$ $= 0.9-2$  non.dim.) giving $k_{s1} = 2.7 - 6 \times 10^5$ g/cm/s$^2$ and $k_{s3} = 0.9 - 2 \times 10^5$ g/cm/s$^2$. }
\end{figure}

\begin{figure}[ht]
\centering
\includegraphics[scale=0.35]{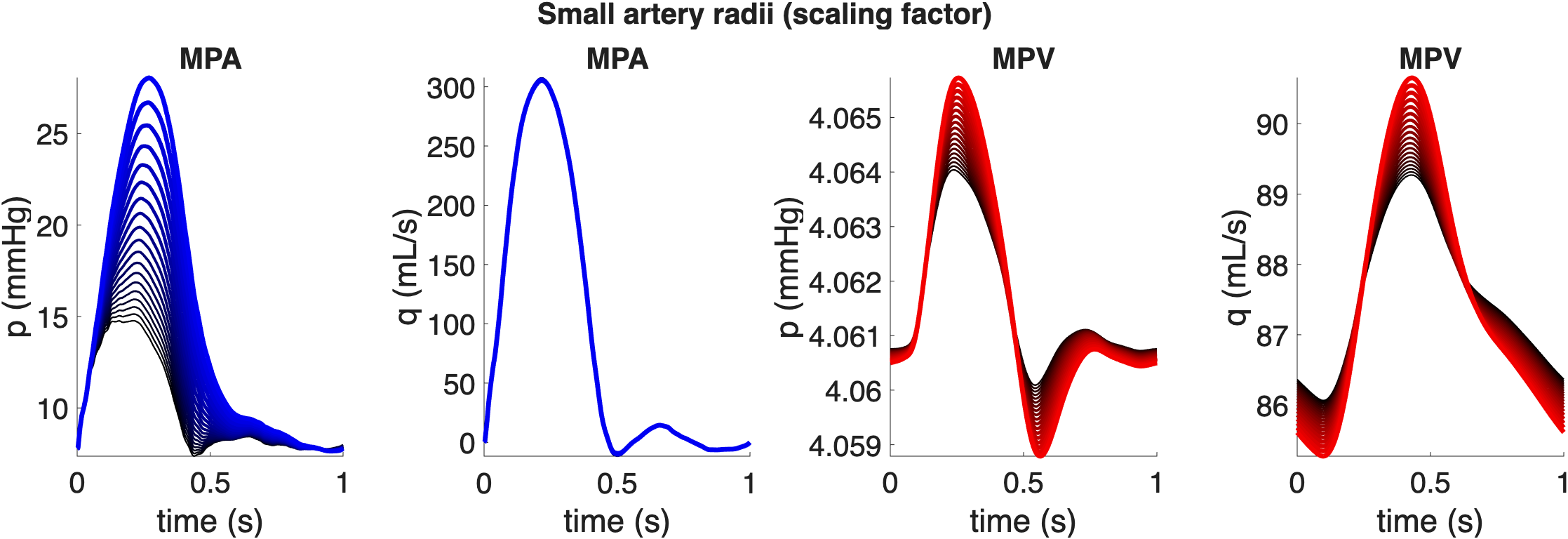}
\caption{Parameter variations: effects on pressure and flow wave forms to variations in 
small artery radii scaling ($r_{sa} = 1.2-0.85$  non.dim.). }
\end{figure}

\begin{figure}[ht]
\centering
\includegraphics[scale=0.35]{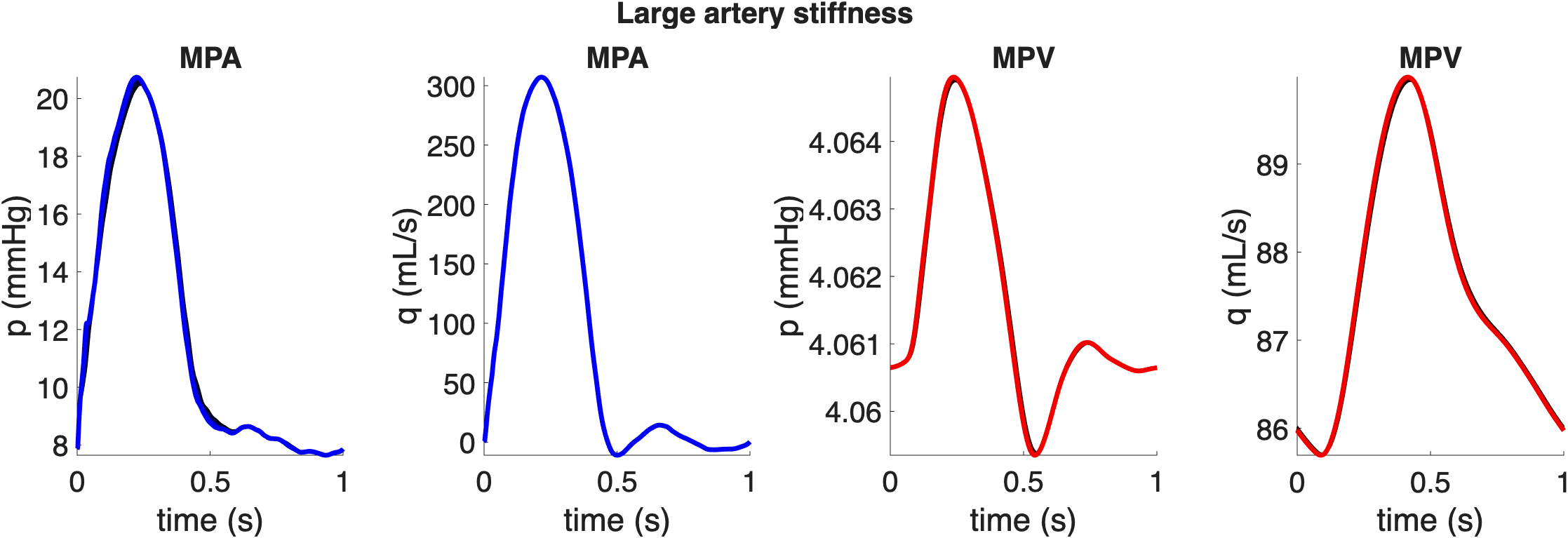}
\caption{Parameter variations: effects on pressure and flow wave forms to variations in 
large artery stiffness. (LAstiff $=3.6-8 \times 10^5$ g/cm/s$^2$). }
\end{figure}

\begin{figure}[ht]
\centering
\includegraphics[scale=0.35]{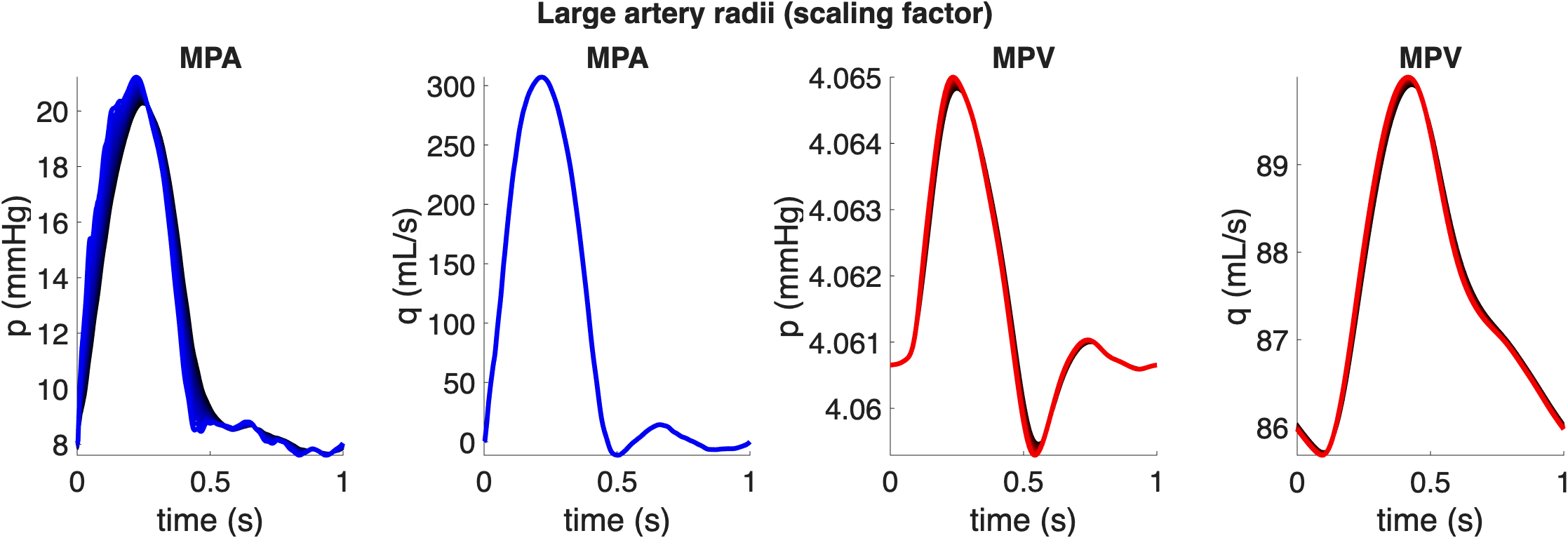}
\caption{Parameter variations: effects on pressure and flow wave forms to variations in large artery radii scaling ($r_a = 1.2-0.7$  non.dim.). }
\end{figure}

\clearpage
\newpage

\section{Parameter variation - effects on mean flow and pressure}
\begin{figure}[ht]
\centering
\includegraphics[scale=0.35]{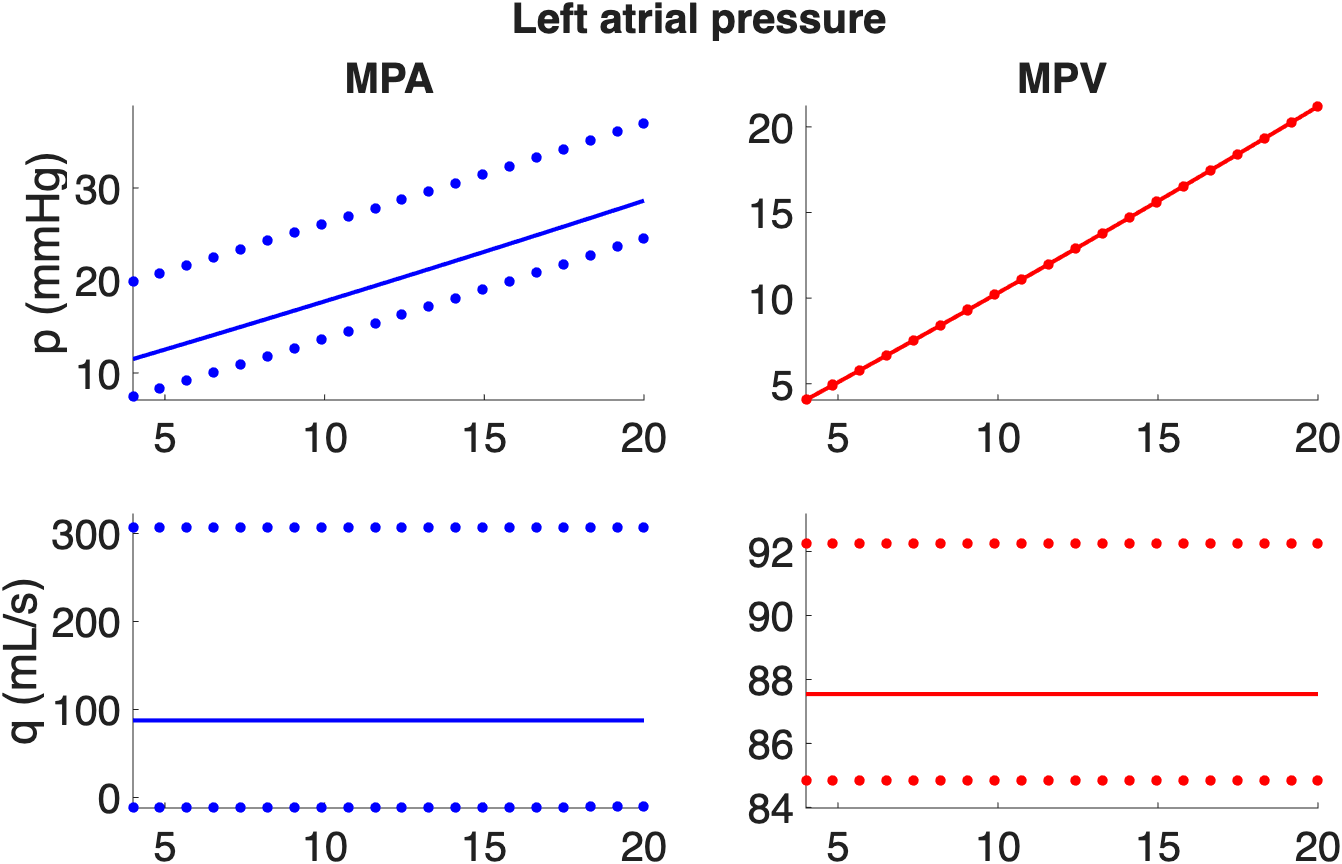}
\caption{Left atrial pressure (LAP $=2-20$ mmHg).} 
\label{fig:LAPmean}
\end{figure}

\begin{figure}[ht]
\centering
\includegraphics[scale=0.35]{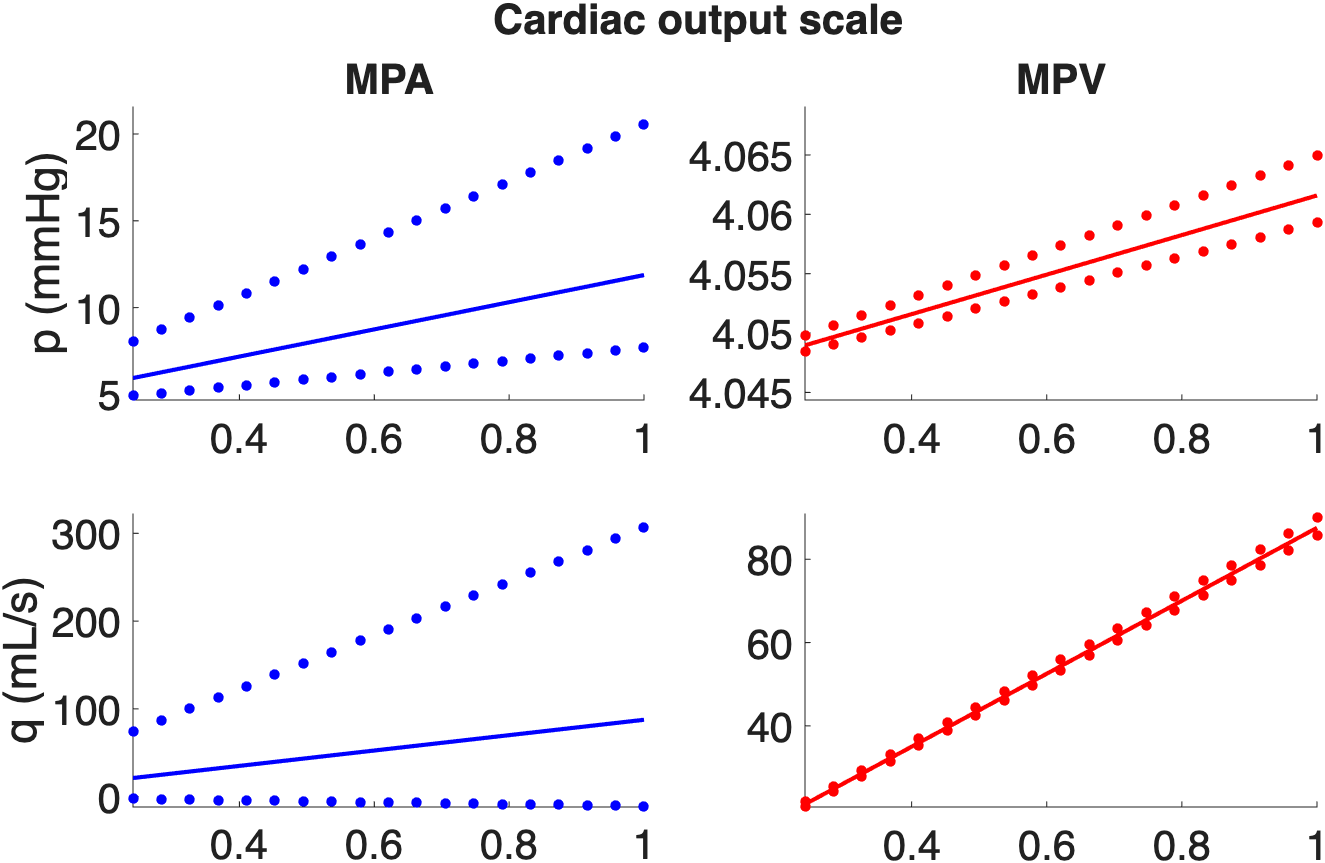}
\caption{Cardiac output scale (CO$_s$ $= 1-0.2$ non.dim)} 
\label{fig:COmean}
\end{figure}

Figures \ref{fig:LAPmean}--\ref{fig:COmean} show impact on mean arterial and venous flow and pressure from varying parameters one-at-the-time. For each figure in the arteries, the solid blue line shows mean value and the dashed lines the pulse values, whereas in the veins mean values are depicted with solid red liens and pulse values by dashed lines.  For each parameter varied, the range over which they are varied is listed in Table 3 (in the main manuscript). Note the cardiac cycle length and fluid dynamics parameters including boundary layer thickness $\delta$, viscosity $\mu$, and density $\rho$ are not varied.


\begin{figure}[ht]
\centering
\includegraphics[scale=0.35]{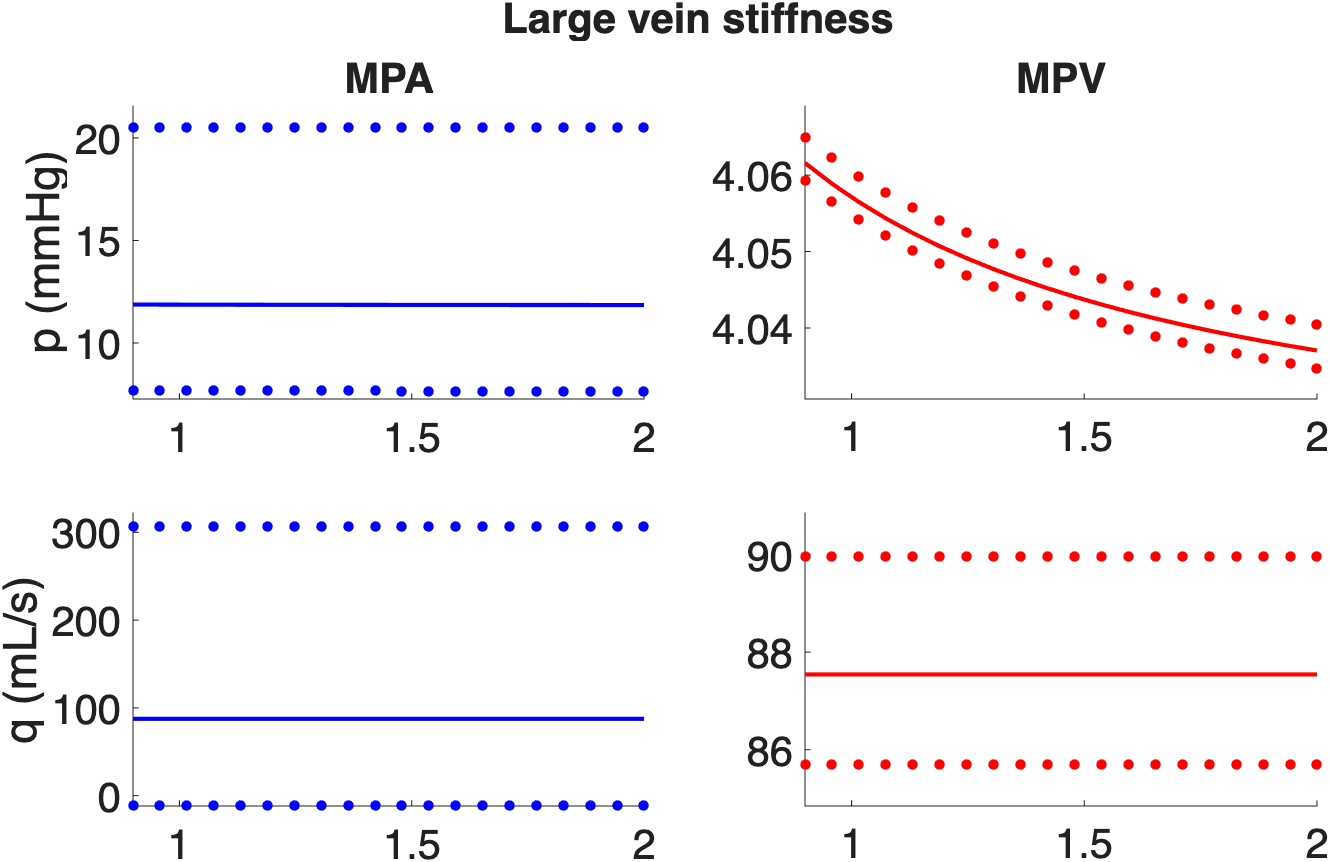}
\caption{Large vein stiffness (LVstiff $= 3.6-8 \times 10^5$ g/cm/s$^2$).} 
\end{figure}  

\begin{figure}[ht]
\centering
\includegraphics[scale=0.35]{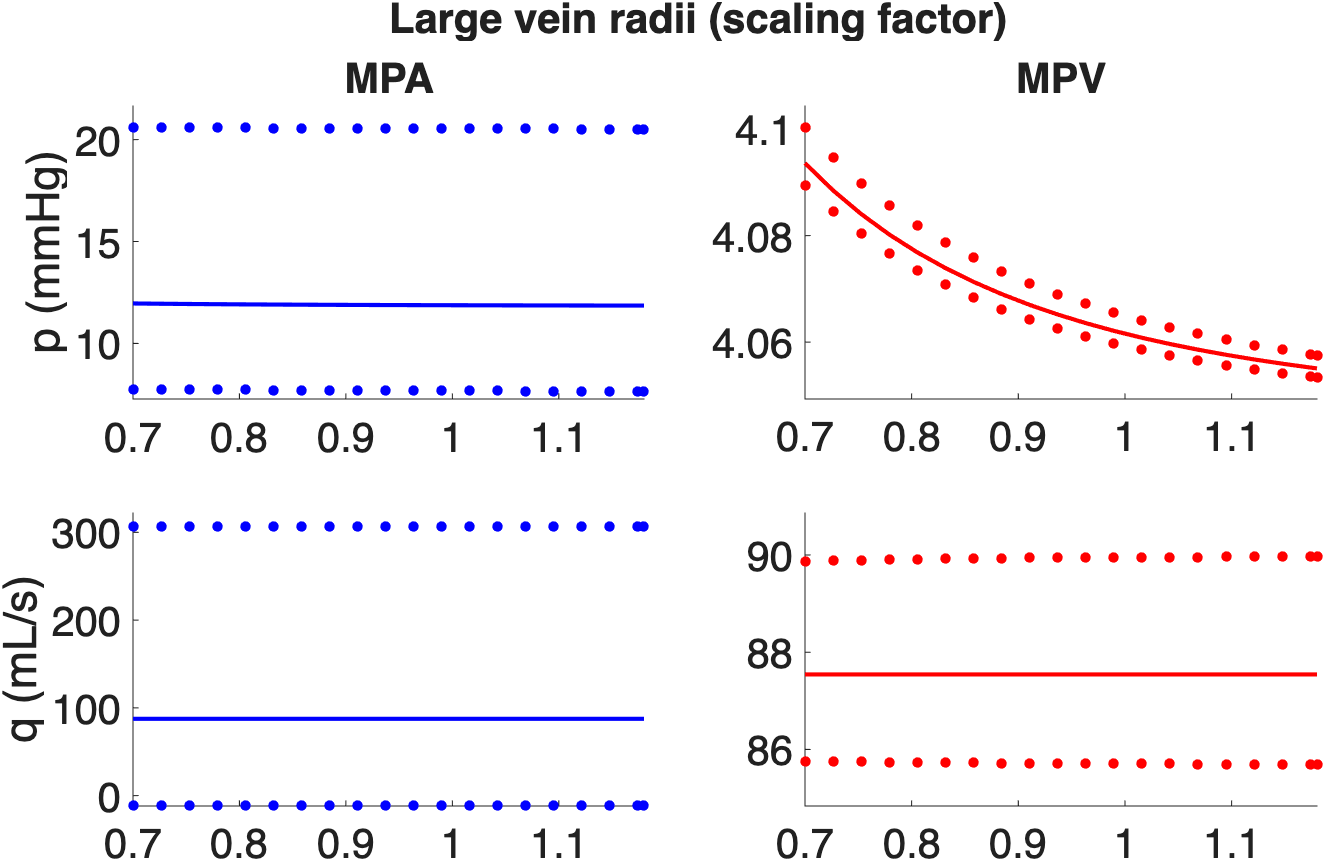}
\caption{Large vein radii scaling factor (rv $= 1.2-0.7$ non.dim.).} 
\end{figure}

\begin{figure}[ht]
\centering
\includegraphics[scale=0.35]
{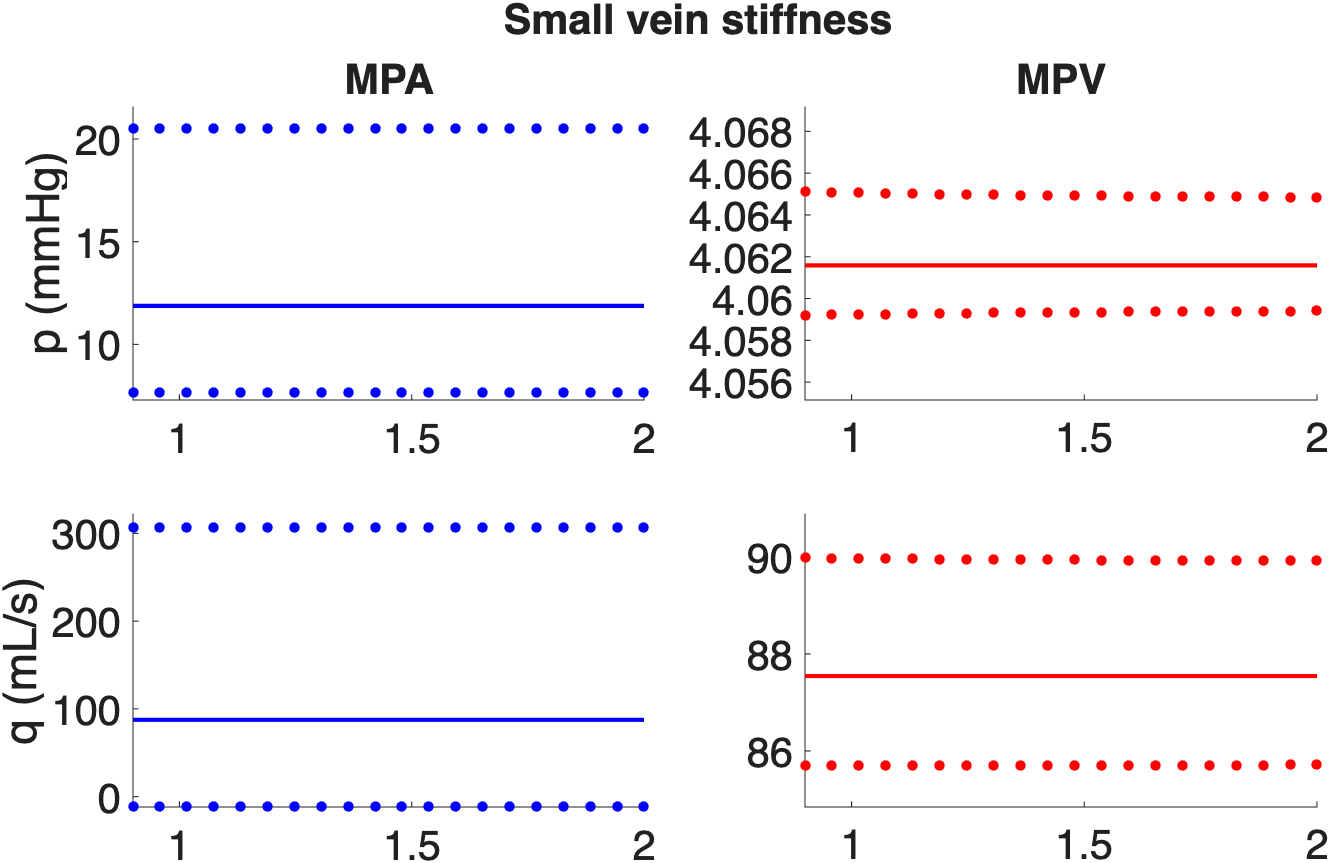}
\caption{Small vein stiffness factor used to multiply both $k_{1s}$ and $k_{3s}$ (SVstiff$_s = 0.9-2$) non.dim.} \end{figure}  

\begin{figure}[ht]
\centering
\includegraphics[scale=0.35]{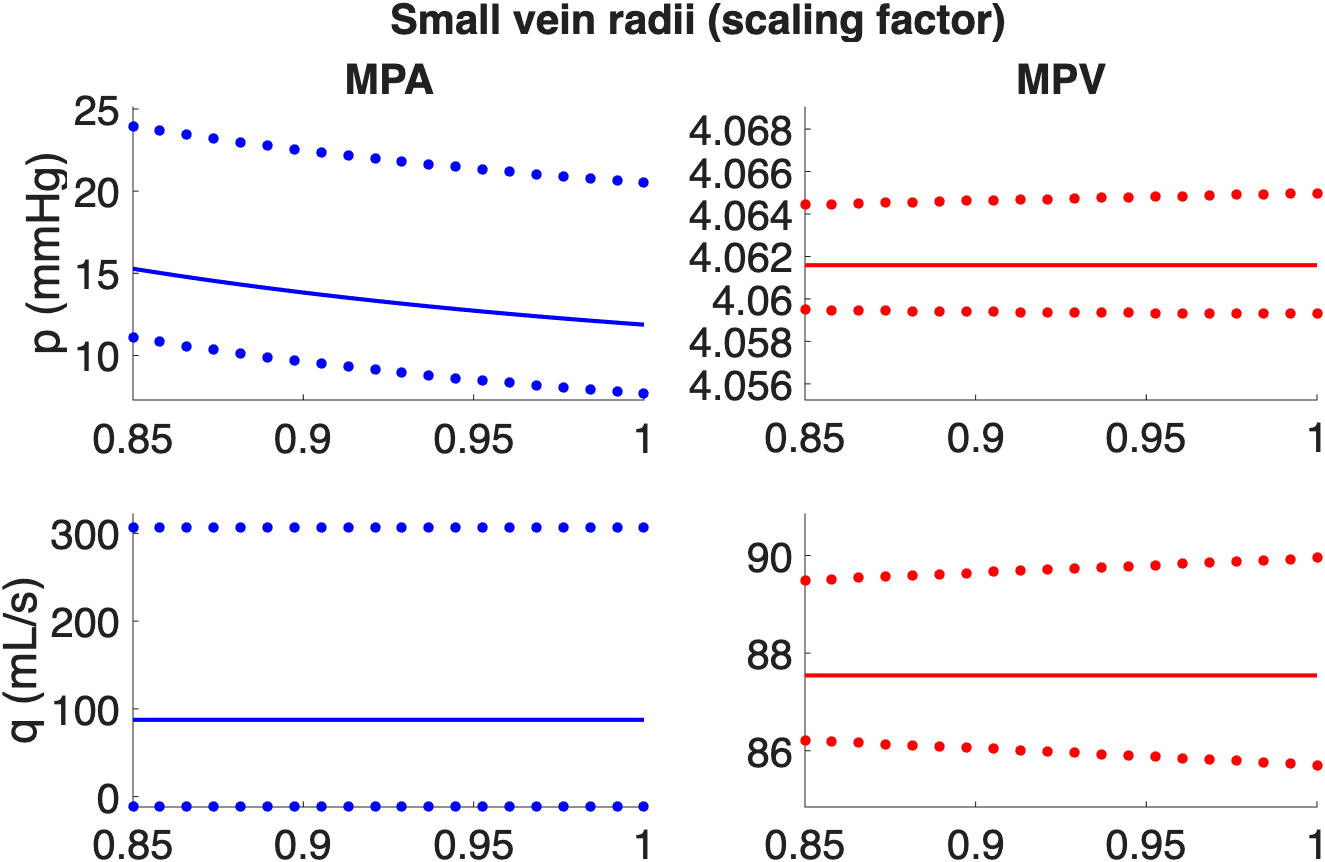}
\caption{Small vein radius scaling  factor (rsv$_s = 1-0.85$ non.dim.).} 
\end{figure} 

\begin{figure}[ht]
\centering
\includegraphics[scale=0.35]{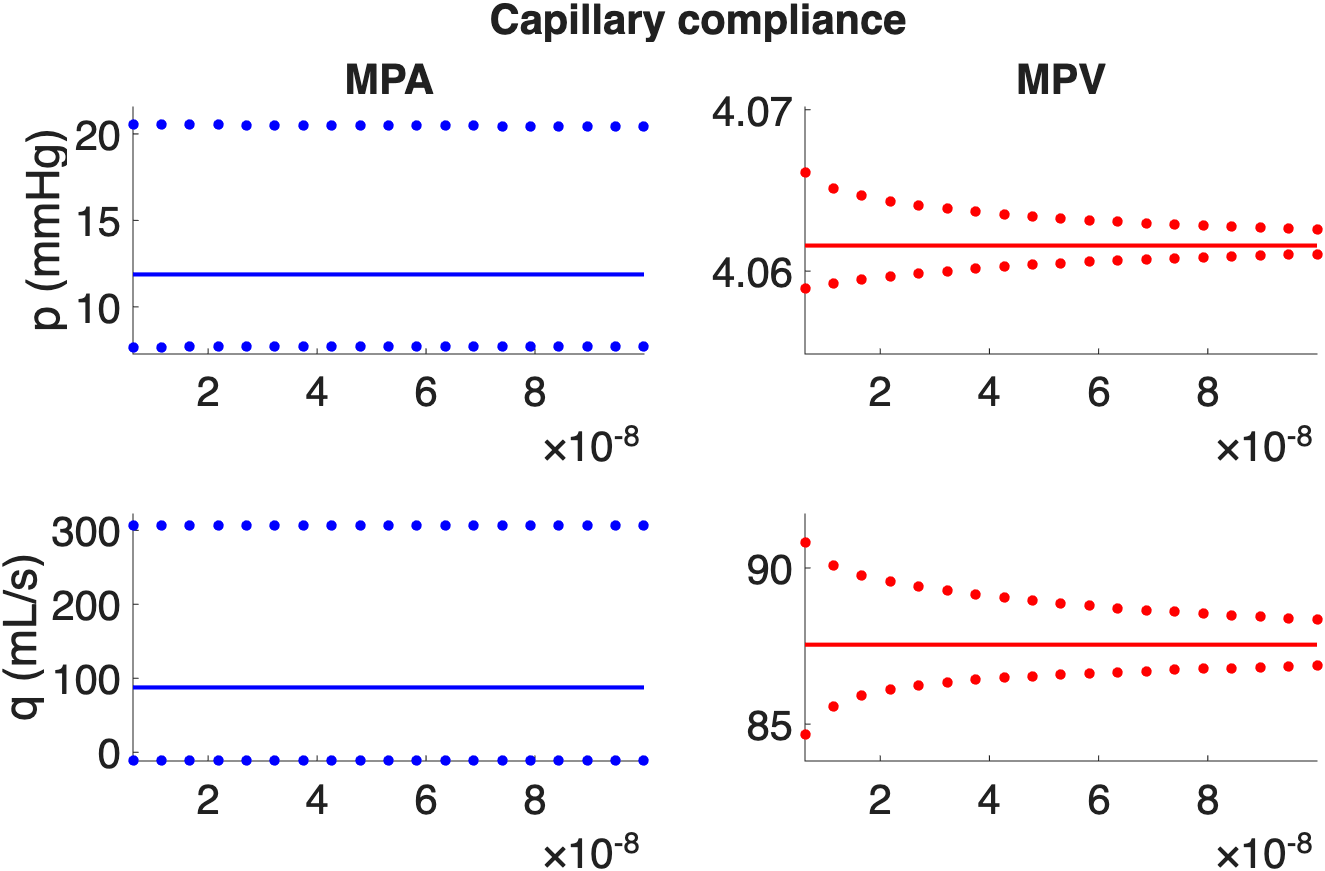}
\caption{Capillary compliance (capComp $= 1 \times 10^{-7}-1 \times 10^{-9}$ (cm s)$^2$/g).} 
\end{figure} 

\begin{figure}[ht]
\centering
\includegraphics[scale=0.35]{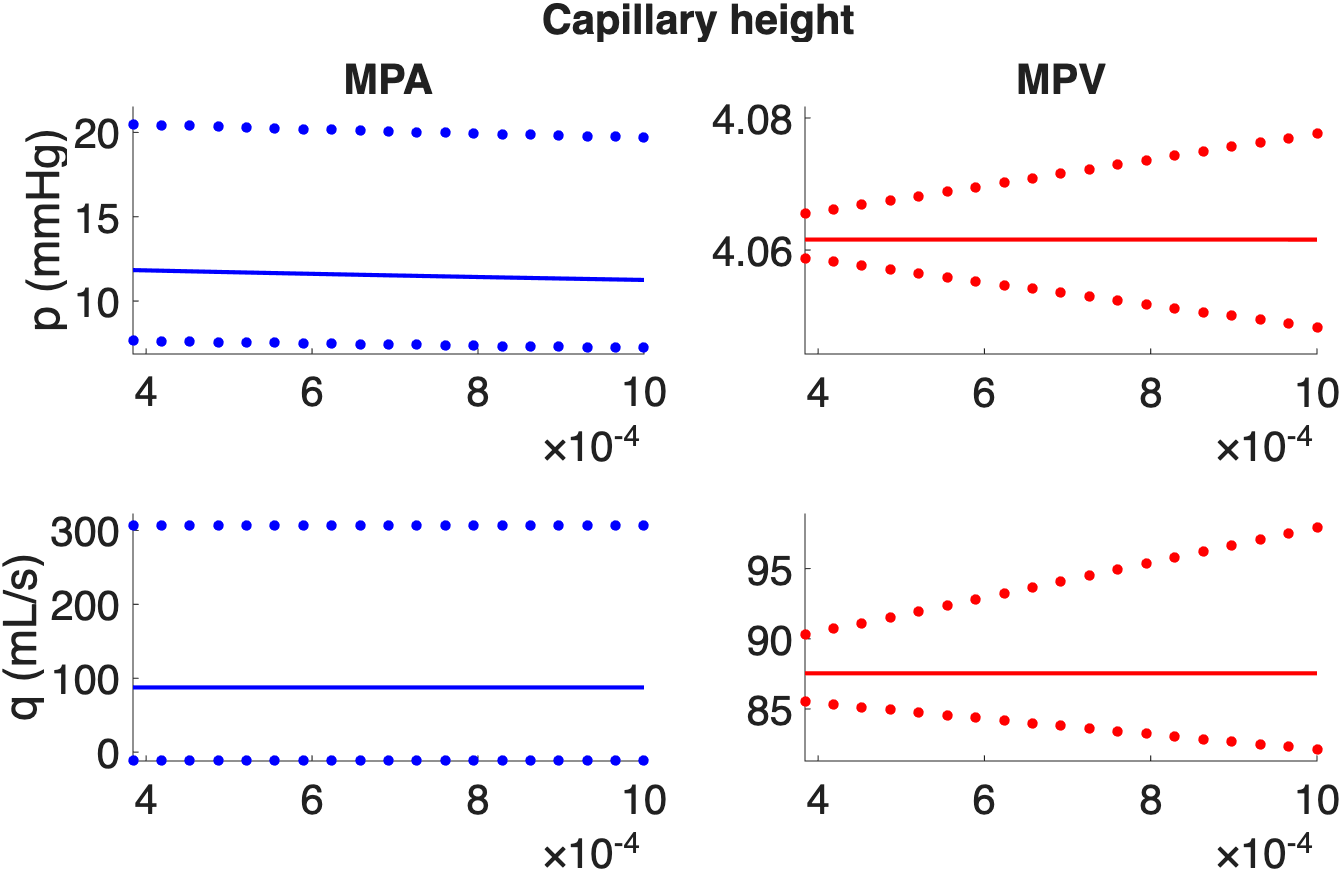}
\caption{Capillary height ($h = 0.00035-0.001$ cm).}
\end{figure} 

\begin{figure}[ht]
\centering
\includegraphics[scale=0.35]{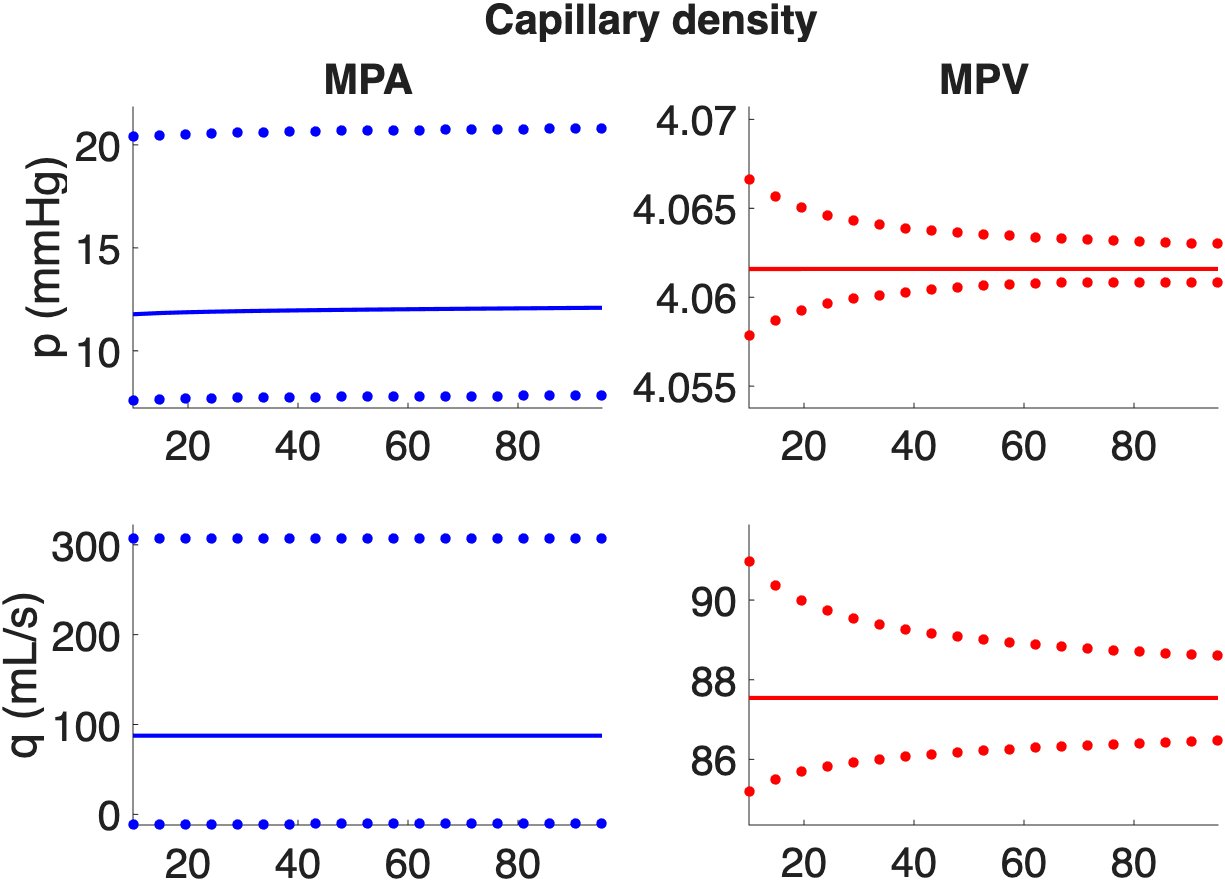}
\caption{Density of capillary posts ($\kappa = 10-100$ non.dim.).}
\end{figure} 

\begin{figure}[ht]
\centering
\includegraphics[scale=0.35]{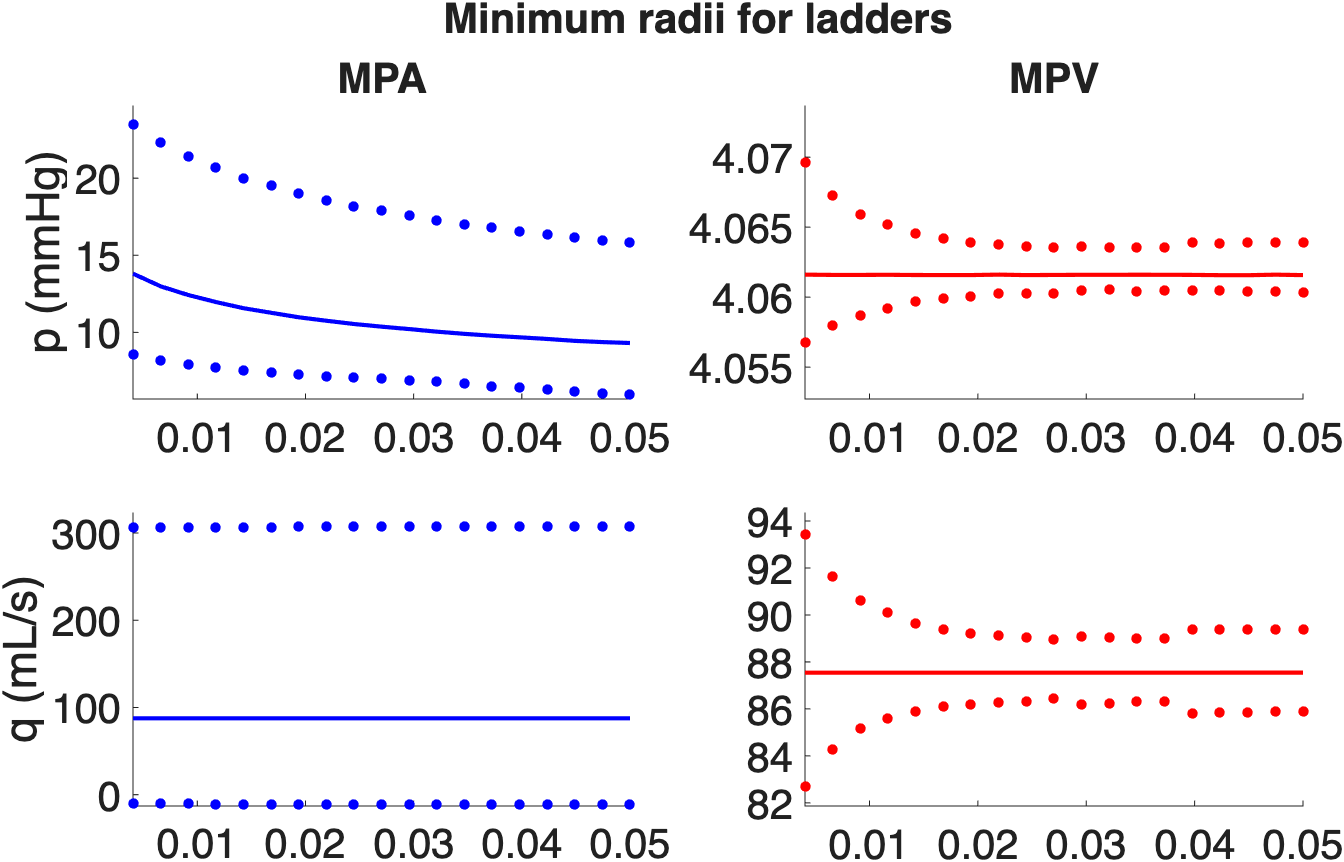}
\caption{Minimum radius at which ladders are connecting arteries and veins ($r_{\text{ladder}} = 100-5 \ r_{\text{min}}$ cm)} 
\end{figure} 

\begin{figure}[ht]
\centering
\includegraphics[scale=0.35]{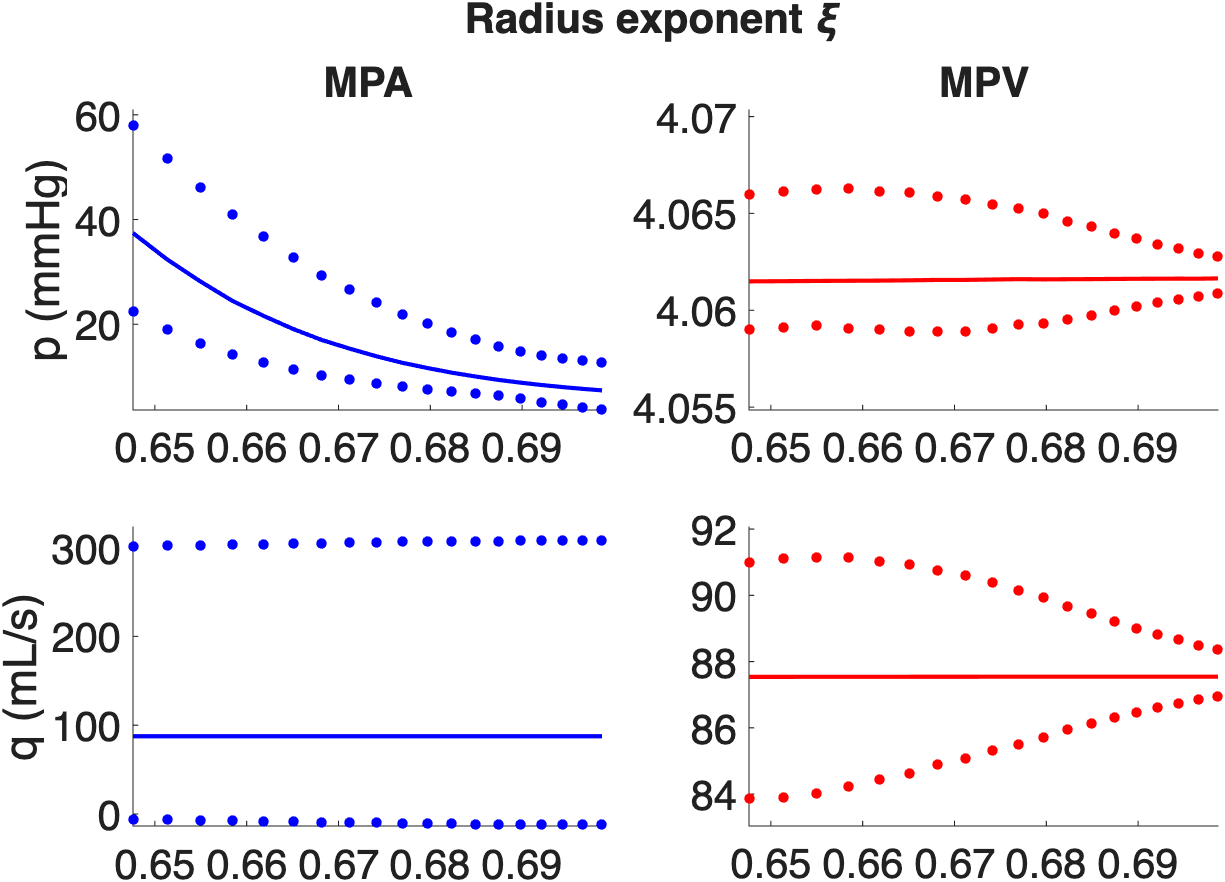}
\caption{Radius exponent $\xi$ ($\xi = 2.3-3.2$ non.dim.)} 
\end{figure} 

\begin{figure}[ht]
\centering
\includegraphics[scale=0.35]{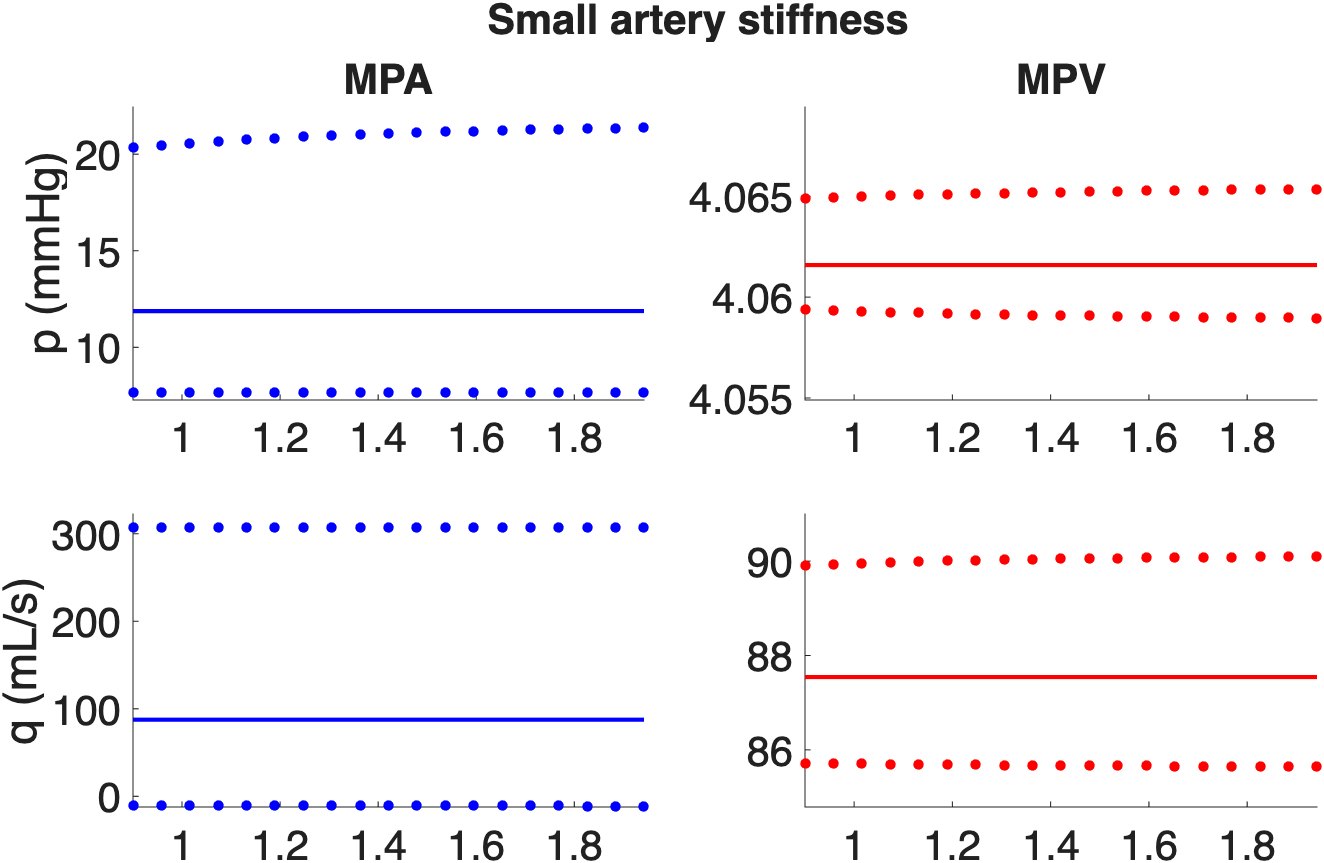}
\caption{Small artery stiffness scaling $k_{s1}$ and $k_{s3}$ (SAstiff $= 0.9-2$ non.dim.).} 
\end{figure} 

\begin{figure}[ht]
\centering
\includegraphics[scale=0.35]{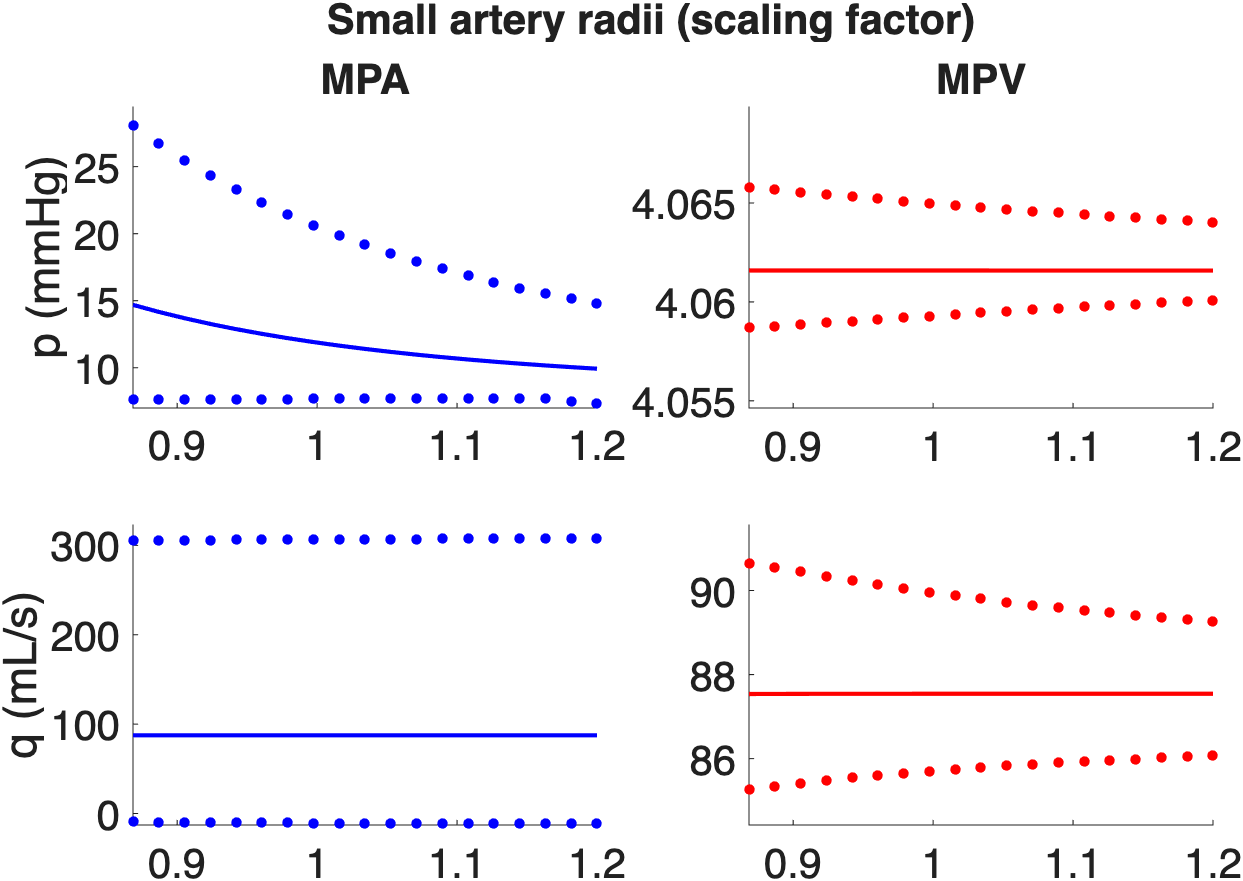}
\caption{Small artery radius scaling factor (rsa $= 1.2-0.5$ non.dim.).} 
\end{figure} 

\begin{figure}[ht]
\centering
\includegraphics[scale=0.35]{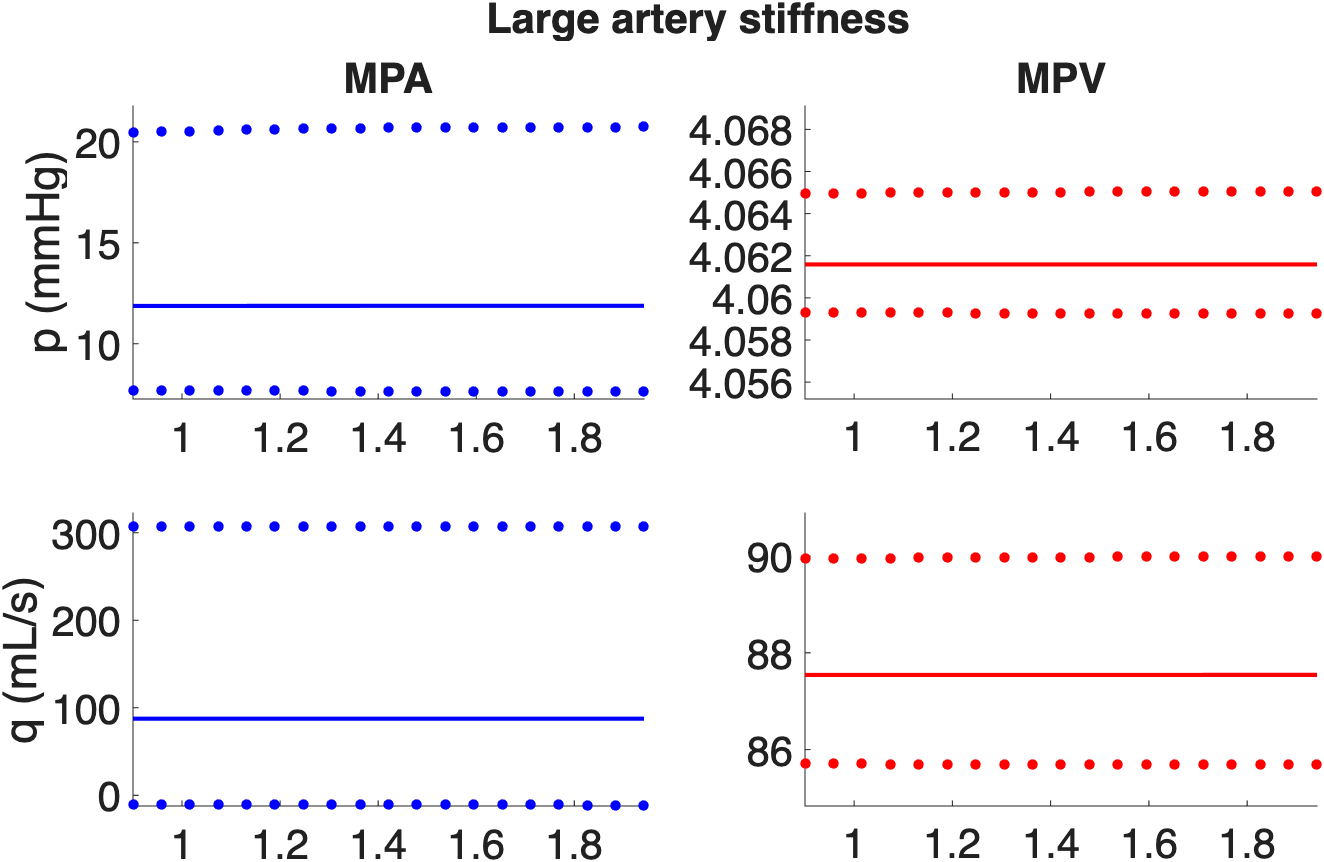}
\caption{Large artery stiffness (LartStiff $= 3.6-8 \times 10^5$ g/cm/s$^2$).} 
\end{figure} 

\begin{figure}[ht]
\centering
\includegraphics[scale=0.35]{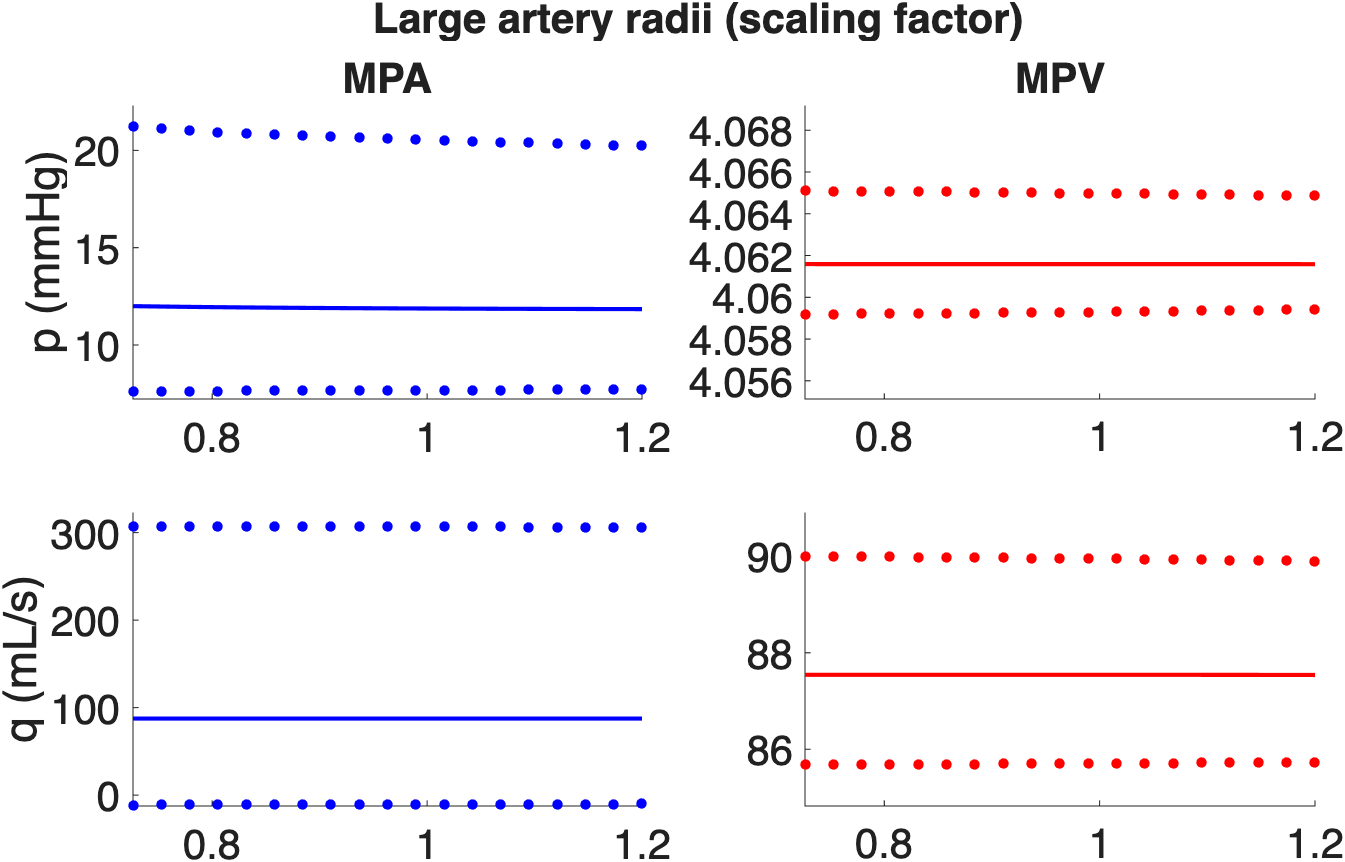}
\caption{Large artery radius scale (ra$_s$ $= 1.2-0.7$ non.dim.).} 
\end{figure} 

